\documentclass[fleqn,usenatbib]{mnras}

\usepackage{newtxtext,newtxmath}

\usepackage[T1]{fontenc}

\DeclareRobustCommand{\VAN}[3]{#2}
\let\VANthebibliography\thebibliography
\def\thebibliography{\DeclareRobustCommand{\VAN}[3]{##3}\VANthebibliography}

\usepackage{graphicx}	
\usepackage{amsmath}	
\usepackage[normalem]{ulem}

\newcommand{\mysim}{\mathord{\sim}}

\newcommand{\myapprox}{\mathord{\approx}}

\defcitealias{Argon_2007}{A07}
\defcitealias{Humphreys_2008}{H08}
\defcitealias{Humphreys_2013}{H13}
\defcitealias{Reid_2019}{R19}
\usepackage{threeparttable}
\usepackage{booktabs}
\usepackage{orcidlink}

\title[Distance to NGC 4258]{The Distance to NGC 4258 from Individual Maser Component Tracking}

\author[van der Boom \& Kushnir]{
Daniella van der Boom$^{\orcidlink{0009-0008-4872-3116} 1}$\thanks{E-mail: daniella.van-der-boom@weizmann.ac.il}
and Doron Kushnir$^{1}$
\\
$^{1}$Dept. of Particle Phys. \& Astrophys., Weizmann Institute of Science, Rehovot 76100, Israel
}

\date{Accepted XXX. Received YYY; in original form ZZZ}

\pubyear{\the\year{}}

\begin{document}
\label{firstpage}
\pagerange{\pageref{firstpage}--\pageref{lastpage}}
\maketitle

\begin{abstract}
We present a reanalysis of the water maser system in NGC 4258 to reassess its geometric distance, commonly reported as $\myapprox7.6\,\, \text{Mpc}$ with percent-level accuracy, a key anchor in extragalactic distance ladder calibrations and recent determinations of the Hubble constant. We introduce a method that relies exclusively on tracking individual maser components, rather than assuming a single averaged trajectory as in previous works, thereby avoiding arbitrary data averaging that can bias interpretations of the disk’s geometry and dynamics. This approach requires spatially resolved individual maser components; consequently, the majority of observational epochs were excluded, as they lack sufficient spatial resolution to localize the maser position. We track individual maser components across multiple epochs and introduce an efficient marginalization method over nuisance parameters (angular radius and azimuth of each maser), reducing the number of free parameters from hundreds to 14. Our analysis reveals that the current observational cadence is insufficient to reliably track the individual masers, which our method relies on, given their intrinsic variability. Across a range of maser selections and model configurations, inferred distances span $\myapprox6.7$ to $\myapprox8.1\,\text{Mpc}$, demonstrating significant sensitivity to how our method selects individual masers. Even visually and statistically robust fits can differ by several standard deviations, reflecting ambiguity in component identification across sparsely sampled epochs. We evaluate the impact of observational cadence on tracking fidelity and distance precision, and show that high-cadence monitoring is needed for our method to track individual masers and produce a robust anchor for cosmology.
\end{abstract}





\section{Introduction}

The recent SH0ES determination of the Hubble constant \citep[$H_0=73.04\pm1.04\,\textrm{km}\,\textrm{s}^{-1}\,\textrm{Mpc}^{-1}$;][]{Riess_2022} deviates significantly, by $\myapprox5\sigma$, from the value inferred by the \textit{Planck} collaboration \citep[$H_0=67.4\pm0.5\,\textrm{km}\,\textrm{s}^{-1}\,\textrm{Mpc}^{-1}$;][]{Planck_2020}, prompting considerable interest in possible extensions to the standard cosmological model \citep[see][for a review]{DiValentino_2021}. The SH0ES distance ladder is based on the period–luminosity relation of  Cepheids \citep[][]{Leavitt_1912}, which are situated in Type Ia supernova host galaxies along with other anchor galaxies with absolute distance measurements, including the Milky Way (MW), Large and Small Magellanic Clouds (LMC and SMC), and NGC 4258. Among these, NGC 4258 is unique in that the same observational methods and reduction techniques are used for observing both the Cepheids in this galaxy and those in the SN host galaxies, thereby minimizing cross-calibration uncertainties. In contrast, comparisons between distant host Cepheids, typically at distances of  $\myapprox15-50\,\textrm{Mpc}$, and much closer anchors like the MW or LMC involve brightness differences exceeding 10 magnitudes. Such a large dynamic range could introduce subtle systematics due to differences in observational techniques or data reduction pipelines.

To mitigate these concerns, \citet{Kushnir_2024} proposed a restricted subset of the SH0ES sample that uses NGC 4258 as the sole geometric anchor, thereby avoiding cross-anchor systematics related to photometric calibration, metallicity corrections, and dust treatment. This approach has a competitive statistical uncertainty of $\mysim$2.5\%, sufficient to preserve a $\mysim$3$\sigma$ tension with the \textit{Planck} value while reducing systematic uncertainty in the Cepheid calibration. However, this method introduces a critical trade-off: the resulting value of $H_0$ becomes heavily dependent on the accuracy of the geometric distance to NGC 4258—the very quantity under investigation in this work. 

The distance to NGC 4258 is determined by observing water megamasers in a nearly edge-on accretion disk surrounding the galaxy’s central supermassive black hole. Since an efficient maser amplification is only obtained in regions of the disk where line-of-sight (LOS) velocity coherence is maintained, masers are observed either near the galaxy's systemic velocity of $\mysim470\,\mathrm{km}\,\textrm{s}^{-1}$ \citep[\textit{systemic masers};][]{Claussen_1984} or at high  Doppler shifts of $\mysim\pm1000\,\mathrm{km}\,\mathrm{s}^{-1}$ relative systemic \citep[\textit{high-velocity masers};][]{Nakai_1993}. 

The systemic masers exhibit a linear relationship between LOS velocity and projected position, consistent with circular Keplerian orbits in a narrow annular region. Crucially, systemic masers also display measurable LOS accelerations, which enable a geometric determination of the galaxy’s distance. This is because the degeneracy between black hole mass ($M_{\mathrm{BH}}$) and distance ($D$) in Keplerian motion, where velocity scales as $v \propto \sqrt{M_{\mathrm{BH}}/D}$ and acceleration as $a \propto M_{\mathrm{BH}}/D^2$, can be broken by simultaneously fitting velocities and accelerations. However, because amplification is an exponential process, even small fluctuations in the seed radiation can lead to large variations in the observed maser intensity and spectral structure, complicating the identification and tracking of individual components across epochs.

The high-velocity masers are located symmetrically along the disk’s midline. These features arise from gas with the largest LOS component of the orbital velocity. Because their impact parameter is approximately equal to the orbital radius, they exhibit negligible LOS acceleration. Although they do not contribute directly to resolving the $M_{\mathrm{BH}}$–$D$ degeneracy, high-velocity masers provide strong constraints on the disk geometry and orientation. Together, the systemic and high-velocity maser features enable a direct, geometric measurement of the distance to NGC 4258.

The first direct confirmation that masers trace rotational motion came from VLBI imaging by \citet{Greenhill_1995a}, based on 1984 observations, which revealed that the masers lie in a thin, edge-on disk and follow a Keplerian rotation curve. Subsequent studies \citep[e.g.,][]{Haschick_1994, Greenhill_1995b, Nakai_1995, WatsonWallin_1994} measured LOS accelerations, enabling the first geometric distance estimates to NGC 4258. \citet{Herrnstein_1999} provided the first robust geometric distance determination by combining measurements of maser sky positions, LOS velocities, and accelerations, yielding a distance of $7.2 \pm 0.3\,\text{Mpc}$. VLBA observations conducted between 1994 and 1997, at 4–8 month intervals, monitored 20–35 systemic masers per epoch. However, individual maser tracking was limited by their dense clustering in both position and velocity space. Instead, \citet{Herrnstein_1999} inferred the bulk rotational motion and acceleration by collectively analyzing the maser ensemble. This analysis calculates the geometric distance to NGC 4258 by applying a warped disk model to position and velocity data to infer the underlying disk parameters with Bayesian techniques \citep{Herrnstein_1997phd}. A comprehensive review of these foundational results and their implications is provided in \citet{Moran_2000}.

\citet[][hereafter H13]{Humphreys_2013} determined a geometric distance of $7.60 \pm 0.17 \,\text{(stat.)} \pm 0.15\, \text{(sys.)} \, \text{Mpc}$ using VLBA observations from 1997 to 2000, originally presented in \citet[][hereafter A07]{Argon_2007}. The method for measuring LOS accelerations was described in \citet[][hereafter H08]{Humphreys_2008} and conducted over a ten-year period, primarily utilizing data from \citetalias{Argon_2007} and \cite{Bragg_2000}. In \citetalias{Humphreys_2013}, the angular radius ($r$) and azimuth ($\phi$) of each maser spot were treated as free parameters in a global disk model fit using a Metropolis–Hastings MCMC algorithm. \citet{Riess_2016} repeated the fit with a longer MCMC chain, increasing the number of samples from $10^{7}$ to $10^{9}$ in three independent strands (starts at $0\%$ and $\pm10\%$ of the \citetalias{Humphreys_2013} results). This modification slightly changed the inferred distance and reduced sensitivity to the “initial-conditions” systematic, yielding $D = 7.54 \pm 0.17\,\text{(stat.)} \pm 0.010\,\text{(sys.)}\,\text{Mpc}$. \citet[][hereafter R19]{Reid_2019} further refined the technique and dataset by introducing error floors as additional free parameters alongside the disk parameters. This enhancement significantly reduced both statistical and systematic uncertainties, yielding a distance of $7.576 \pm 0.082 \, \text{(stat.)} \pm 0.076\,\text{(sys.)} \,\, \text{Mpc}$. This measurement is widely regarded as the most reliable geometric distance to NGC~4258 currently available and is frequently adopted in determinations of $H_0$.

While the claimed $\mysim1\%$ statistical accuracy on the distance to NGC~4258 is impressively small, it should be interpreted with care in light of the observational challenges involved. First, the maser disk is observed nearly edge-on, which constrains our knowledge of the location of masers in the disk. Second, only a small fraction, $\mysim1\%$, of each maser's $\myapprox800$-year orbital period is observed. This limited orbital coverage severely constrains the ability to precisely constrain all orbital parameters. For comparison, even for the Galactic Center's supermassive black hole, where full stellar orbits are resolved, a $\myapprox3\%$ ($\myapprox3.6\sigma$) discrepancy exists between independent distance estimations of $8178\pm13\,(\text{stat.})\pm22\,(\text{sys.})\, \text{pc}$  
\citep[][]{GRAVITY_2019} and $7946\,\pm50\,(\text{stat.})\pm32\,(\text{sys.})\,\text{pc}$ \citep[][]{Tuan_2019}.  

In this work, we present a reanalysis of the water maser system in NGC 4258 to revisit its geometric distance. While previous studies have relied on averaging heterogeneous observations, our approach instead tracks individual maser components to maintain physical consistency. We find that the existing observational cadence is insufficient for robust identification across epochs. We focus on the determination of statistical uncertainties; therefore, throughout this work, any reference to "uncertainties" or "errors" refers to statistical uncertainties unless explicitly stated otherwise. 
Note that the statistical uncertainties reported from \citetalias{Humphreys_2013} and \citetalias{Reid_2019} include the uncertainty inflation applied in those studies via their respective $\sqrt{\chi_\nu^2}$ factors.
 
We begin by reviewing previous distance determinations, particularly those by \citetalias{Humphreys_2013} and \citetalias{Reid_2019} (Section \ref{sec:related_work}). We reproduce their fits under various configurations using our new fitting framework, which reduces the dimensionality of the previous model, thereby validating our parameter estimation method. We show in Section~\ref{sec:caveats} that previous analyses introduced a critical simplification---the single-orbit approximation---which is neither supported by the raw data nor justified by physical modeling. It also ignores line broadening of maser features with full width at half maximum up to $2~\text{km s}^{-1}$ \citep{Varshalovich_2006}, which may result in oversampled data and underestimated uncertainties. To overcome these issues, we construct a new dataset (Section~\ref{sec:D_estimation}) by directly identifying and tracking individual maser components, explicitly avoiding arbitrary averaging and the assumptions it entails. This revised dataset incorporates measured linewidths into the error model, assigns a unique position to each maser feature, and maintains temporal consistency across epochs. Additionally, we introduce an efficient method to marginalize over the nuisance parameters (the angular radius and azimuth within the disk of each maser), reducing the number of free parameters from many hundreds to 14. This reduction significantly improves sampling efficiency and enhances convergence reliability.

We explore various combinations of maser selection criteria and model fitting parameters. Configurations that meet a stringent goodness-of-fit threshold yield inferred distances spanning $\myapprox6.7$ to $\myapprox8.1\,\text{Mpc}$. These results highlight the sensitivity of the inferred disk geometry and distance estimates to selection methodology. Even fits that are statistically robust (with associated uncertainties between $\myapprox0.07$ and $\myapprox0.19\,\rm{Mpc}$) and visually convincing exhibit differences of several $\sigma$ across setups--underscoring the intrinsic ambiguity of maser tracking over extended and sparsely sampled time baselines. To mitigate this sensitivity, we assess the impact of observational cadence on tracking fidelity (Section~\ref{sec:obs_suggestion}). Our analysis suggests that monitoring 10 systemic maser components over $\mysim40$ epochs with a 10-day cadence (spanning $\mysim14$ months) would enable distance measurements with $\mysim$2\% uncertainty. We summarize and contextualize our findings in Section~\ref{sec:summary}.

\section{Available Observations and Modern Distance Measurements} 
\label{sec:related_work}

This section reviews the observational data, dataset construction, modeling framework, and fitting results used by \citetalias{Humphreys_2013} and \citetalias{Reid_2019} to estimate the geometric distance to NGC 4258. These studies draw on a decade of observations, spanning 1994 to 2004, collected through a range of radio observatories and observing programs. Section~\ref{subsec:Observations_overview} summarizes the key datasets employed in their analyses. Section~\ref{subsec:binned_dataset} describes how \citetalias{Humphreys_2013} and \citetalias{Reid_2019} combine these observations into a single averaged dataset, along with the assumptions underlying this approach. The warped disk model and fitting procedure used to derive distance estimates are outlined in Sections~\ref{subsec:disk_model} and~\ref{subsec:fitting}, respectively. Finally, in Section~\ref{subsec:reproduced results}, we reproduce their results with the same dataset, primarily to validate our fitting scheme.

\subsection{Observations overview} 
\label{subsec:Observations_overview}

Table~\ref{tab:obs_years} summarizes the primary observational datasets used in \citetalias{Humphreys_2013} and \citetalias{Reid_2019}, including program codes, instruments, channel spacing, sensitivities, and references. The observations span the period from 1994 to 2004 and comprise multi-epoch campaigns conducted with the VLBA, VLA, Effelsberg 100-m telescope, and the Green Bank Telescope (GBT). Many epochs combine data from multiple facilities to achieve both high spatial resolution and sensitivity. Raw VLBA and VLA data are publicly available from the NRAO Science Data Archive\footnote{\url{https://data.nrao.edu}}. In addition, post-processed data corresponding to entries 7–10 in Table~\ref{tab:obs_years} are provided in Table 5 of \citetalias{Argon_2007}. These observations offer significantly improved sensitivity---down to a few mJy---enabling the spatial localization of maser emission within the sub-parsec disk (see below), in addition to the construction of integrated spectra across the full velocity range, which all datasets permit. This high-quality dataset forms the basis for the subsequent analysis presented in Section~\ref{sec:D_estimation}\footnote{\label{fn:public_data}Entry 11 in Table~\ref{tab:obs_years}, consisting of three high-sensitivity GBT epochs from 2003 April 10 to December 8, is excluded from our analysis due to the lack of publicly available, post-processed data and to preserve homogeneity in sensitivity and instrumental setup across the dataset.}.

\textbf{A07 Dataset:} \citetalias{Argon_2007} presents position and velocity measurements of maser spots in the sub-parsec disk of NGC 4258, based on 18 epochs observed between March 6, 1997, and August 12, 2000 (entries 7–10 in Table~\ref{tab:obs_years}). The observations span a velocity range of $\mysim2000\,\mathrm{km}\,\mathrm{s}^{-1}$, divided across several overlapping basebands. While these overlaps ensure continuous coverage of the systemic, redshifted, and blueshifted maser components, they introduce alignment challenges that are addressed in Section~\ref{subsubsec:identify_peaks}. For each epoch, the data were divided into velocity channels, each yielding a two-dimensional ($XY$) sky map. Maser spots were identified by fitting Gaussian profiles to bright, compact features in these maps. Although the fitting procedure permitted multiple Gaussian components per channel to account for possible spectral blending, inspection of the published \citetalias{Argon_2007} data shows that, in practice, each velocity channel in each epoch contains at most one spatial position. To minimize false detections, only peaks exceeding a $6\sigma$ threshold were retained if they appeared in at least three consecutive channels and had fitted sizes smaller than twice the synthesized beam. This procedure resulted in the identification of 14,291 maser spots, which are fully characterized in Table 5 in \citetalias{Argon_2007}: for each spot the table gives the spectral‐channel index, classical and relativistic LSR velocities, fitted peak flux density and its uncertainty, east–west and north–south offsets from the $510\, \text{km s}^{-1}$ (relativistic) reference spot (with uncertainties), the fitted Gaussian full width at half maximum (FWHM) along its major and minor axes (and uncertainties), the Gaussian position angle (and its uncertainty), and the baseband idntifier.

\begin{table*}
	\centering
	\caption{Summary of Primary Observations Since 1994 for Maser Identification, adapted from \citetalias{Humphreys_2008}}
	\label{tab:obs_years}
	\begin{tabular}{cllcccc} 
		\hline
		& Program Code & Dates (\# Epochs) & Instrument(s) & $\Delta v$ [$\mathrm{km}\,\mathrm{s}^{-1}$] & Sensitivity [mJy per $\Delta v$] & References \\
		\hline
		1 & BM19 & 1994 Apr 19 & B & 0.21 & 43 & [1], [2], [3], [5] \\
		2 & AG448 & 1995 Jan 7 -- 1997 Feb 10 (17) & A & 0.33 & 15--60 & [2], [5] \\
		3 & BM36a,b & 1995 Jan 8, May 29 & B & 0.21 & 30, 80 & [1], [2], [3], [5] \\
		4 & ... & 1995 Mar 16 -- Jun 25 (5) & E & 0.33 & 85--110  & [1], [2], [5] \\
		5 & BM56a,b & 1996 Feb 22, Sep 21 & B & 0.21 & 20, 25 & [2], [3], [5] \\
		6 & BH25.x & 1997 Mar 07 -- Apr 07 (3) & B + A + E + G & 0.44 & ... & [3] \\ 
		\textsuperscript{\textdagger}7 & BM56c & 1997 Mar 6 & B + A + E & 0.21 & 4.7 & [4], [5] \\
		\textsuperscript{\textdagger}8 & BM81a,b & 1997 Oct 1, 1998 Jan 27 & B + A + E & 0.21 & 4.1, 5.0  & [4], [5] \\
		\textsuperscript{\textdagger}9 & BM112.x & 1998 Sep 5 -- 2000 May 4  (16) & B + A + E (2); B (14) & 0.21, 0.42 & 3.5--6.3 & [4], [5] \\
		\textsuperscript{\textdagger}10 & BG107 & 2000 Aug 12 & B + A + E & 0.42 & 7.2 & [4], [5] \\
		11 & ... & 2003 Apr 10 -- Dec 8 (3) & G & 0.21 & 2.9 & [5] \\
		12 & AH847 & 2004 May 21 & A & 0.21 & 20 & [5] \\
		\hline
	\end{tabular}
    
    \medskip
    \raggedright
    \small
    \textit{Notes.} Instruments: B = VLBA, A = VLA, E = Effelsberg 100 m, G = GBT. \\
    Program codes, dates, channel spacing ($\Delta v$), and sensitivity: taken directly from reference [5], except for entry 6, taken directly from [3]. The notation '.x' in program codes denotes multiple sub-codes (e.g., BM112a, BM112b, etc.) that were used as part of the observing campaign.  \\
    References: 
    [1] \cite{Herrnstein_1997phd};
    [2] \cite{Bragg_2000};  
    [3] \cite{Herrnstein_2005};  
    [4] \citetalias{Argon_2007};  
    [5] \citetalias{Humphreys_2008} \\ 
    \textsuperscript{\textdagger} Used in the subsequent analysis presented in Section~\ref{sec:D_estimation}
\end{table*}

\subsection{The Construction of an Averaged Dataset (The Single-Orbit Assumption)} 
\label{subsec:binned_dataset}

As described above, \citetalias{Argon_2007} provides position and velocity measurements of individual maser spots. Independently, \citetalias{Humphreys_2008} presents velocity and acceleration measurements derived from the integrated spectra of a broader set of observations encompassing all entries in Table~\ref{tab:obs_years}. Since these two datasets were constructed separately, \citetalias{Humphreys_2013} combines them by binning the data in velocity space using a bin width of $\Delta v=1\,\mathrm{km}\,\mathrm{s}^{-1}$. This yields a unified dataset of positions, velocities, and accelerations---referred to hereafter as the \textit{averaged dataset}---which is used in the disk modeling. This procedure effectively imposes a single-orbit assumption, in which each velocity is associated with a unique position and acceleration; this assumption is discussed further in Section~\ref{subsec:single_orbit}. Moreover, this process does not account for the intrinsic line widths of the maser features. When the bin width is smaller than the line broadening, and the non-independence of the signal between adjacent bins is not explicitly modeled, this results in spectral oversampling and an overestimation of measurement precision. This issue is discussed further in Section~\ref{subsec:thermal_width}. Below, we describe the construction of the \citetalias{Humphreys_2008} dataset and the averaged dataset in greater detail.

\begin{itemize}

\item \textbf{H08 Dataset -- acceleration-velocity relation:} Acceleration measurements reported in \citetalias{Humphreys_2008} are derived from 51 epochs collected between 1994 and 2004, covering the full set of observations listed in Table~\ref{tab:obs_years}, and relying primarily on data from \citet{Bragg_2000} and \citetalias{Argon_2007}. The integral spectrum in each epoch was decomposed into Gaussian line profiles, and the drift in centroid velocity between epochs was used to determine acceleration. Each maser component was characterized at each epoch by its Doppler velocity, linear velocity drift, line width, and amplitude. The number of Gaussian components was constrained such that the residuals remained below the $5\sigma$ level. Accelerations for the high-velocity maser features fall within the range $-0.7$ to $+0.7\,\mathrm{km}\,\mathrm{s}^{-1}\,\mathrm{yr}^{-1}$, while systemic features exhibit accelerations between $6.1$ and $10.8\,\mathrm{km}\,\mathrm{s}^{-1}\,\mathrm{yr}^{-1}$. Approximately 55 Gaussian components were used to fit the systemic portion of the spectra. The \citetalias{Humphreys_2008} dataset provides only Doppler velocities and accelerations; positional information is not included, as it was not constructed from spatially resolved observations. Measurements for the red- and blueshifted masers are listed in Tables 2 and 3 of \citetalias{Humphreys_2008}, respectively. Only a binned version (in $10\,\mathrm{km}\,\mathrm{s}^{-1}$ intervals) of the systemic maser data is provided in Table 4 of \citetalias{Humphreys_2008}. These measurements were later binned by \citetalias{Humphreys_2013}, see below. 
 \item \textbf{H13 Dataset -- averaged position-velocity-acceleration relation:} Since the acceleration measurements were derived independently of the positional data, \citetalias{Humphreys_2013} combined the two by binning them into common velocity intervals. Specifically, position and velocity data from \citetalias{Argon_2007} were binned using a velocity bin width of $\Delta v=1\,\mathrm{km}\,\textrm{s}^{-1}$, and the same binning scheme was applied to the velocity and acceleration data from \citetalias{Humphreys_2008}. For each velocity bin, the east–west ($X$) and north–south ($Y$) positions, and where available the line-of-sight accelerations ($a_{\text{los}}$), were averaged using inverse-variance weighting, and the uncertainty of each binned value was taken as the standard error of the weighted mean (effectively assuming independent measurements). The uncertainties for the velocities were uniformly set to $0.01\,\mathrm{km\,s^{-1}}$, representing neither the bin width nor the actual measurement precision. The averaged dataset includes 370 maser entries, but a 3$\sigma$ clipping reduced it to 358 entries. Of these, 151 (95 systemic) has acceleration data. The dataset is not publicly available and was kindly provided by M. J. Reid.
 
\end{itemize}

\subsection{The Warped Disk Model} 
\label{subsec:disk_model}

The warped disk model, adopted in \citetalias{Humphreys_2013} and \citetalias{Reid_2019}, is the basis for geometric distance estimation in these studies. It includes 14 global parameters: the distance to the galaxy ($D$), the black hole mass ($M_{\text{BH}}$), the systemic velocity of the galaxy ($v_0$), the position of the black hole on the sky ($X_0$, $Y_0$), the orbital eccentricity ($e$), the periapsis angle ($\omega_0$, $\omega_1$), the inclination angle ($i_0$, $i_1$, $i_2$), and the position angle of the disk ($\Omega_0$, $\Omega_1$, $\Omega_2$). In addition, each maser entry is assigned two nuisance parameters—the angular radius ($r$) and azimuth ($\phi$) within the disk—resulting in a total of $2N$ additional parameters, where $N$ is the number of maser entries in the dataset. \citetalias{Reid_2019} further extended the model by introducing five additional free parameters representing error floors for eastward and northward positions, systemic velocities, red/blue-shifted velocities, and accelerations. A detailed description of the warped disk model is provided in Appendix~\ref{appendix:warped}.

\subsection{Fitting program and results} 
\label{subsec:fitting}

To estimate the distance to NGC 4258, \citetalias{Humphreys_2013} and \citetalias{Reid_2019} employed the warped disk model described in Section~\ref{subsec:disk_model}, fitting it to the averaged dataset introduced in Section~\ref{subsec:binned_dataset}. The fitting was performed using a Markov Chain Monte Carlo (MCMC) approach based on the Metropolis–Hastings algorithm. The model included the global disk parameters\footnote{\citetalias{Humphreys_2013} fit include 13 global disk parameters, omitting the second-order inclination warp term $i_2$, while \citetalias{Reid_2019} included all 14 global disk parameters.}, as well as nuisance parameters $(r, \phi)$ representing the polar disk coordinates of each maser entry in the averaged dataset---adding $716-740$ additional parameters, depending on whether the initial or the 3$\sigma$-clipped averaged dataset was used.

\citetalias{Reid_2019} treated five error floor parameters as free: two for positional uncertainties ($X$, $Y$), two for LOS velocity uncertainties (for systemic and high-velocity masers), and one for accelerations. In contrast, \citetalias{Humphreys_2013} fixed the error floors to 0.02 mas for $X$, 0.03 mas for $Y$, $1\,\mathrm{km}\,\textrm{s}^{-1}$ for LOS velocities, and $0.3\,\mathrm{km}\,\textrm{s}^{-1}\,\textrm{yr}^{-1}$ for accelerations. Both studies adopted flat priors for all parameters.

From this fitting procedure, \citetalias{Humphreys_2013} derived a distance of $7.596 \pm 0.167\,\mathrm{Mpc}$, with a reported reduced chi-squared of $\chi^2_\nu \approx 1.4$. By allowing the error floors to vary, \citetalias{Reid_2019} reduced the statistical uncertainty by nearly a factor of two, obtaining $7.576 \pm 0.082\,\mathrm{Mpc}$ with a reported reduced chi-squared of $\chi^2_\nu \approx 1.2$. \citetalias{Reid_2019} interpreted this result as an improvement, noting that optimizing the error floor parameters reduced the impact of unquantified systematics and yielded a fit with $\chi^2_\nu$ closer to unity. This distance measurement is widely regarded as the most reliable in the current literature and is frequently adopted in determinations of $H_0$. \\

\subsubsection{The Statistical Uncertainties}
\label{subsec:stat_errs}

A rough estimate for the magnitude of the distance statistical uncertainties can be obtained by assuming that the distance is primarily constrained by the acceleration measurements of systemic masers. In this framework, each maser component follows a Keplerian orbit, which can be approximated as
\begin{equation}
    v_{\text{los}} - v_0 \approx \sqrt{\frac{GM_{\text{BH}}}{rD}} \cos{\phi} \approx \sqrt{\frac{GM_{\text{BH}}}{r^3 D}} (X-X_0),
\label{eq:v_x_relation}
\end{equation}
with the acceleration following
\begin{equation}
    a = \frac{GM_{\text{BH}}}{r^2 D^2}.
    \label{eq:a_kepler}
\end{equation}
Combining Equations~\eqref{eq:v_x_relation} and~\eqref{eq:a_kepler} we obtain an expression for the distance:
\begin{equation}
    D = \frac{(v_{\text{los}} - v_0)^2}{(X-X_0)^2} \frac{r}{a}.
    \label{eq:D_kepler}
\end{equation}
The relative uncertainty $R$ in the distance for a given maser entry $i$ can then be estimated as 
\begin{equation}
    R_i^2 \approx 
    \frac{4 \sigma_{v,i}^2}{(v_{\text{los,i}} - v_0)^2} + 
    \frac{\sigma_{a,i}^2}{a_i^2} + 
    \frac{4\sigma_{X,i}^2}{(X_i - X_0)^2},
\label{eq:R_i_nocorr}
\end{equation}
where uncertainties in $X_0$, $v_0$, and $r$, as well as cross-correlations between parameters, are neglected, as they are much smaller than the other terms. For typical values $\sqrt{(v_{\text{los},i} - v_0)^2}\sim30\,\text{km}\,\text{s}^{-1}$, $\sqrt{(X_i - X_0)^2}\sim 0.15\,\text{mas}$, $a_i\sim8\,\text{km s}^{-1}\,\text{yr}^{-1}$ and adopting the error floors $\sigma_{X/v/a}$ used or fitted by \citetalias{Humphreys_2013} and \citetalias{Reid_2019}, we obtain $R \sim 25\%$ and $R \sim 5\%$. 

The overall relative distance uncertainty can then be estimated using the inverse-variance weighted average:
\begin{equation}
    \frac{\hat{\sigma}_D^2}{\hat{D}^2} = \frac{1}{\sum_{i} \frac{1}{R_{i}^2}}.
\label{eq:sigma_D_D}
  \end{equation}
Applying this formalism with the \citetalias{Humphreys_2013} error floors and the adopted $X_0$ and $v_0$ values yields a distance uncertainty of $2.43\%$, compared to $2.22\%$ obtained by \citetalias{Humphreys_2013}. Using the \citetalias{Reid_2019} error floors, we obtain a distance uncertainty of $0.78\%$, compared to $1.08\%$  obtained by \citetalias{Reid_2019}. The simple estimate $\hat{\sigma}_D / \hat{D} \approx R / \sqrt{N}$, using the typical values of $R$ discussed above for \citetalias{Humphreys_2013} and \citetalias{Reid_2019} with $N \sim 100$, yields relative distance uncertainties of $\mysim2.5\%$ and $\mysim0.5\%$, respectively.

\subsection{Reproducing H13 and R19 Results}
\label{subsec:reproduced results}

In this section, we attempt to reproduce the results of \citetalias{Humphreys_2013} and \citetalias{Reid_2019}, using the clipped-averaged dataset, primarily to validate our fitting scheme against the referenced results. We fit the averaged dataset under three configurations: using \citetalias{Humphreys_2013}'s fixed error floors, using \citetalias{Reid_2019}’s best-fit error floors, and allowing the error floors to vary freely. While the resulting fits exhibit varying levels of agreement with the data, none achieves a statistically ideal reduced chi-squared value. This outcome underscores the challenges associated with the averaged approach, which are further discussed in Section~\ref{sec:caveats}.

We employ a Bayesian nested sampling algorithm (\texttt{dynesty}) to fit the warped disk model. Unlike \citetalias{Humphreys_2013} and \citetalias{Reid_2019}, who sampled over the full parameter space, we marginalize over the nuisance parameters $(r, \phi)$ for each maser entry, given a set of global disk parameters. Each maser entry is assumed to be independent of the others, and the marginalization is performed by numerically integrating the likelihood over a restricted region around the maximum-likelihood estimates. Flat priors are adopted for all fitted quantities. Further implementation details and our fitting results are provided in Section~\ref{subsec:fitting_method} and Appendix~\ref{appendix:fitting_results}, respectively, while the main findings are summarized below.

\subsubsection{Fixed H13 error floors}
\label{subsec:H13_results}

Using the fixed error floors adopted by \citetalias{Humphreys_2013}, our reanalysis yields a distance of $D = 7.277 \pm 0.166\,\mathrm{Mpc}$, compared to their original result of $7.596 \pm 0.167\,\mathrm{Mpc}$ ($\myapprox1.3\sigma$ difference). The corresponding black hole mass in our analysis is $M_{\mathrm{BH}} = (3.83 \pm 0.09) \times 10^{7}\,M_\odot$, slightly lower than the value reported by \citetalias{Humphreys_2013}, $M_{\mathrm{BH}} = (4.00 \pm 0.09) \times 10^{7}\,M_\odot$ ($\myapprox0.1\%$ difference in the ratio $M_{\mathrm{BH}}/D$ between the two analyses). We also note discrepancies in several geometric parameters: the disk center $X_0 = -0.204 \pm 0.005\,\mathrm{mas}$ in \citetalias{Humphreys_2013} fit versus $X_0 = -0.154 \pm 0.006\,\mathrm{mas}$ in our analysis, and the first-order warp in the periapsis angle, $\omega_1 = 60 \pm 10^{\circ}$ in their model compared to our result of $116 \pm 4^{\circ}$. While \citetalias{Humphreys_2013} reported $\chi^2_\nu \approx 1.4$, we obtain a much-lower reduced chi-squared value of $\chi^2_\nu \approx 0.4$ for 495 degrees of freedom (although the $v_\text{los}$ values are binned, they—together with $X$, $Y$, and $a$—are treated as measured quantities), which is statistically implausible. This suggests that \citetalias{Humphreys_2013} adopted error floors may be overly conservative. 

We attribute the discrepancy in recovered distance to degeneracies in nuisance parameters, particularly azimuthal degeneracy in the high-velocity masers, where $v_{\mathrm{los}} \propto \cos\phi$ and the redshifted (blueshifted) masers cluster near $\phi \approx 0^\circ$ ($\phi \approx 180^\circ$). We suspect that \citetalias{Humphreys_2013} did not fully explore this parameter space in their MCMC analysis. To test this, we used a non-marginalized approach: for each set of global disk parameters and for each maser, we evaluate the likelihood at the $(r_i, \phi_i)$ values that maximize it. Initial guesses are based on $(X,Y)$ positions, refined using the full $(X,Y,v,a)$ likelihood. This avoids convergence to local maxima and recovers $D = 7.58 \pm 0.21\,\mathrm{Mpc}$ (with $\chi^2_\nu \approx 0.6$) as well as $M_{\mathrm{BH}} = (4.01 \pm 0.11) \times 10^{7}\,M_\odot$, $X_0 = -0.203 \pm 0.010\,\mathrm{mas}$, and $\omega_1 = 56 \pm 18^\circ$. This method likely mimics the behavior of the \citetalias{Humphreys_2013} sampler, which is appropriate when parameter degeneracies are weak, but it is less robust than marginalization, which integrates over nuisance parameters and properly accounts for multi-modal, non-Gaussian likelihood surfaces.

\subsubsection{Fixed R19 error floors}
As a first step toward reproducing the \citetalias{Reid_2019} results, we test their adopted best-fit error floors.
Using these error floors, our analysis yields a distance of $D=7.551\pm0.072\,\text{Mpc}$, consistent with \citetalias{Reid_2019} distance estimate of $7.576 \pm 0.082\,\mathrm{Mpc}$. However, the corresponding reduced chi-squared of $\chi^2_\nu\approx1.4$ (495 degrees-of-freedom) is notably higher than the value of $\chi_\nu^2\approx1.2$ reported by \citetalias{Reid_2019}. The probability of obtaining $\chi^2_\nu > 1.4$ by chance is approximately $9.1 \times 10^{-9}$, indicating that the model is statistically incompatible with the data.

\subsubsection{Free Error Floors}
Using error floors as free parameters, we obtain a distance of $7.547\pm0.079\,\text{Mpc}$, consistent with the distance reported by \citetalias{Reid_2019}, along with the additional disk parameters. We result in a reduced chi-squared value of $\chi^2_\nu \approx 1.2$ (490 degrees of freedom)--comparable to their reported value. This configuration yields the $\chi^2_\nu$ value closest to unity among the three tested, indicating an improvement in the overall fit quality; however, the use of the reduced chi-squared statistic as a measure of goodness-of-fit is non-trivial in this case. In such fits, the residuals and the total uncertainties (which include the fitted error floor) are no longer independent, which changes the behavior of the likelihood and invalidates the standard assumptions that give a chi-squared distribution \citep[see sections~6.3 and 7.4 in][]{Cowan_1998} 

In addition, the resulting fitted error floors appear somewhat inconsistent. For high-velocity masers, the fitted velocity error floor exceeds the bin width, suggesting unresolved velocity features, despite the VLBA velocity resolution being $0.2-0.4\,\textrm{km}\,\textrm{s}^{-1}$, which is finer than the bin width. Conversely, for systemic masers, the fitted error floor is smaller than the bin width. It is therefore unclear why the high-velocity masers would appear unresolved, while systemic masers yield fitted error floors smaller than the bin width. These results hint that the fitted error floors could be influenced by residual systematics or data characteristics, rather than physical measurement uncertainties.


\section{Caveats in H13 and R19 Distance Determinations} 
\label{sec:caveats}

The analyses of \citetalias{Humphreys_2013} and \citetalias{Reid_2019} introduce several limitations that may affect the accuracy and reliability of the inferred distance measurement. One of the most significant issues arises from the averaging procedure, which introduces a key assumption: that each LOS velocity corresponds to a unique sky position and a unique acceleration. This implies that each maser component follows a distinct trajectory in both position–velocity and velocity–acceleration space, with only Gaussian scatter around it—an assumption we refer to as the \textit{single-orbit approximation}. In Section~\ref{subsec:single_orbit}, we show that this assumption is not supported by either physical models or the raw data. Differences in calibration between the \citetalias{Argon_2007} and \citetalias{Humphreys_2008} datasets---which are combined in the averaging process (see Section~\ref{subsec:binned_dataset})---may further contribute to systematic uncertainties.

Additionally, the averaging procedure does not account for the intrinsic line widths of the maser features. In Section~\ref{subsec:thermal_width}, we further discuss this point, and in Appendix~\ref{appendix:line_widths} we demonstrate that both theoretical expectations and observational data indicate that maser line widths are on the order of a few $\mathrm{km}\,\mathrm{s}^{-1}$. Applying velocity bin widths smaller than the intrinsic line width,  while treating the resulting bins as independent measurements of distinct maser components, may lead to oversampling and an overestimation of the measurement precision.

\subsection{The Single-Orbit Assumption}
\label{subsec:single_orbit}

The single-orbit assumption underlying the averaging procedure appears inconsistent with the known physical structure of the system. As shown by \citet{Herrnstein_2005}, a warped disk naturally forms a "bowl" shape along its near edge, which explains the observed narrow confinement of the systemic masers. In this model, the brightest systemic emission is expected to originate near the bottom of the bowl, confined to a narrow declination band on the sky corresponding to the vertical thickness of the masering layer. Adopting the warped disk geometry from \citetalias{Reid_2019} and a measured thickness of $\myapprox5\,\mu\text{as}$ from \citetalias{Argon_2007}, the bowl at azimuth $\phi = 270^\circ$ spans radii from approximately $3.4$ to $4.1\,\text{mas}$. This implies that systemic maser amplification can arise over a range of radii, rather than being confined to a single orbital path.
The relative optical depth, following \citet{Gray_2012}, is given by
\begin{equation}
\tau(X, v_{\text{los}}) \sim \int \exp \left[-\frac{1}{2}\left(\frac{v_{\text{los}} - v'_{\text{los,model}}}{\Delta v_D}\right)^2\right] dZ,
\end{equation}
where $\Delta v_D$ is the Doppler width, and $v'_{\text{los,model}}$ is the model LOS velocity along the LOS coordinate $Z$ at fixed sky position $X$. The integration is bounded by the upper and lower edges of the bowl. This expression yields the optical depth up to a constant normalization factor, i.e., it is relative to the maximum value in ($X$,$v_\text{los}$) parameter space. The optical depth varies by a factor of $\sim$$3$–$5$ across different $v_{\text{los}}$ values, leading to relative amplification factors of $e^{\tau} \sim 20-150$, assuming a uniform background intensity. These values are consistent with the observed range of the maser fluxes (typically $\sim$100) reported in \citetalias{Argon_2007}. However, for fixed $v_{\text{los}}$ values, the optical depth remains relatively constant across the bowl, indicating that maser amplification arises over a range of radii--in contrast to the single-orbit assumption. 

\begin{figure}
    \centering
    \includegraphics[width=0.90\linewidth]{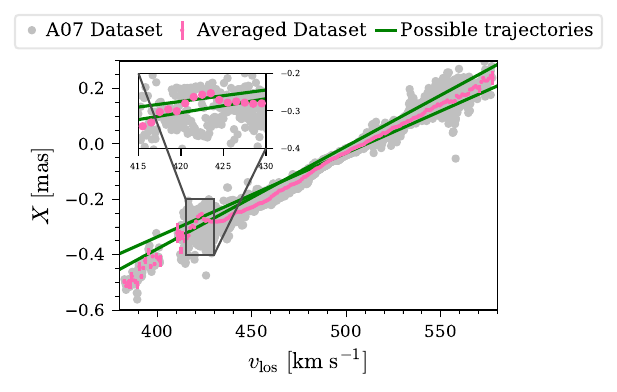}
    \caption{Comparison between the \citetalias{Argon_2007} dataset (grey) and the averaged dataset from \citetalias{Humphreys_2013} (pink), showing the east–west offset, $X$, as a function of LOS velocity, $v_{\text{los}}$. Although the averaged dataset includes uncertainties---typically around $1\,\text{$\mu$as}$---the error bars are barely visible, even after incorporating the $1.6\,\text{$\mu$as}$ error floors adopted by \citetalias{Reid_2019}. Green lines show synthetic trajectories of two maser components, computed using initial angular radii of $3.5$ and $4.0\,\text{mas}$ (within the efficient amplification region), based on the disk parameters from \citetalias{Reid_2019}. These trajectories demonstrate that multiple $X$ values can correspond to the same $v_{\text{los}}$. The inset in the upper left zooms in on the region near  $v_{\text{los}} \approx 425\,\textrm{km}\,\textrm{s}^{-1}$, where two distinct $X$ values are observed in the \citetalias{Argon_2007} data. This example highlights a case in which the averaged dataset fails to capture the true spread of the measurements, and the assumption of a Gaussian distribution is clearly invalid.}
    
    \label{fig:xv_sys}
\end{figure}

To illustrate the inconsistency of the single-orbit assumption, Figure~\ref{fig:xv_sys} compares the averaged dataset (pink) with the original \citetalias{Argon_2007} measurements (gray) in the systemic velocity region. The averaged dataset reports extremely small uncertainties in $X$, typically around $1\,\text{$\mu$as}$, which remain nearly unchanged even after incorporating \citetalias{Reid_2019}'s fitted error floor of $1.6\,\text{$\mu$as}$. These uncertainties are barely visible in the figure. The green lines show two example maser component trajectories in the $X$–$v_{\text{los}}$ plane, computed using the disk parameters from \citetalias{Reid_2019}, as described in Appendix~\ref{appendix:trajectories}. Both trajectories share an initial azimuth of $260^\circ$ but differ in their initial angular radii, at $3.5$ and $4.0\,\text{mas}$. The fact that these trajectories yield different $X$ values at the same $v_{\text{los}}$ highlights the core issue. For instance, near $v_{\text{los}} \approx 425\,\mathrm{km\,s}^{-1}$ (see inset in the upper left of Figure~\ref{fig:xv_sys}), the radial separation of $\myapprox 0.5\,\text{mas}$ between the trajectories leads to a positional offset of $\Delta X \approx 50\,\mu\text{as}$. In contrast, the averaged data, with an effective uncertainty of only $\Delta X_\text{avg} \approx 2\,\mu\text{as}$, implicitly assumes that the maser locations are confined to a radial range of just $\myapprox 0.02\,\text{mas}$, a factor of $\sim$25 narrower than what the warped disk model permits.
The \citetalias{Argon_2007} dataset itself contradicts the single-orbit assumption: again near $v_{\text{los}} \approx 425\,\mathrm{km\,s}^{-1}$, two distinct $X$ values are observed for the same velocity. In this region, the averaged dataset fails to capture the true positional spread, and the assumption of a Gaussian distribution within each velocity bin clearly breaks down. Consequently, the averaged data may misrepresent the underlying structure and kinematics of the disk.

\begin{figure*}
    \centering
    \includegraphics[width=0.9\linewidth]{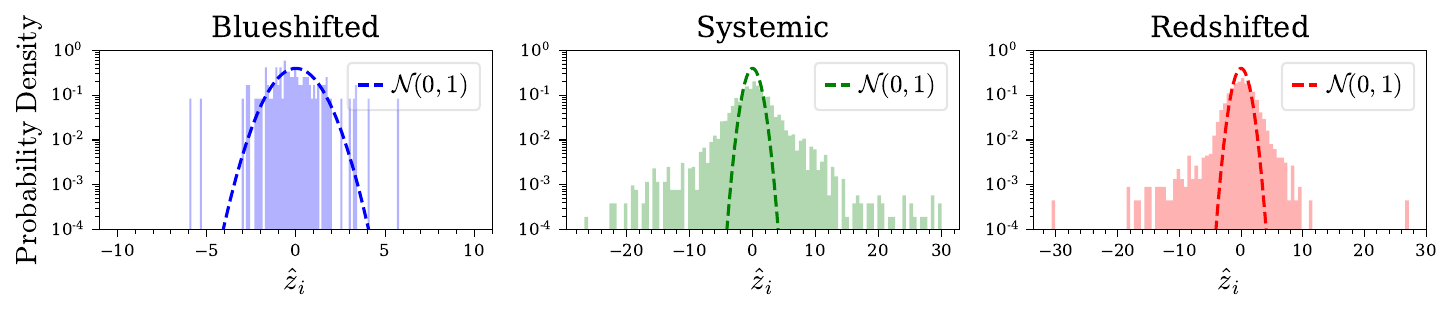}
    \caption{Distribution of normalized residuals $\hat{z}_i$ (defined in Equation~\eqref{eq:z_i}) for the blueshifted (left, blue), systemic (middle, green), and redshifted (right, red) masers, comparing $X$ positions from the averaged and \citetalias{Argon_2007} datasets. The dashed curves represent the standard normal distribution, $\mathcal{N}(0, 1)$, expected under the Gaussian assumptions of the averaging procedure. In all three cases, the observed residuals exhibit significant deviations from this model, indicating that the averaged dataset does not accurately capture the true distribution of positional measurements.}
    \label{fig:xraw_xbin}
\end{figure*}

To test the Gaussian assumption quantitatively across the entire velocity space, we define the normalized residual:
\begin{equation}
\label{eq:z_i}
\hat{z}_i = \frac{\Delta_i}{\sqrt{\text{Var}(\Delta_i)}},
\end{equation}

where $\Delta_i = X_{\text{A07},i} - X_{\text{avg},i}$, is the residual between the east–west position $X_{\text{A07},i}$ from the \citetalias{Argon_2007} dataset, and the corresponding averaged value $X_{\text{avg},i}$. The variance of the residuals $\ \mathrm{Var}(\Delta_i)$ must account for the correlation between the individual measurement $X_{\text{A07},i}$ and its corresponding averaged value $X_{\text{avg},i}$. The averaged value $i$ is the weighted mean of the measurements in that bin, $X_{\text{avg},i} = \sum_j w_j X_{\text{A07},j} / \sum_j w_j$, where the weights $w_j=\sigma_{\text{A07},j}^{-2}$. Its variance is simply the inverse of the weight sum, $\sigma_{X,\text{avg},i}^2=1/\sum_jw_j$. Those determine the covariance between the individual measurement $X_{\text{A07},i}$ and its averaged value $X_{\text{avg},i}$ (the weighted mean), 
\begin{equation}
    \text{Cov}(X_{\text{A07},i}, X_{\text{avg},i}) = \frac{w_i}{\sum_j w_j} \sigma_{X,\text{A07},i}^2 = \sigma_{X,\text{avg},i}^2, 
\end{equation}
assuming that the individual measurements $X_{\text{A07}}$ are independent. Consequently, the variance of the residual reduces to, 
\begin{equation}
    \mathrm{Var}(\Delta_i) = \sigma_{X,\text{A07},i}^2 - \sigma_{X,\text{avg},i}^2. 
    \label{eq:var_res}
\end{equation}
The uncertainties of the averaged values incorporate \citetalias{Reid_2019} fitted error floors. If the averaged dataset is consistent with the \citetalias{Argon_2007} data under a Gaussian error model, the normalized residuals $\hat{z}_i$ should follow a standard normal distribution, $\mathcal{N}(0,1)$. In the limit of large-number statistics--where each bin contains many \citetalias{Argon_2007} data points--we would expect $\sigma_{X,\text{avg},i}^2$ to be negligible. However, we retain this term, as some velocity bins contain fewer than five \citetalias{Argon_2007} points, making the term non-negligible. For some bins in the blueshifted maser regime, the variance of the residuals is negative as the averaged error exceeds the corresponding \citetalias{Argon_2007} error. This occurs even when the error floors are not included; therefore, those bins are excluded. Figure~\ref{fig:xraw_xbin} shows the distribution of $\hat{z}_i$ values for the blueshifted (left, blue), systemic (middle, green), and redshifted (right, red) masers. In all cases, the distributions deviate significantly from a standard normal distribution, with the redshifted and systemic masers exhibiting heavy tails -- residuals that extend well beyond the expected range. This allows us to clearly reject the hypothesis that $\hat{z}_i \sim \mathcal{N}(0,1)$ and further confirms the inconsistency between the averaged and \citetalias{Argon_2007} data. Together, these findings indicate that the single-orbit assumption is not supported by the underlying data and may lead to biased interpretation of the disk geometry and dynamics. 

In Appendix~\ref{appendix:synthetic_binned} we assess how the single-orbit approximation introduced by the averaging procedure affects both the recovered distance and the resulting $\chi^2$ distribution. When following both error-floor methodologies similar to those of \citetalias{Humphreys_2013} and \citetalias{Reid_2019}, the former results in a high probability of biased distance estimates, while both methods exhibit reduced chi-squared ($\chi^2_\nu$) values greater than 1. This suggests that the averaging procedure itself introduces an unrecognized systematic bias. \citetalias{Humphreys_2013} discusses adopting a $1\%$ systematic uncertainty to account for unmodeled spiral-disk structure; although not stated explicitly, such a term can affect both the distance estimate and the resulting chi-squared statistic. Our analysis shows explicitly that a similar effect can also arise solely from the averaging procedure.

\subsection{Line Widths}
\label{subsec:thermal_width}
Masers in velocity–flux spectra exhibit finite linewidths. \citet{Varshalovich_2006} showed that the FWHM of these features can reach up to $2\, \text{km s}^{-1}$, as a result of thermal broadening and the hyperfine structure of the transition. The observed linewidth is also affected by the optical depth: higher optical depth causes the line to become narrower.  The finite linewidth implies that adjacent velocity channels generally contain overlapping contributions from the same physical emission and therefore are not independent measurements of distinct maser components. However, in the averaged dataset, these channels were treated as independent, which can artificially upweight low-S/N line wings and inflate the apparent number of independent maser components. Empirically, the integrated spectra contain fewer than five blueshifted masers, approximately ten redshifted masers, and at most 40 systemic masers—far fewer than the $\mysim 370$ independent data points implied by the averaged dataset—indicating significant oversampling.

Additionally, blending between masers is likely, especially among systemic features that originate from similar regions of the disk. This effect was not accounted for in the previous analysis. For these reasons, we adopt a component-level approach rather than per-channel averaging-one that explicitly recognizes that adjacent channels can trace the same physical maser component and allows for the possibility of blended features. Supporting evidence, including an analysis of the redundancy between adjacent velocity channels in the averaged and \citetalias{Argon_2007} datasets and a demonstration of positional blending among systemic masers, is presented in Appendix~\ref{appendix:line_widths}. 

\section{Reanalysis of NGC 4258 maser data} 
\label{sec:D_estimation}

In this section, we reanalyze the NGC~4258 maser data without invoking the unjustified single-orbit assumption (see Section~\ref{subsec:single_orbit}) and while explicitly accounting for the intrinsic linewidths of the maser features, which were previously neglected (see Section~\ref{subsec:thermal_width}). To avoid relying on these assumptions, we must track individual maser components across multiple epochs. This requires VLBI data with resolved maser positions, enabling the reconstruction of each component’s trajectory over time. We therefore restrict our analysis to the \citetalias{Argon_2007} dataset\footref{fn:public_data}, which reduces the temporal baseline from approximately 10 years to about 4 years.

The main conclusion from this analysis is that the available observational cadence is insufficient to reliably track individual maser components given their high intrinsic variability. Consequently, we demonstrate that different plausible tracking methods can yield significantly different distance estimates—even if each method independently produces small formal uncertainties. We do not attempt to identify or optimize a “best” tracking method, but rather aim to show that the distance inference is highly sensitive to the underlying assumptions in component identification and tracking.

The construction of maser trajectories from the VLBI data is described in Section~\ref{subsec:new_dataset}, where we specifically incorporate the finite line widths of the masers. In Section~\ref{subsec:fitting_method}, we present our method for fitting the warped disk model used by \citetalias{Humphreys_2013} and \citetalias{Reid_2019} to the reconstructed trajectories. The results of the model fits, including the inferred parameters, are presented in Section~\ref{subsec:results}.

\subsection{Construction of trajectories}
\label{subsec:new_dataset}
The construction of maser trajectories is carried out in two steps. First, we identify candidate maser components in each individual observational epoch. Second, we analyze all candidates across epochs to track components that follow consistent trajectories over time. The outcome of this procedure is a set of trajectories, each comprising $X$ and $Y$ positions, LOS velocities, LOS accelerations, and their associated uncertainties.

\subsubsection{single-epoch candidate maser components}
\label{subsubsec:identify_peaks}

\begin{figure*}
    \centering
    \includegraphics[width=0.8\linewidth]{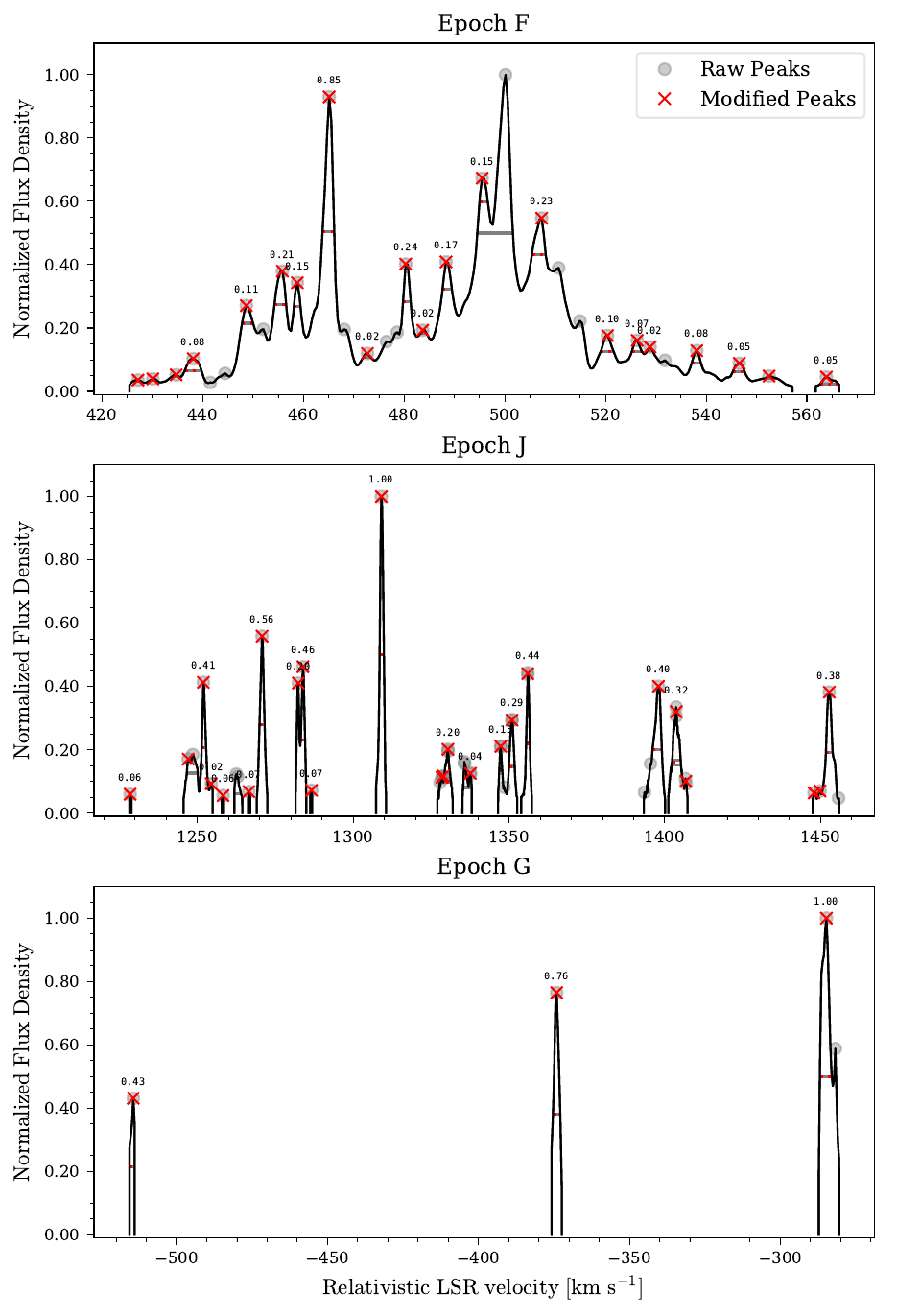}
    \caption{Integrated spectra from \citetalias{Argon_2007} for epoch F (systemic, top), epoch J (redshifted, middle), and epoch G (blueshifted, bottom). The plotted quantity is normalized flux density (i.e., flux density divided by the maximum value in each spectrum) as a function of relativistic LSR velocity. Spectral peaks and their corresponding FWHMs are identified using the procedure described in Section~\ref{subsec:thermal_width}. Gray circle markers ('o') denote the initial set of raw peaks, while red cross markers ('×') indicate the final set of clean, well-isolated peaks. FWHMs are shown as continuous gray lines for the raw peaks and dashed red lines for the final set. Prominence values exceeding 0.02 are labeled above the corresponding peaks in black. To standardize prominence calculations across epochs and velocity ranges, synthetic zero-flux values were added at the baseband edges, ensuring all peaks are evaluated relative to a common zero baseline.}
    
    \label{fig:vF_spec}
\end{figure*}

In each observational epoch, we separately analyze the integrated flux density spectrum as a function of relativistic LSR velocity for each velocity regime. We begin by normalizing each spectrum to its peak flux across all basebands and appending synthetic zero-flux values at the baseband edges. Within each baseband, spectral peaks are identified using the {\tt\string scipy.signal.find\_peaks} function. Figure~\ref{fig:vF_spec} shows examples from three representative epochs—F, J, and G—corresponding to the systemic, redshifted, and blueshifted maser components, respectively. All initially detected peaks are marked with a ‘o’ grey symbol, while the ‘×’ red markers show the final selected set (after refinement). For each peak, we compute its prominence (values shown in black in Figure~\ref{fig:vF_spec}), which quantifies how much the peak stands out relative to its surroundings. Specifically, prominence is defined as the vertical distance between the peak and a reference level. This level is determined by examining the spectrum on both sides of the peak, extending outward until either the signal rises toward a higher peak or the spectrum boundary is reached. The minimum values found in these intervals are compared, and the larger of the two is taken as the reference level. The prominence definition allows an isolated, low-amplitude peak to be more prominent than a taller peak embedded within a broader structure, thereby offering a more robust measure of the distinctiveness of spectral features. The initial normalization ensures that prominence values are consistent across epochs and velocity regimes. Finally, we compute the FWHM for each peak, defined as the width at 50\% of the prominence height (shown as dashed red or continuous gray horizontal line segments in Figure~\ref{fig:vF_spec}). At this stage, each candidate maser component is characterized by three properties: peak velocity, prominence, and FWHM.

Since we cannot exclude a significant contribution of thermal broadening to the FWHM (see Section~\ref{subsec:thermal_width}), each maser feature is assigned a single peak velocity, with its uncertainty estimated from the FWHM using the standard relation $\mathrm{FWHM} = 2\sqrt{2\ln 2}\,\sigma$. We emphasize that this does not imply the line profile is strictly Gaussian; the conversion is adopted as a conventional approximation. We do not estimate the peak uncertainty by dividing $\sigma$ by the number of velocity channels across the feature, as the peak value does not represent a statistical mean, and the line shape may be influenced by blending or intrinsic variability.

Since peaks are identified independently within each baseband, some maser components may be detected multiple times in overlapping spectral regions. Two peaks are considered to correspond to the same component if their velocity difference falls below a defined tolerance. This tolerance is empirically estimated by computing the median channel spacing in each baseband and taking twice the maximum value across all basebands. When a match is identified, the peak properties—such as velocity, flux, FWHM, and prominence—are averaged. We note that averaging the overlapping basebands before peak identification would be problematic, because those do not share an identical velocity channel grid. Directly averaging misaligned channels can introduce additional noise and potentially create artificial peaks in the subsequent peak-finding process. Finally, the assigned velocity of each peak is taken as the relativistic LSR velocity value from \citetalias{Argon_2007} (for the same epoch) that is closest to the averaged peak velocity. The difference between the assigned and averaged velocities is added in quadrature to the uncertainty $\sigma$.


To ensure that only clean, isolated components are included in the analysis, we exclude peaks that show signs of contamination from nearby features, such as overlapping shoulders or blended profiles. Specifically, we verify within each baseband that a peak’s FWHM does not extend into the region of another peak. For instance, the feature near $500\,\textrm{km}\,\textrm{s}^{-1}$ in Figure~\ref{fig:vF_spec} is excluded due to significant overlap with a neighboring line. Such cases are removed to retain only well-separated, uncontaminated components.

Final quality checks are performed using the FWHM region of each peak, evaluated across the full spectrum rather than within individual basebands. Peaks whose FWHM span includes only a single data point are considered unreliable and are discarded. To further confirm that each peak is isolated, we apply a Gaussian smoothing filter to the spectrum using the {\tt\string scipy.ndimage.gaussian\_filter} function with a standard deviation of one channel, corresponding to the separation between adjacent spectral bins. We examine the filtered spectrum for the presence of any other peaks within the FWHM interval. This filtering step is critical, as overlapping basebands often contain noise artifacts that can result in spurious detections. It also helps eliminate cases where duplicate peaks across bands may have been missed during the merging process described earlier.

We next determine the $X$ and $Y$ positions for each candidate maser component. Since position data are provided for each velocity channel, we extract the $X$ and $Y$ values corresponding to the peak velocity and compute the positional widths $\Delta X$ and $\Delta Y$ as the difference between the maximum and minimum positions across the FWHM range. These widths are converted to standard deviations $\sigma$, using the same relation as before $\text{FWHM} = 2\sqrt{2\ln 2}\sigma$. The resulting \,$\sigma$ values are then added in quadrature to the random positional uncertainties of the peak velocity. These uncertainties are listed in \citetalias{Humphreys_2013} and derived from $\theta_\text{beam}/(2\,\times\,\text{S/N} )$, where $\theta_\text{beam}$ is the size of the synthetic beam. At this stage, each peak is characterized by a single position and an associated positional uncertainty. We reiterate that, because thermal broadening likely contributes significantly to the FWHM (see Section~\ref{subsec:thermal_width}), we assign a single, point-like position to each maser component. This approach avoids the oversampling issue inherent in the averaged dataset.


A complete summary of all identified peaks—including both the initial detections and the refined catalog, along with their associated positional data—is available in the dataset provided at \url{bit.ly/400pIcn}.


\subsubsection{Identify individual maser components across multiple observational epochs}

The next step in the analysis is to identify individual maser components across multiple observational epochs. To do this, we employ the RANdom SAmple Consensus (RANSAC) algorithm \citep{Fischler_1981}, assuming linear drifts in both velocity and sky position. This approach enables us to establish a physical connection between position and acceleration for each maser component, avoiding the artificial mixing of signals introduced by averaging data from independent sources. The method proceeds as follows: two data points are randomly selected, and independent linear fits are applied to $v_\text{LOS}$, $X$, and $Y$ as functions of time. Additional data points are classified as inliers if they lie within a specified uncertainty-based distance threshold from all three fitted trends. These inliers are then grouped as a candidate maser component. The distance thresholds are defined relative to the uncertainties in $X$, $Y$, and $v_\text{LOS}$ for each point. A group is retained only if the slopes of the fitted lines fall within physically plausible ranges: LOS accelerations must lie between $6$ and $10\,\mathrm{km}\,\mathrm{s}^{-1}\,\mathrm{yr}^{-1}$ for systemic masers and between $-1$ and $1\,\mathrm{km}\,\mathrm{s}^{-1}\,\mathrm{yr}^{-1}$ for high-velocity masers; proper motion components in $X$ and $Y$ must be within $-1$ to $1\,\mathrm{mas}\,\mathrm{yr}^{-1}$. This procedure is repeated $10^5$ times to identify the largest valid group. Once identified, if the group size meets the minimum inlier criterion, the members of this group are removed from the dataset, and the process is repeated iteratively on the remaining data until no further groups are found.

\begin{figure}
    \centering
    \includegraphics[width=0.8\linewidth]{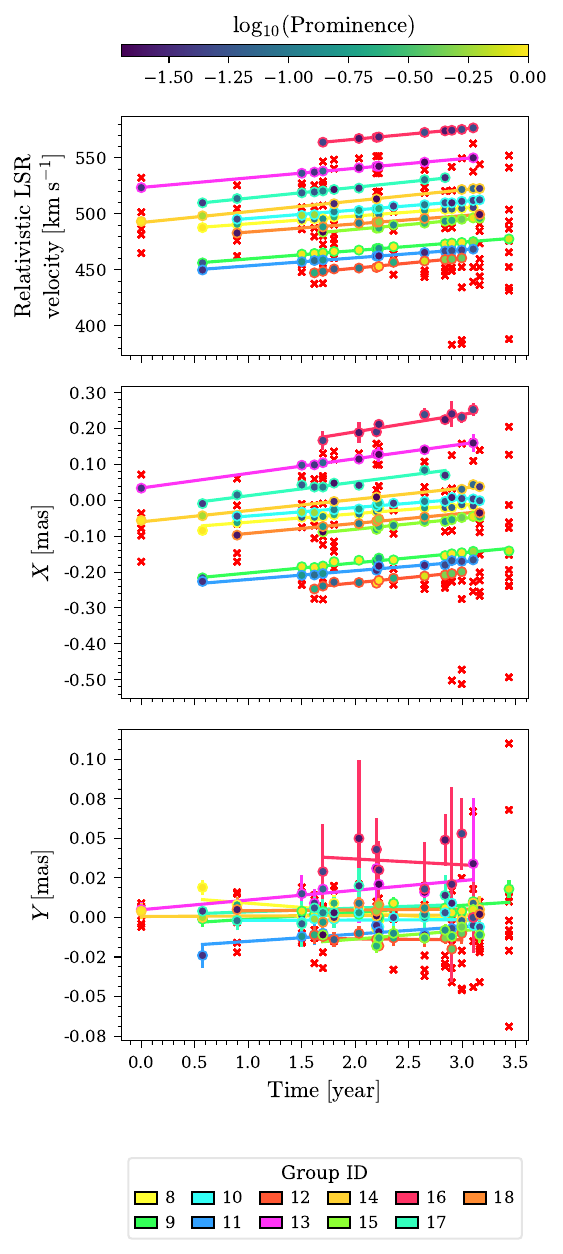}
    \caption{Results of the RANSAC algorithm, as described in Section~\ref{subsec:new_dataset}, for the dataset corresponding to entry \#8 in Table~\ref{tab:results_newdata}. The top panel shows LOS velocity as a function of time; the middle and bottom panels display east–west ($X$) and north–south ($Y$) sky-plane positions, respectively, also plotted against time. Red crosses indicate observations rejected by the algorithm. Colored lines and marker edges correspond to individual maser components identified by RANSAC, each assigned a group ID consistent with the identifiers in the public catalog\footref{fn:url}. Marker face color encodes $\log_{10}$(prominence).}
    \label{fig:ransac}
\end{figure}

To assess the robustness of the RANSAC algorithm for maser identification, we repeat the full procedure 100 times. Groups with at least 70\% membership overlap across runs are merged, and only those that both appear in all 100 iterations and contain more than three components are retained as robust detections. For each group, we assign a single acceleration value with an associated uncertainty and compute the cross-correlation coefficient $\rho_{va}$ between its LOS velocity measurements and corresponding accelerations. Figure~\ref{fig:ransac} displays the LOS velocity (top) and sky positions (middle and bottom) of systemic maser features as a function of time, along with the results of a representative RANSAC run (dataset \#8 in Table~\ref{tab:results_newdata}). Colored lines and matching marker edges denote maser groups identified by RANSAC, with marker face color indicating $\log_{10}$(prominence). Each group is labeled with its unique group ID, which matches the identifiers listed in our public catalog\footnote{\url{bit.ly/400pIcn}\label{fn:url}}. Red crosses represent ungrouped outliers. A high outlier fraction is observed at the beginning and end of the observational timeline—for example, only one inlier is found in the final epoch of dataset \#8—reflecting the increasing uncertainty in fitted velocity estimates near the edges of the time baseline.

\subsubsection{obtained trajectories dataset}
\label{subsec:obtained_trajectories}

We generated multiple datasets using different parameter choices in the maser component selection and model fitting process. The aim is to evaluate the robustness of the results with respect to reasonable variations in data filtering criteria. The following parameters were explored:

\begin{enumerate}
\item \textbf{Peak Width:} Two approaches were tested for filtering peaks based on their velocity linewidths. The first approach retains all detected peaks, regardless of width. The second retains only peaks with widths up to $3\,\text{km\,s}^{-1}$ or $2\,\text{km\,s}^{-1}$. As explained in Section~\ref{subsec:thermal_width}, the widths of maser features resulting from thermal broadening and hyperfine structure are typically up to $2\,\text{km\,s}^{-1}$. We also consider the possibility of additional broadening caused by other physical processes such as blending, turbulence, or Zeeman splitting (though the latter is expected to be negligible, \citealt{Herrnstein_1998a}).

\item \textbf{Prominence Threshold:} Peaks were filtered based on their prominence using three methods. The first applied no threshold, retaining all detected peaks regardless of prominence. The second applied fixed thresholds of 0.02 and 0.05 in absolute units. The third used a relative criterion: peaks were retained only if they exceeded their surrounding baseline by at least $5\sigma_F$, where $\sigma_F$ is the flux uncertainty reported in \citetalias{Argon_2007}. This approach helps eliminate low-significance features and noise artifacts.

\item \textbf{RANSAC Distance Threshold:} This parameter defines the maximum allowable deviation of a data point from a fitted linear trend for it to be considered an inlier. The distance is expressed in units of $\sigma_{X/Y/v}$, incorporating both velocity and positional uncertainties. Adjusting this threshold controls the strictness of component grouping. In our analysis, we consistently used a threshold of $1\sigma_v$ for velocity, and either $1\sigma_{X/Y}$ or $3\sigma_{X/Y}$ for positions.

\item \textbf{RANSAC Minimum Inliers:} We vary the minimum number of inliers required for a group to be accepted as a valid maser component. For blueshifted masers, a minimum of 3 inliers was used; for redshifted and systemic masers, thresholds of 5 and 8 inliers were tested.
\end{enumerate}

We aim to assess the robustness of both the method and the dataset. Ideally, fitting different subsets of the same data should yield consistent results. 
Thus, we generate 40 configurations obtained by combining different RANSAC settings with the prominence and linewidth selection criteria described above. For RANSAC, we adopt four parameter sets: one using a $3\sigma_{X/Y}$ positional threshold, with a minimum of 8 inliers for both systemic and redshifted maser components, and three using a stricter $1\sigma_{X/Y}$ threshold with inlier combinations for systemic and redshifted components: (8,8), (8,5), and (5,5). For the prominence selection, we considered four cases: no threshold, fixed thresholds of 0.02 and 0.05 (in absolute units), and a relative threshold of $5\sigma_F$, where $\sigma_F$ is the flux uncertainty reported in \citetalias{Argon_2007}. The no-threshold case was paired only with the $0-2\,\mathrm{km\,s^{-1}}$ linewidth limit, while the other three prominence criteria were combined with all three linewidth limits. Datasets containing fewer than five identified systemic maser groups were excluded from further analysis, as they fail to yield robust distance estimates across multiple fitting runs. This filtering resulted in 17 retained datasets, as detailed in Table~\ref{tab:results_newdata}. For each configuration, the table reports the total number of single-epoch maser components used, along with the number of identified maser groups in each velocity regime (systemic, redshifted, and blueshifted). The corresponding sub-datasets are provided in digital format at \url{bit.ly/400pIcn}. 

The results are found to be quite sensitive to the choice of selection criteria. We interpret this sensitivity as a consequence of the limited observational cadence, which is insufficient to reliably track individual masers given their high intrinsic variability (see discussion in Section~\ref{sec:obs_suggestion}).

\begin{table*}
\centering
\caption{Summary of Analysis Parameters and Results}
\label{tab:results_newdata}
\renewcommand{\arraystretch}{1.3}
\begin{tabular}{c|cc|c|cc|c|ccc|cc|cc}
\hline
  &
  Prom. &
  Width &
  DisThresh. &
  R$_{\rm inl}$ & S$_{\rm inl}$ &
  N &
  R$_{\rm grp}$ & B$_{\rm grp}$ & S$_{\rm grp}$ &
  $D$ [Mpc] &
  $M_{\text{BH}}\; [10^7\; M_\odot]$ & 
  $\chi^2_\nu$ &
  $dof$ \\
\hline
$\star$1 & $...$ & $0-2$ & 1 & 5 & 5 & 91 & 8 & 1 & 6 &  $8.091^{+0.208}_{-0.153}$ &  $4.234^{+0.105}_{-0.080}$ & 1.05 & 168 \\
{\color[HTML]{FF0000} 2} & {\color[HTML]{FF0000} $...$} & {\color[HTML]{FF0000} $0-2$} & {\color[HTML]{FF0000} 3} & {\color[HTML]{FF0000} 8} & {\color[HTML]{FF0000} 8} & {\color[HTML]{FF0000} 86} & {\color[HTML]{FF0000} 4} & {\color[HTML]{FF0000} 2} & {\color[HTML]{FF0000} 5} &  {\color[HTML]{FF0000} $7.346^{+0.087}_{-0.082}$} &  {\color[HTML]{FF0000} $3.901^{+0.049}_{-0.047}$} & {\color[HTML]{FF0000} 0.60}\textsuperscript{\textdagger} & {\color[HTML]{FF0000} 158} \\
3 & 0.02 & $0-3$ & 1 & 5 & 5 & 128 & 8 & 2 & 10 &  $7.166^{+0.112}_{-0.100}$ &  $3.777^{+0.057}_{-0.054}$ & 1.13 & 242 \\
4 & 0.02 & $0-3$ & 3 & 8 & 8 & 151 & 7 & 2 & 8 &  $7.205^{+0.097}_{-0.098}$ &  $3.797^{+0.055}_{-0.054}$ & 1.05 & 288 \\
5 & 0.02 & $...$ & 1 & 5 & 5 & 148 & 11 & 3 & 9 &  $7.022^{+0.101}_{-0.102}$ &  $3.719^{+0.055}_{-0.053}$ & 0.78 & 282 \\
6 & 0.02 & $...$ & 1 & 5 & 8 & 126 & 11 & 3 & 5 &  $6.737^{+0.108}_{-0.101}$ &  $3.564^{+0.059}_{-0.055}$ & 0.79 & 238 \\
7 & 0.02 & $...$ & 1 & 8 & 8 & 80 & 3 & 3 & 5 &  $6.825^{+0.097}_{-0.111}$ &  $3.64^{+0.052}_{-0.059}$ & 1.22 & 146 \\
$\star$8 & 0.02 & $...$ & 3 & 8 & 8 & 204 & 8 & 3 & 11 &  $7.182^{+0.091}_{-0.086}$ &  $3.843^{+0.050}_{-0.048}$ & 1.14 & 394 \\
9 & 0.05 & $...$ & 1 & 5 & 5 & 119 & 11 & 3 & 5 &  $7.419^{+0.187}_{-0.187}$ &  $3.958^{+0.097}_{-0.100}$ & 0.82 & 224 \\
10 & 0.05 & $...$ & 3 & 8 & 8 & 140 & 7 & 3 & 5 &  $6.783^{+0.122}_{-0.117}$ &  $3.604^{+0.065}_{-0.063}$ & 0.95 & 266 \\
11 & $5\sigma_F$ & $0-2$ & 1 & 5 & 5 & 75 & 6 & 1 & 5 &  $8.061^{+0.088}_{-0.076}$ &  $4.221^{+0.057}_{-0.041}$ & 1.41 & 136 \\
12 & $5\sigma_F$ & $0-3$ & 1 & 5 & 5 & 125 & 7 & 2 & 10 &  $7.347^{+0.070}_{-0.077}$ &  $3.869^{+0.038}_{-0.041}$ & 0.94 & 236 \\
13 & $5\sigma_F$ & $0-3$ & 3 & 8 & 8 & 186 & 7 & 2 & 12 &  $7.297^{+0.074}_{-0.076}$ &  $3.849^{+0.048}_{-0.044}$ & 1.01 & 358 \\
14 & $5\sigma_F$ & $...$ & 1 & 5 & 5 & 153 & 10 & 3 & 10 &  $7.309^{+0.110}_{-0.106}$ &  $3.865^{+0.060}_{-0.057}$ & 0.87 & 292 \\
15 & $5\sigma_F$ & $...$ & 1 & 5 & 8 & 136 & 10 & 3 & 7 &  $7.31^{+0.115}_{-0.115}$ &  $3.869^{+0.063}_{-0.063}$ & 0.84 & 258 \\
16 & $5\sigma_F$ & $...$ & 1 & 8 & 8 & 95 & 3 & 3 & 7 &  $7.359^{+0.112}_{-0.097}$ &  $3.915^{+0.059}_{-0.052}$ & 1.12 & 176 \\
{\color[HTML]{FF0000} 17} & {\color[HTML]{FF0000}$5\sigma_F$} & {\color[HTML]{FF0000}$...$} & {\color[HTML]{FF0000} 3} & {\color[HTML]{FF0000} 8} & {\color[HTML]{FF0000} 8} & {\color[HTML]{FF0000} 229} & {\color[HTML]{FF0000} 8} & {\color[HTML]{FF0000} 3} & {\color[HTML]{FF0000} 13} &  {\color[HTML]{FF0000} $7.155^{+0.061}_{-0.060}$}  & {\color[HTML]{FF0000} $3.84^{+0.035}_{-0.034}$}  & {\color[HTML]{FF0000} 1.31}\textsuperscript{\textdagger} & {\color[HTML]{FF0000} 444} \\
\hline
\end{tabular}

\medskip
\raggedright
\small
\textit{Notes.} Prom. = Prominence threshold; Width = Peak width $[\mathrm{km}\,\mathrm{s}^{-1}]$; 
DisThresh. = RANSAC XY distance threshold (in units of $\sigma_{X,Y}$); 
inl. = RANSAC minimum inliers for red/systemic masers (blueshifted masers used a fixed threshold of 3); 
N = Number of single-epoch maser components; Grp. = Number of groups; 
$D$ = Estimated distance to NGC~4258. Red color indicates a statistically less likely fit, where the probability of obtaining a smaller or larger $\chi^2_{\nu}$ by chance is less than $10^{-3}$. \\
$^\star$ Entry \#8 is a representative dataset that appears in Figure~\ref{fig:ransac}, Figure~\ref{fig:xv_sys_traj}, and Appendix~\ref{appendix:traj_results}. Entry \#1 appears as an additional example in Appendix~\ref{appendix:traj_results}. \\
$^\dagger$ Indicates a statistically less likely fit (marked also in red).
\end{table*}

\begin{figure}
    \centering
    \includegraphics[width=0.9\linewidth]{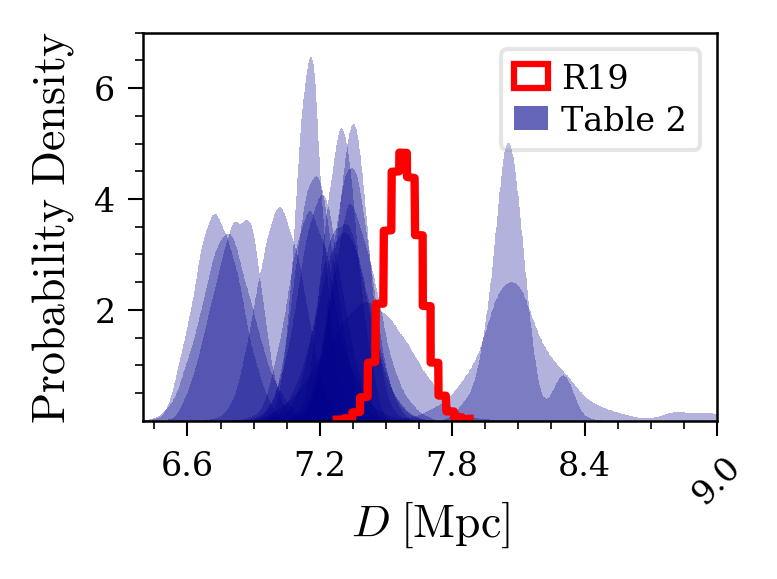}
    \caption{Posterior distributions (blue shaded regions) of the distance-fitting results from Table~\ref{tab:results_newdata}, excluding statistically less likely fits. The red curve shows the distance posterior from \citetalias{Reid_2019} for comparison.}
    \label{fig:hist_results_table2}
\end{figure}

To summarize, for each parameter configuration, we generate a dataset of maser component trajectories. Each trajectory contains multiple epochs with measurements of $X$ and $Y$ positions, LOS velocities and accelerations, their associated uncertainties, and the velocity–acceleration cross-correlation values. These trajectories incorporate maser linewidths into the error model, ensure that all fitted quantities are based on consistently tracked components, and avoid the single-orbit assumption. Representative examples of maser component trajectories from selected datasets are presented in Appendix~\ref{appendix:traj_results}.

\subsection{Fitting Program}
\label{subsec:fitting_method}
While \citetalias{Humphreys_2013} employed a Markov Chain Monte Carlo (MCMC) algorithm based on the Metropolis–Hastings method, our analysis utilizes the \texttt{dynesty} nested sampling framework \citep{Speagle_2020,Koposov_2023,Skilling_2004,Skilling_2006,Higson_2018,Feroz_2009,Neal_2003,Handley_2015a,Handley_2015b}. 
The primary distinction lies in our treatment of the nuisance parameters $(r, \phi)$. Whereas \citetalias{Humphreys_2013} included these as free parameters alongside the global parameters of the warped disk model (see Section~\ref{subsec:disk_model}), we instead reduce the dimensionality of the problem by marginalizing over the nuisance parameters. Consequently, as in the case of the clipped–averaged dataset, this approach reduces the number of free parameters from many hundreds to 14. This reduction significantly improves sampling efficiency and enhances convergence reliability. As discussed in Section~\ref{subsec:H13_results}, treating the nuisance parameters as free parameters can impede proper exploration of the parameter space when strong degeneracies are present.

The likelihood for a given set of global parameters $\mathbf{G}$ is computed by integrating over the nuisance parameter space:
\begin{equation}
\mathcal{L}(\mathbf{d}|\mathbf{G}) =
\int \mathcal{L}(\mathbf{d}|\mathbf{G}, \boldsymbol{r}, \boldsymbol{\phi}) \pi(\boldsymbol{r}, \boldsymbol{\phi}) d\boldsymbol{r} d\boldsymbol{\phi},
\end{equation}
where $\mathbf{d}$ is the set of observations and $\pi(\boldsymbol{r}, \boldsymbol{\phi})$ denotes the prior on the nuisance parameters. Assuming independent observations $d_i$, the total likelihood becomes a product over individual integrals:
\begin{equation}
\mathcal{L}(\mathbf{d}|\mathbf{G}) = \prod_i \int \mathcal{L}(d_i | \mathbf{G}, r_i, \phi_i) \pi(r_i, \phi_i) dr_i d\phi_i.
\end{equation}
Rather than integrating over the full $(r_i, \phi_i)$ domain, we perform numerical marginalization on a $200\times200$ grid centered on the maximum-likelihood values $(r_{i,\text{max}}, \phi_{i,\text{max}})$ for each data point. The integration domain is restricted to $(r_{i,\text{max}} \pm 2\,\text{mas}, \phi_{i,\text{max}} \pm 1.5^\circ)$ for systemic masers, and to $(r_{i,\text{max}} \pm 0.5\,\text{mas}, \phi_{\text{high}} \pm 20^\circ)$ for high-velocity masers. The nominal azimuth $\phi_{\text{high}}$ is taken to be $0^\circ$ for redshifted and $180^\circ$ for blueshifted components, reflecting the strong degeneracy in these regimes. To verify that the numerical marginalization has converged, we performed the following tests: (i) we repeated the analysis with tighter bounds for the systemic masers, 
$(r_{i,\text{max}} \pm 1\,\text{mas},\, \phi_{i,\text{max}} \pm 0.75^\circ)$, and for the high-velocity masers, 
$(r_{i,\text{max}} \pm 0.25\,\text{mas},\, \phi_{\text{high}} \pm 20^\circ)$, obtaining consistent global-parameter estimates; (ii) we visually inspected the $(r_i, \phi_i)$ likelihood surfaces for the resulting global-disk parameters to ensure that all prominent likelihood peaks lie within the adopted integration windows; and (iii) using the resulting global-disk parameters, we increased the grid resolution and confirmed the stability of the likelihood result.

The individual log-likelihood term $\ln \mathcal{L}(d_{i}|\mathbf{G}, r_{i}, \phi_{i})$ is given by:
\begin{equation}
\begin{split}
\ln \mathcal{L}(d_{i}\mid\mathbf{G}, r_{i}, \phi_{i})
= {} & \,-\tfrac{1}{2}\,(d_{i} - \mu_{i})^{T}\,\Sigma_{i}^{-1}\,(d_{i} - \mu_{i}) \\
     & \quad - \ln \sqrt{\det \Sigma_{i}}\,.
\end{split}
\label{eq:LogLgi}
\end{equation}
The observed data vector $d_{i}$, the model prediction $\mu_{i}$, and the covariance matrix $\Sigma_{i}$ are:
\begin{equation}
\label{eq:d_Sigma}
\begin{split}
d_{i} &= 
\begin{pmatrix}
X_{\text{obs,{i}}} \\
Y_{\text{obs,{i}}} \\
v'_{\text{obs,{i}}} \\ 
a_{\text{obs,{i}}} \\ 
\end{pmatrix},
\quad
\mu_{i} = 
\begin{pmatrix}
X_{\text{model,{i}}} \\
Y_{\text{model,{i}}} \\
v'_{\text{model,{i}}} \\ 
a_{\text{model,{i}}} \\ 
\end{pmatrix},
\\
\Sigma_{i} &= 
\begin{pmatrix}
\sigma_{X,i}^2 & 0 & 0 & 0 \\
0 & \sigma_{Y,i}^2 & 0 & 0 \\
0 & 0 & \sigma_{v,i}^2 & \rho_{va, i} \sigma_{v,i} \sigma_{a,i} \\
0 & 0 & \rho_{va,i} \sigma_{v,i} \sigma_{a,i} & \sigma_{a,i}^2
\end{pmatrix}.
\end{split}
\end{equation}
Note that single-maser components that are grouped together have the same acceleration value and uncertainty, and the velocity-acceleration correlation is captured via $\rho_{va}$ in the off-diagonal terms.
We assign flat priors to all 14 global parameters. We restrict the maximum-likelihood values of the angular radius $r_{i,\text{max}}$ to the range $[1, 11]$ mas for high-velocity masers and $[1, 7]$ mas for systemic masers. The the maximum-likelihood values of the azimuthal angle $\phi_{i,\text{max}}$ is limited to $[-20^\circ, 20^\circ]$ for redshifted masers, $[160^\circ, 200^\circ]$ for blueshifted masers, and $[260^\circ, 280^\circ]$ for systemic masers. We perform the fit using $10^3$ live points and adopt a stopping criterion of $\Delta \log_{10} \mathcal{Z} = 10^{-2}$, where $\mathcal{Z}$ is the Bayesian evidence. If error floors are added, they should be added in quadrature to the uncertainties $\sigma$. However, unlike previous analyses, we do not include error floors, nor do we treat them as free parameters. Instead, by manually tracking maser components across epochs and incorporating their linewidths into the error model, we reduce the need to account for unquantified uncertainties—as was necessary in the averaged analyses.

To validate the fitting program, we generate a synthetic dataset to test whether it can reliably recover a known input distance. The procedure is described in detail in Appendix~\ref{appendix:synthetic_data}; here, we provide a brief summary. A representative dataset from our analysis—defined by a specific set of selection parameters—is selected, and a synthetic counterpart is constructed with a similar number of identified maser components, the same number of epochs per group, and comparable measurement uncertainties. The synthetic trajectories are generated using the method outlined in Appendix~\ref{appendix:synthetic_data}, based on the global disk parameters from \citetalias{Reid_2019}, with an input distance of $D = 7.6\,\mathrm{Mpc}$. Fitting the warped disk model to approximately 50 synthetic realizations yields an inferred distance of $D=7.60^{+0.14}_{-0.28}\,\mathrm{Mpc}$, in agreement with the input value.

\subsection{Results}
\label{subsec:results}

The results of fitting the warped disk model to the 17 retained datasets from Section~\ref{subsec:obtained_trajectories} are summarized in Table~\ref{tab:results_newdata}. For each configuration, the table lists the inferred distance to NGC~4258, the black hole mass, and the reduced chi-squared value of the fit. 
The reduced chi-squared value is evaluated using $(r_{i,\mathrm{max}}, \phi_{i,\mathrm{max}})$ values of each data entry that maximize the individual log-likelihood (Equation~\eqref{eq:LogLgi}), given the resulting global disk parameters. The distance posterior distributions are shown in blue in Figure~\ref{fig:hist_results_table2}, with the \citetalias{Reid_2019} distance posterior shown in red for comparison. All the results shown are statistically reliable, with the high-density regions likely reflecting datasets that are effectively identical. The relevant information is contained in the overall spread of the distances.
Complete sets of fitted parameters, posterior samples, and derived quantities for each dataset are available in digital format at \url{bit.ly/400pIcn}.

\begin{figure*}
    \centering
    \includegraphics[width=0.9\linewidth]{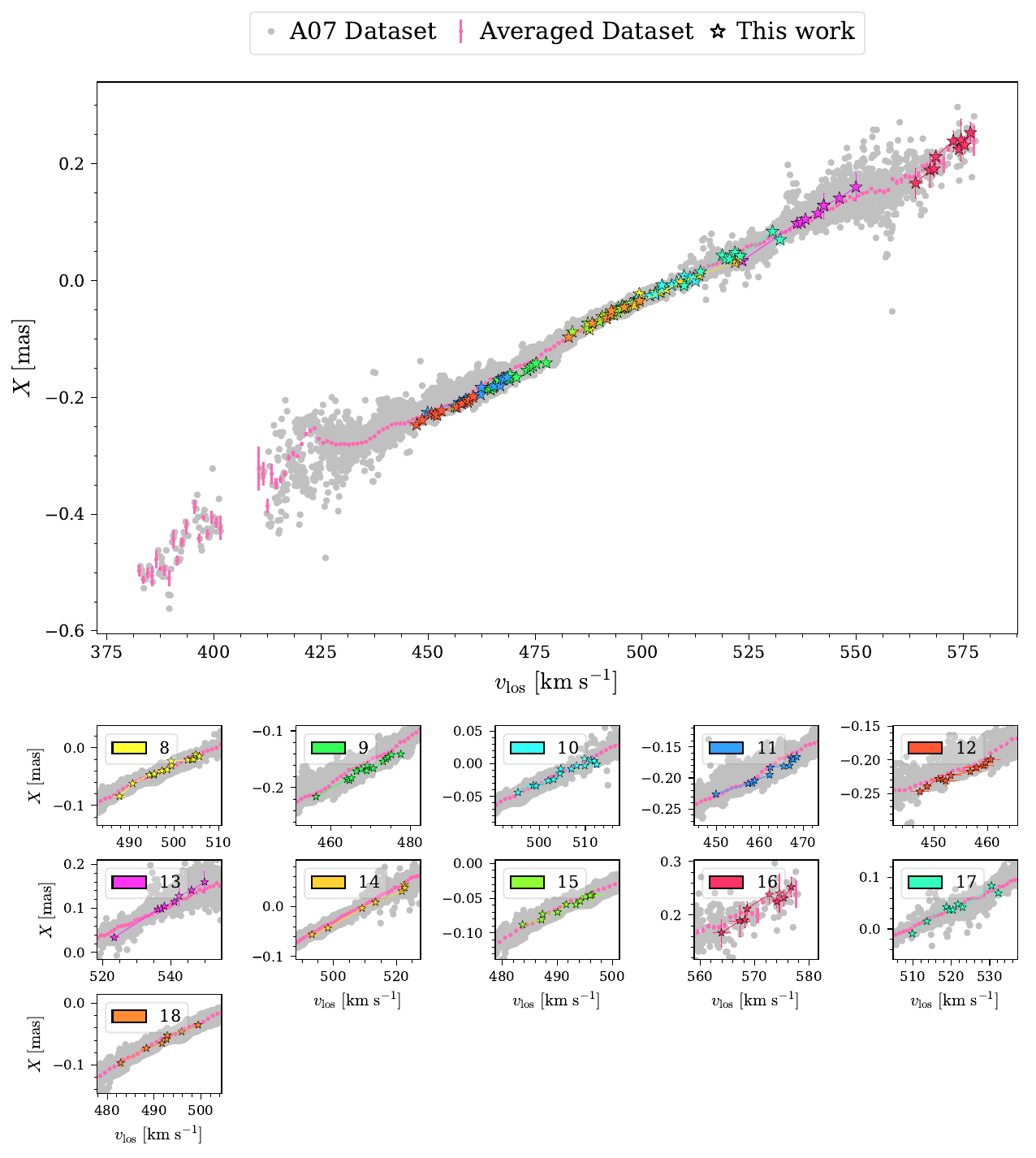}
    \caption{Comparison between a representative set of identified maser components (colored stars) and the averaged dataset from \citetalias{Humphreys_2013} (pink) as well as the original \citetalias{Argon_2007} dataset (gray), plotted in the $X$–$v_{\text{los}}$ plane within the systemic velocity region. The data shown correspond to dataset \#8 in Table~\ref{tab:results_newdata}. Each maser component is plotted with the same color and group ID as in Figure~\ref{fig:ransac}, representing an individual maser feature tracked over time. These group IDs are consistent with those provided in the public digital catalog\footref{fn:url}. Zoomed-in subpanels highlight specific maser trajectories. The reconstructed components do not necessarily align with the structure of the averaged dataset, illustrating key limitations of the averaging approach. In particular, the averaged dataset does not account for the possibility that distinct maser components may overlap in velocity space while remaining spatially separated.}

    \label{fig:xv_sys_traj}
\end{figure*}

\begin{figure*} 
    \centering
    \begin{tabular}{ccc}
        \includegraphics[width=0.32\textwidth]{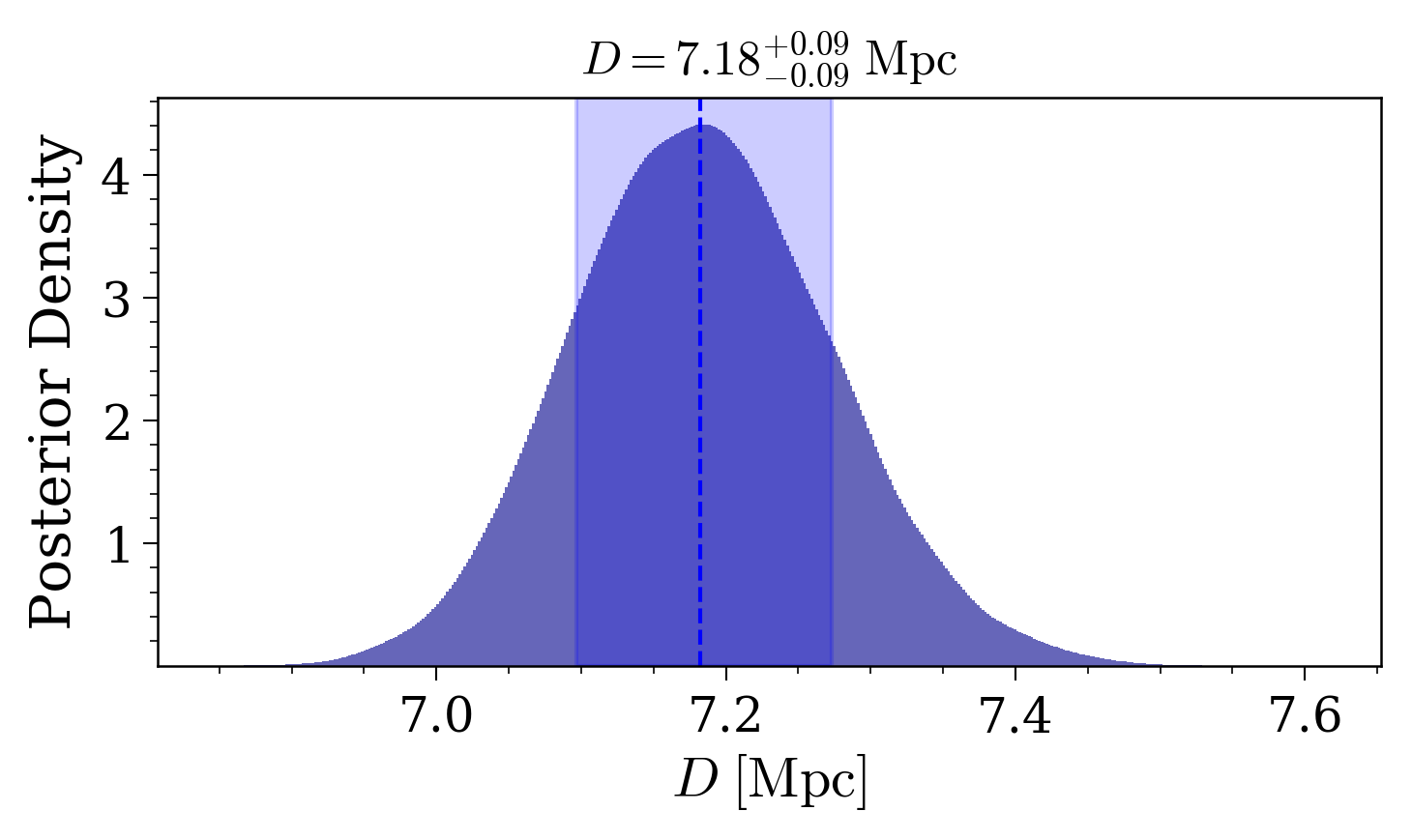} &
        \includegraphics[width=0.32\textwidth]{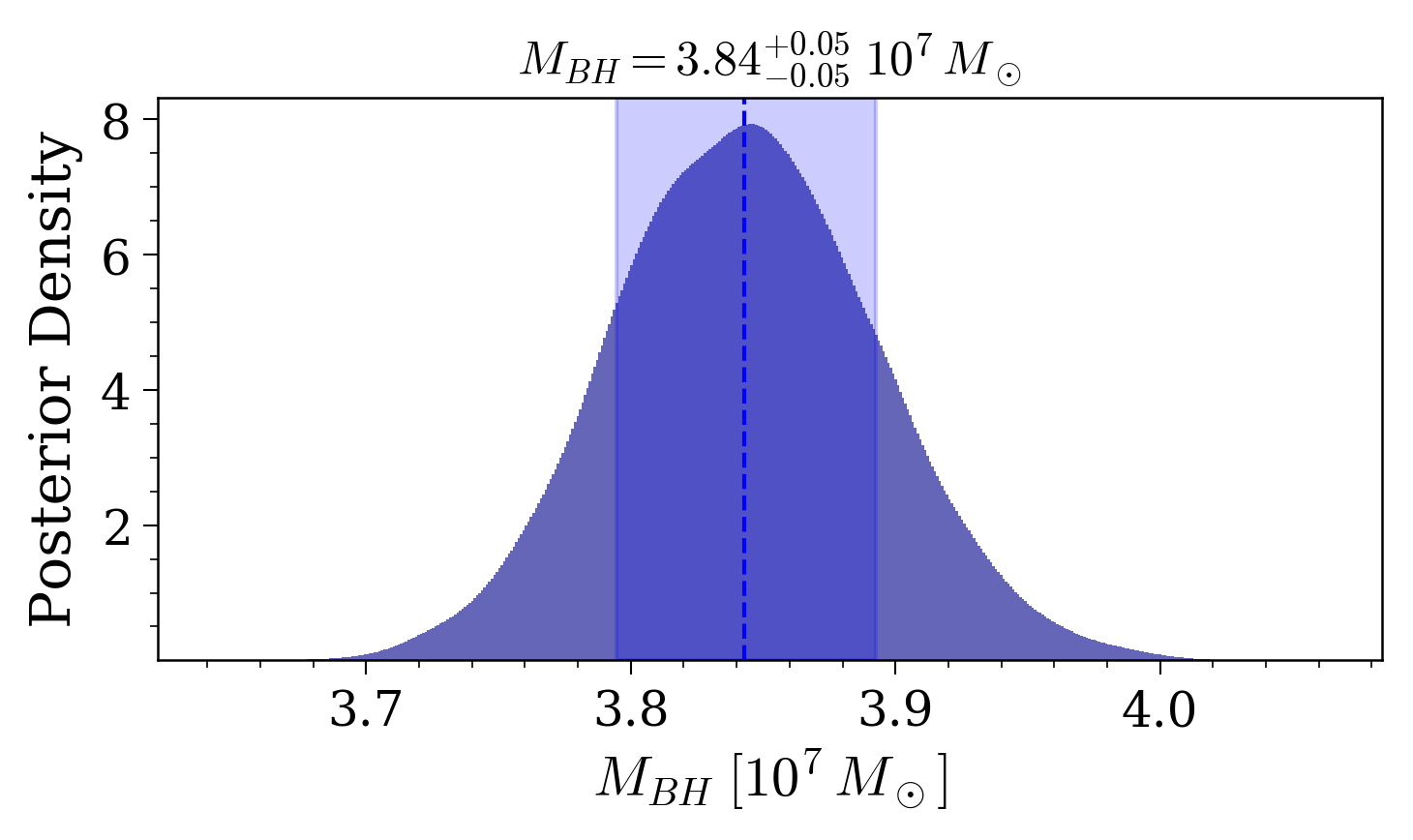} &
        \includegraphics[width=0.32\textwidth]{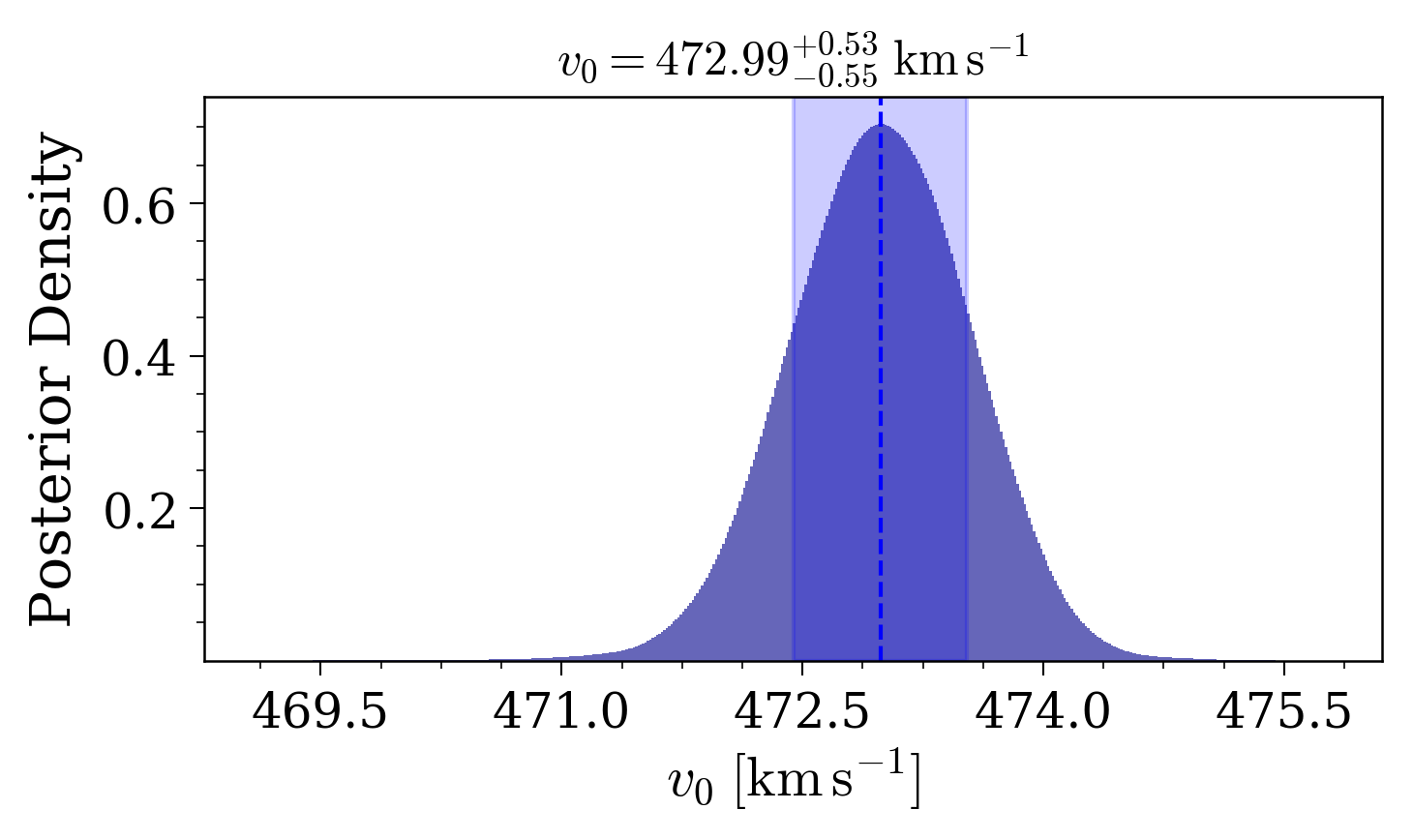} \\
        \includegraphics[width=0.32\textwidth]{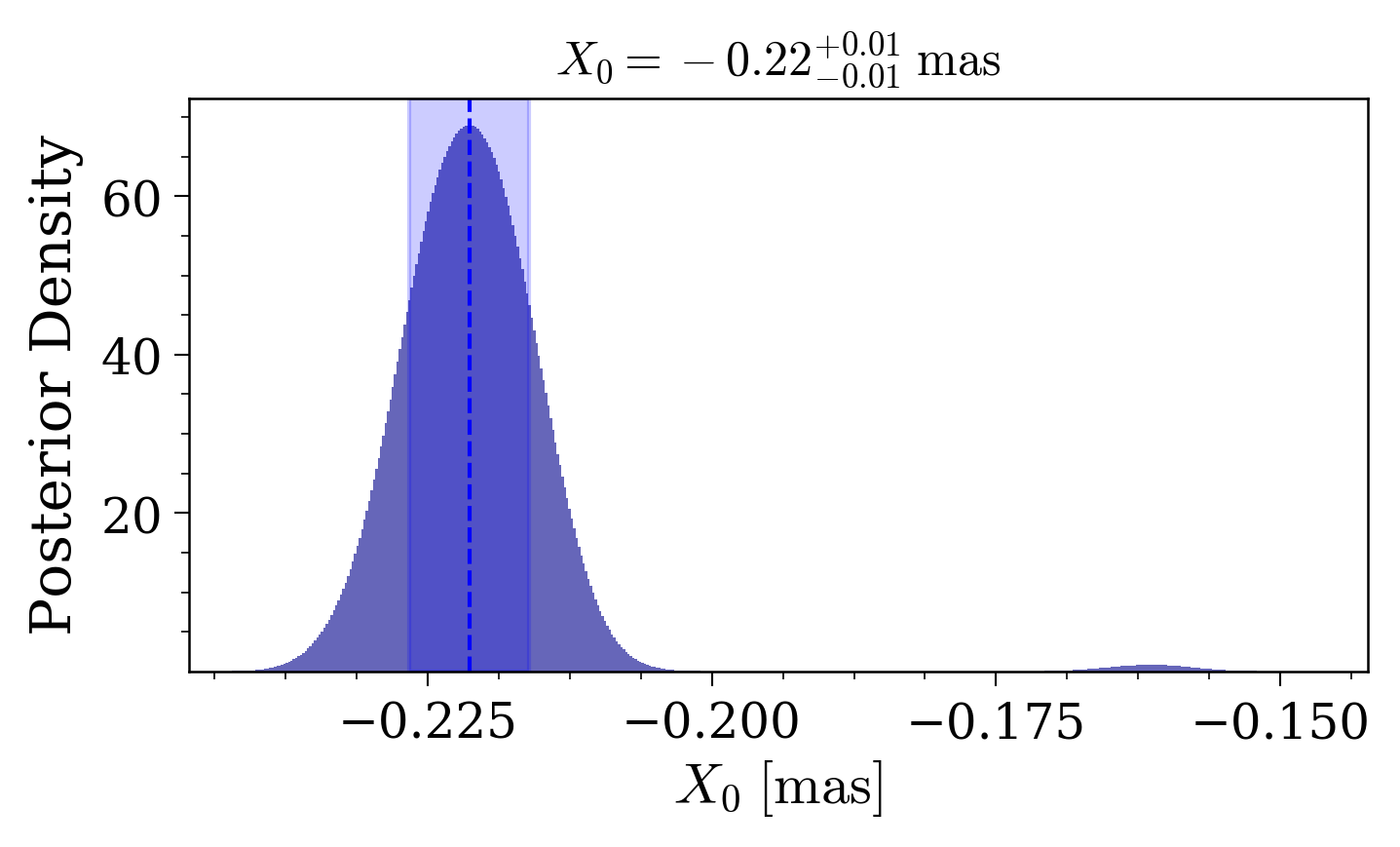} &
        \includegraphics[width=0.32\textwidth]{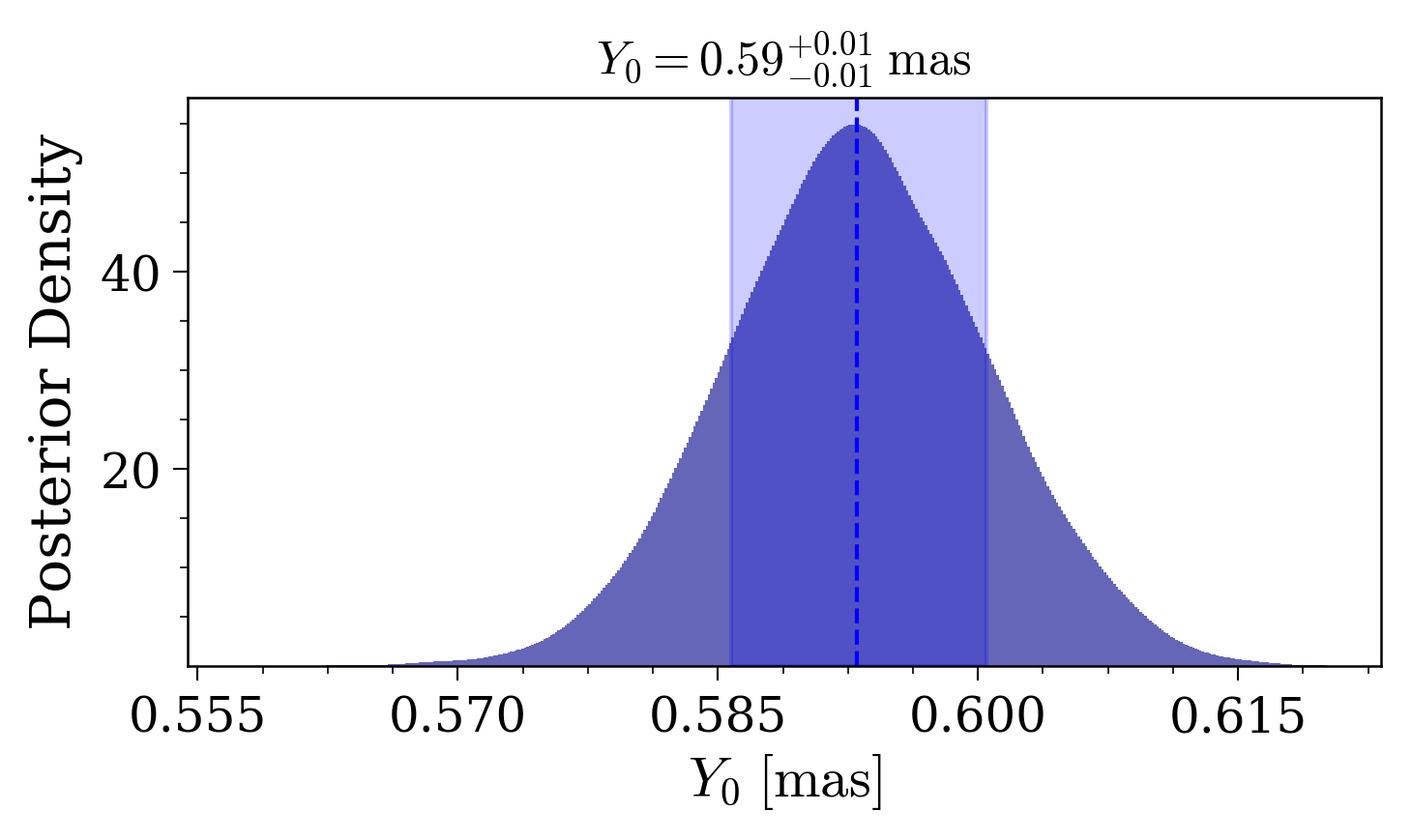} &
        \includegraphics[width=0.32\textwidth]{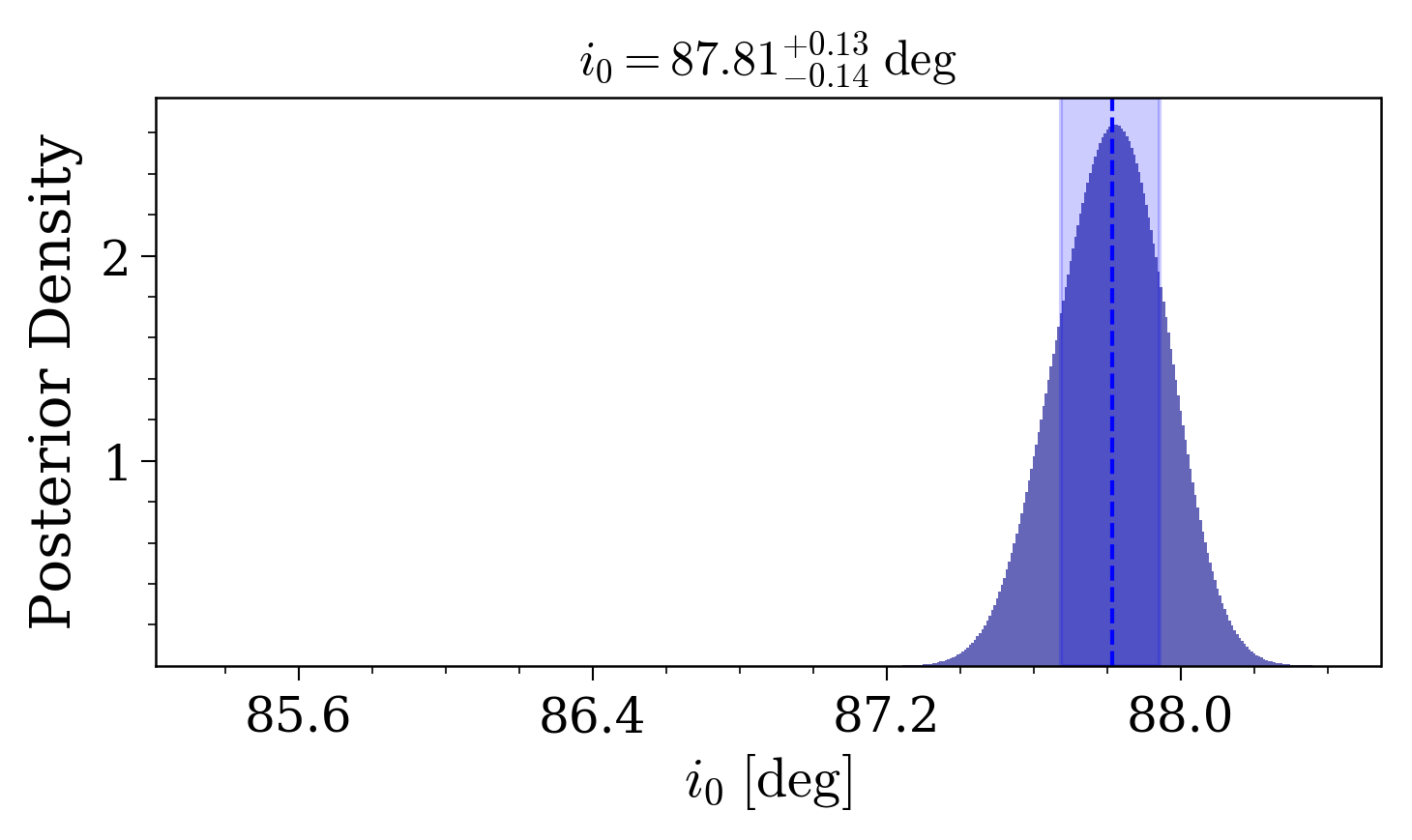} \\
        \includegraphics[width=0.32\textwidth]{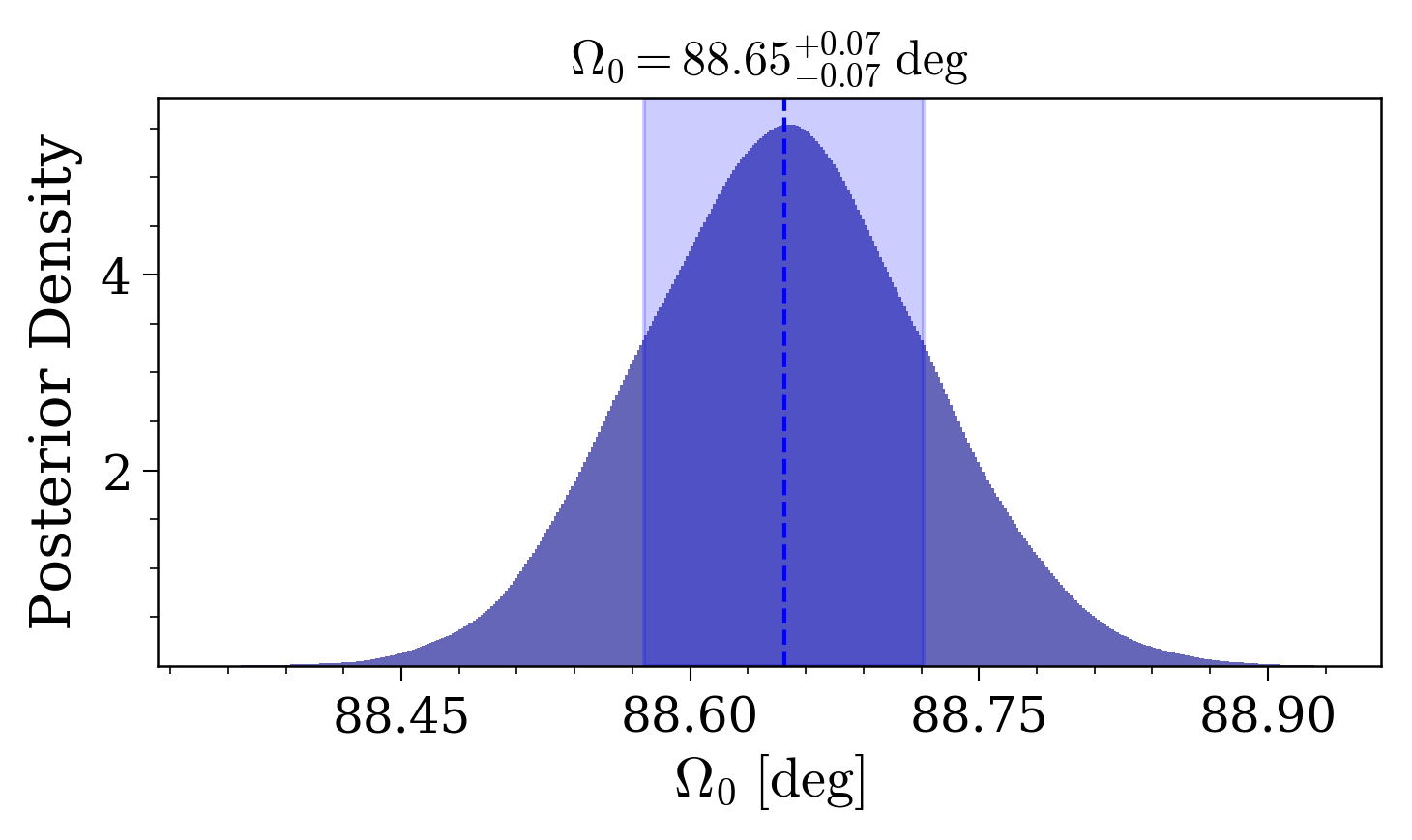} &
        \includegraphics[width=0.32\textwidth]{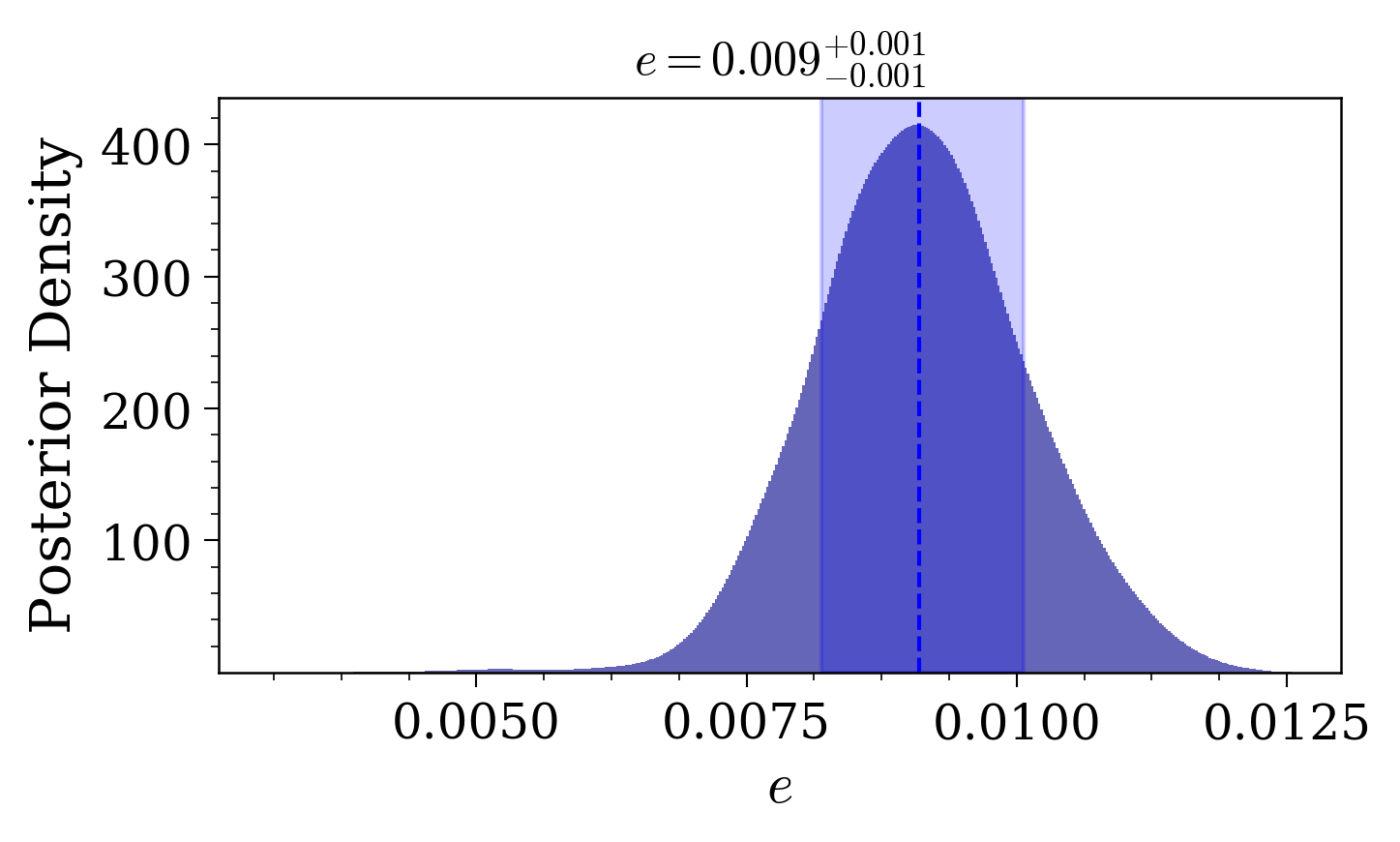} &
        \includegraphics[width=0.32\textwidth]{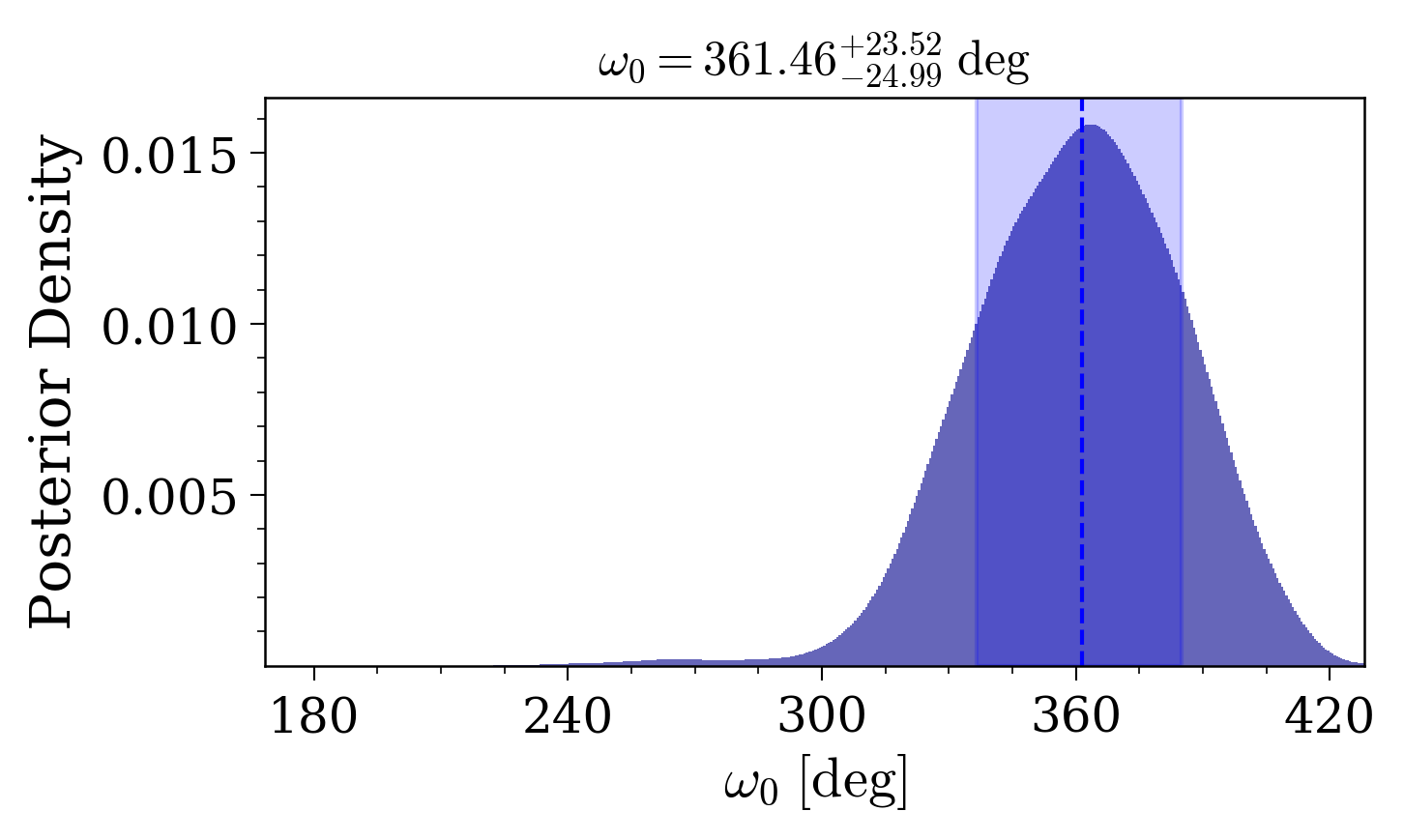} \\
        \includegraphics[width=0.32\textwidth]{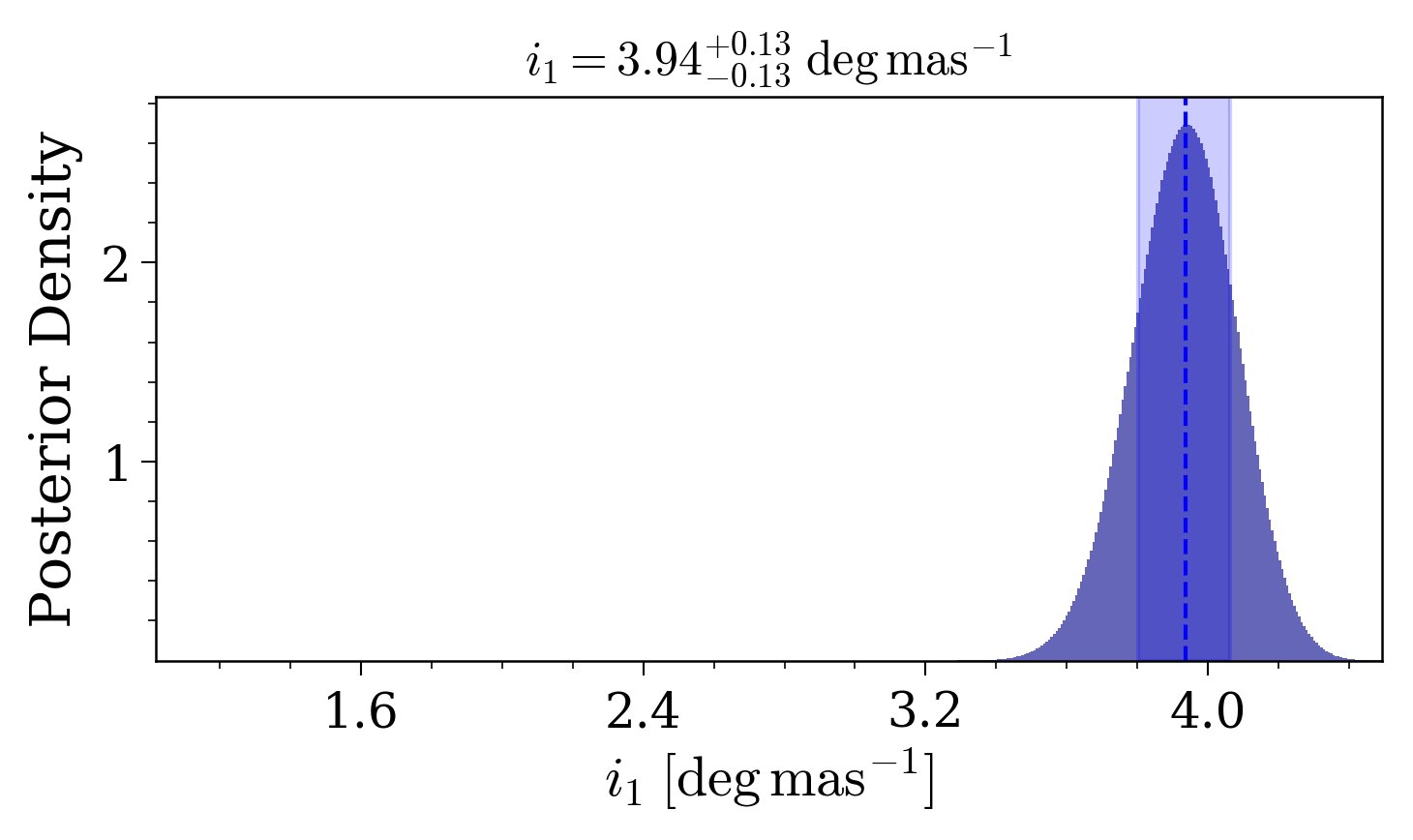} &
        \includegraphics[width=0.32\textwidth]{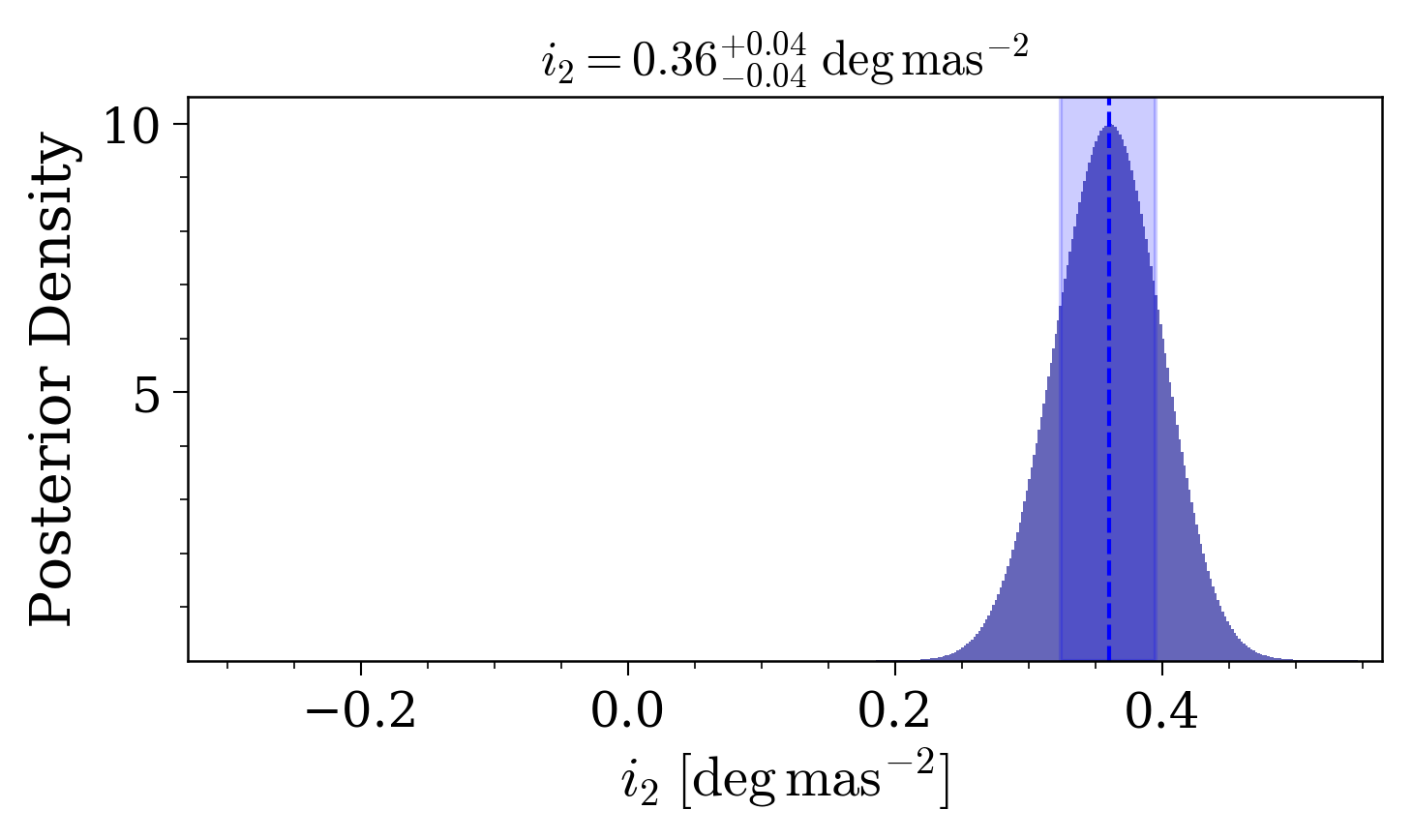} &
        \includegraphics[width=0.32\textwidth]{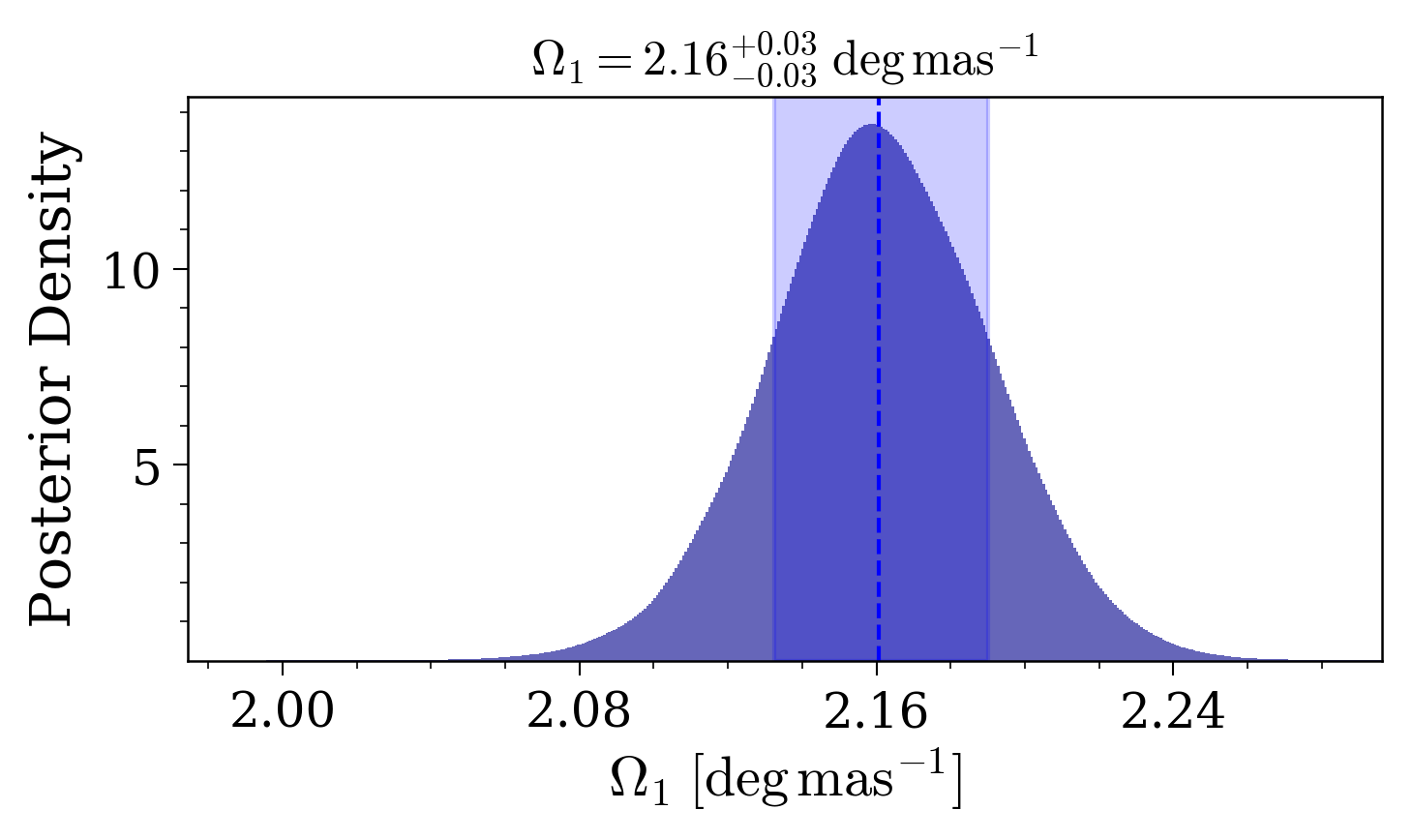} \\
        \includegraphics[width=0.32\textwidth]{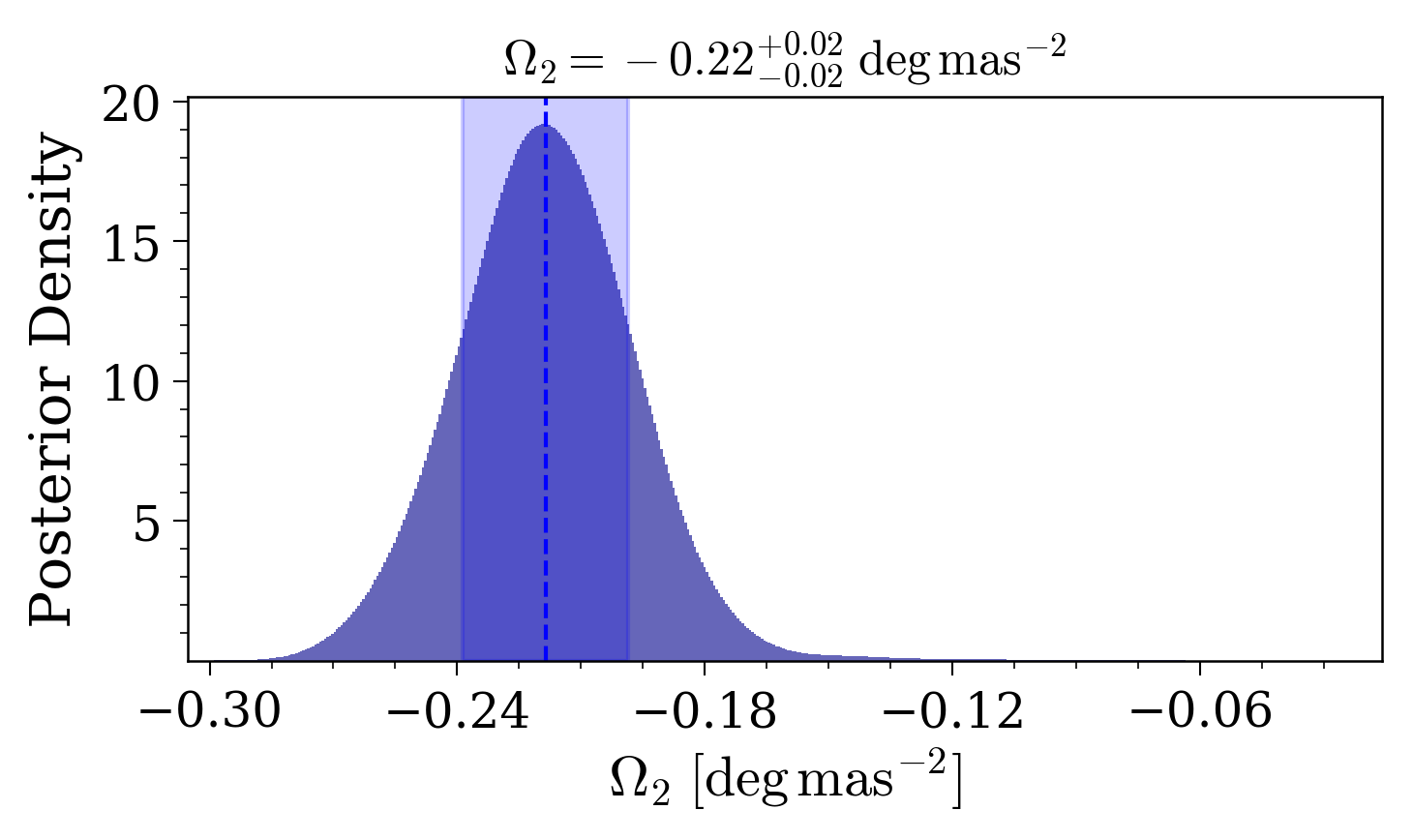} &
        \includegraphics[width=0.32\textwidth]{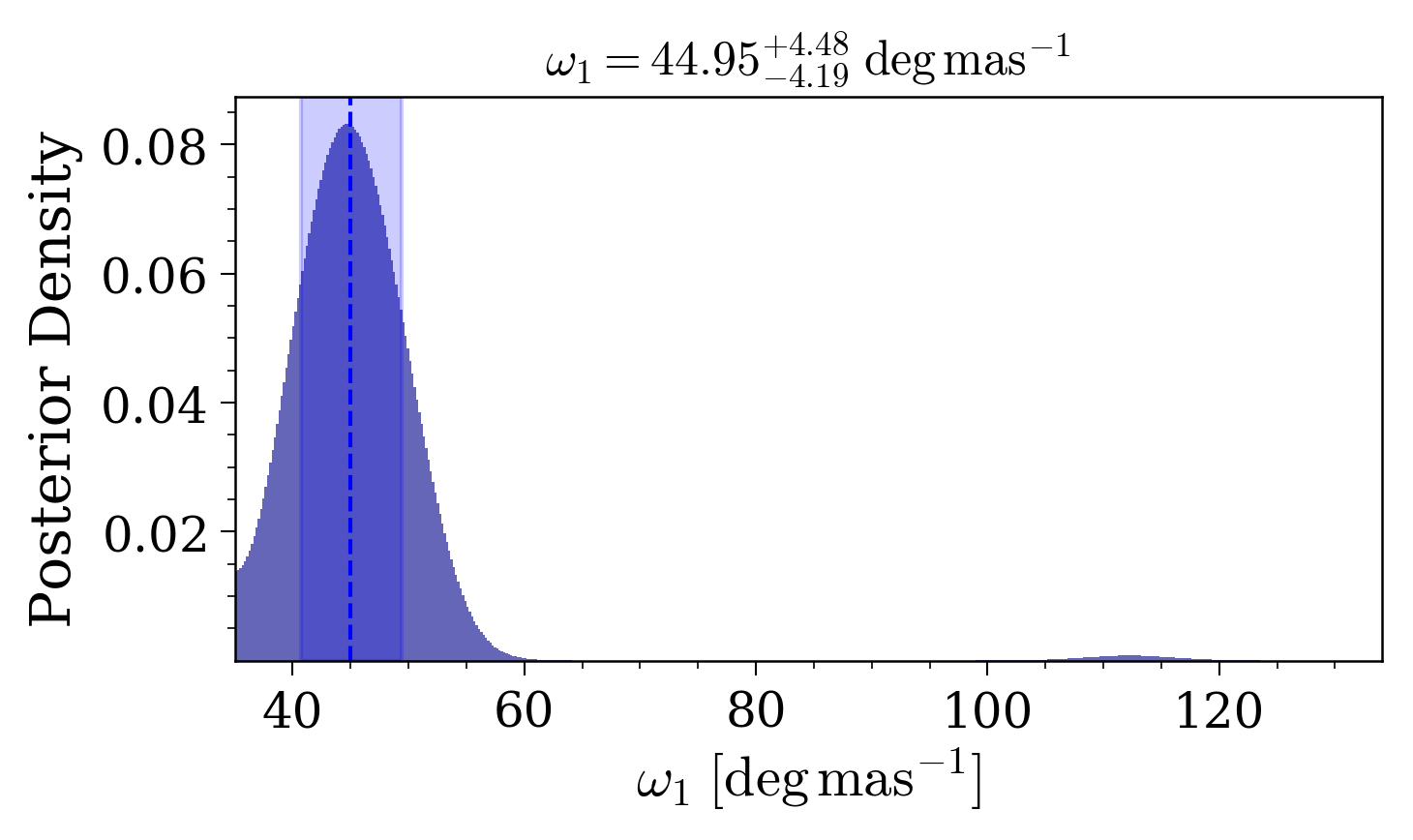} & 
    \end{tabular}
    \vspace{-10pt} 
    \caption{Posterior distribution of the global disk parameters for the representative dataset \#8 of Table~\ref{tab:results_newdata}. The blue dashed line marks the median of each posterior distribution, while the shaded regions correspond to the 16th and 84th percentiles. The parameter values and their associated uncertainties shown in the panel titles are derived from these percentiles.}
    \label{fig:posterior_representive}
\end{figure*}

Figure~\ref{fig:xv_sys_traj} displays a representative dataset (colored stars) on the $X$–$v_{\text{los}}$ plane in the systemic velocity region, overlaid with the averaged dataset (pink) and the original \citetalias{Argon_2007} measurements (gray). This dataset corresponds to entry~\#8 in Table~\ref{tab:results_newdata} and is the same one shown in Figure~\ref{fig:ransac}. Each maser component trajectory is shown in a different color, with zoomed-in views provided in the subpanels. The color coding and group ID labels match those used in Figure~\ref{fig:ransac}, and correspond to the identifiers listed in our digital catalog\footref{fn:url}. As illustrated, the identified maser components do not necessarily align with the averaged dataset, highlighting the limitations of the averaging approach and demonstrating the advantages of direct maser tracking across epochs. The best-fit distance inferred for this dataset is $D = 7.18\pm0.09$ Mpc, with a reduced chi-squared value of $\chi^2_\nu \approx 1.14$ and 394 degrees of freedom, indicating a statistically acceptable fit. 
Further results are provided in Appendix~\ref{appendix:traj_results}, which shows the time evolution of each maser component in position, LOS velocity, and acceleration, along with trajectories computed using samples drawn from the posterior distributions. The posterior distributions of the global disk parameters for this fit are shown in Figure~\ref{fig:posterior_representive}, where the dashed lines and reported parameter values represent the medians, and the shaded regions and quoted uncertainties correspond to the 16th and 84th percentiles.

While the previously discussed dataset yields a statistically robust and visually compelling fit, adopting a different selection criterion can lead to noticeably different disk parameters. For example, dataset~\#1 in Table~\ref{tab:results_newdata}, which includes 91 single-epoch maser components (roughly a third of the number of components in the representative dataset), results in a best-fit distance of $D = 8.091\pm0.18$ Mpc. This estimate has twice the uncertainty of the representative dataset and differs from it by $\myapprox5\sigma$. Nevertheless, the fit remains statistically strong, with a reduced chi-squared value of $\chi^2_\nu \approx 1.05$ and $168$ degrees of freedom. As shown in Appendix~\ref{appendix:traj_results}, this solution is also visually convincing. This example highlights the sensitivity of the inferred disk parameters to the adopted selection criteria: different constraint choices can yield distinct yet equally plausible models. In practice, the flexibility in defining and tracking maser components over a long time baseline introduces considerable ambiguity, and without additional observational constraints, uniquely determining the disk structure becomes challenging.

Using a goodness-of-fit criterion that requires the probability of obtaining a smaller or larger $\chi^2_{\nu}$ by chance to be less than $10^{-3}$, we find that most of the dataset configurations yield statistically acceptable fits, except for two configurations. Among the statistically acceptable configurations, the inferred distances range from $\myapprox6.7$ to $\myapprox8.1\,\text{Mpc}$ with distance uncertainties ranging from $\myapprox0.07$ to $\myapprox0.19\,\rm{Mpc}$.

\subsubsection{The Statistical Uncertainties}
\label{subsec:stat_errs2}
\begin{figure}
    \centering
    \includegraphics[width=0.9\linewidth]{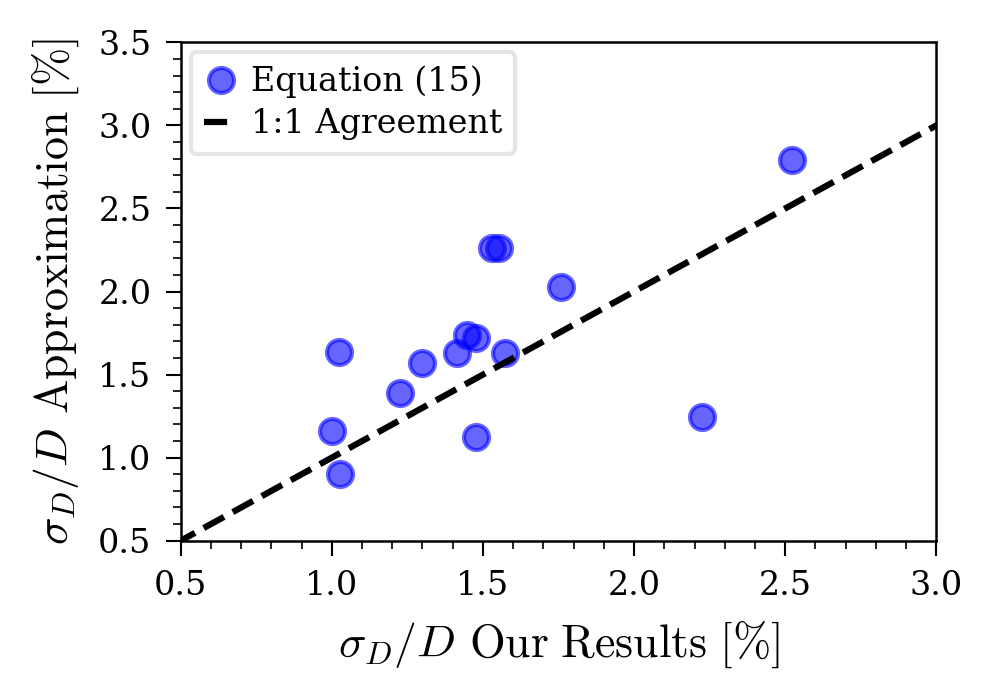}
    \caption{Comparison between the relative distance uncertainties derived from Table~\ref{tab:results_newdata} and the analytical estimates from Equation~\eqref{eq:sigma_D_D}. Each blue circle represents one dataset, plotted with the measured distance uncertainty and the analytical prediction. The black dashed line marks the 1:1 relation, indicating perfect agreement between the analytical approximation and the numerical result. The approximation exhibits reasonable consistency with the uncertainties reported in Table~\ref{tab:results_newdata}.}
    \label{fig:dD_D}
\end{figure}

The precision of the inferred distance can be estimated following the methodology outlined in Section~\ref{subsec:stat_errs}, which relies on the systemic masers. In our analysis, the single-epoch maser components are not independent of one another, whereas the identified tracked maser groups are. Thus, the relative uncertainty of each maser group, $R_g$, can be estimated by using Equation~\eqref{eq:a_kepler}:
\begin{equation}
    R_g^2 \approx \frac{\sigma_{a,g}^2}{a_g^2},
\label{eq:R_j}
\end{equation}
where we assume that the ratio $M_\text{BH}/D$ is well constrained by the other velocity measurements. For example, in dataset $\#8$, using typical values of $a \sim 8.5\,\text{km s}^{-1}\,\text{yr}^{-1}$ and $\sigma_a \sim 0.4\,\text{km s}^{-1}\,\text{yr}^{-1}$, we obtain $R \sim 5\%$ for the grouped components.
To estimate the overall uncertainty in the distance, we combine the uncertainties from time-tracked groups $g$ using the following inverse-variance estimator:
\begin{equation}
    \frac{\hat{\sigma}_D^2}{\hat{D}^2} = \frac{1}{\sum_{g} \frac{1}{R_{g}^2}}.
\label{eq:sigma_D_D}
\end{equation}
assuming that each time-tracked group is independent. In Figure~\ref{fig:dD_D}, we compare the relative uncertainties obtained from Table~\ref{tab:results_newdata} with those estimated using Equation~\eqref{eq:sigma_D_D}. These approximations show reasonable agreement with the measured uncertainties reported in Table~\ref{tab:results_newdata}. For example, the relatively small statistical uncertainty of $\myapprox1.3\%$ for the representative dataset~$\#8$ (comparable to the statistical uncertainties reported by \citetalias{Reid_2019}) is similar to $\myapprox1.4\%$ obtained from Equation~\eqref{eq:sigma_D_D}. The simple estimate $R / \sqrt{N}$, using the typical values of $R$ discussed above with 11 systemic maser groups, yields relative distance uncertainties of $\mysim1.5\%$.

\section{Optimizing Observational Cadence to Improve Distance Accuracy}
\label{sec:obs_suggestion}

The results of Section~\ref{subsec:results} indicate that, while it is possible to achieve a distance precision of $\myapprox0.05$–$0.2\,\mathrm{Mpc}$ for a specific set of maser tracking criteria, the ambiguity in defining and following individual maser components over a long time baseline introduces significant distance uncertainty. Here, we show that this sensitivity arises primarily from the limited observational cadence, which is insufficient to reliably track individual masers given their high intrinsic variability. We then propose an observational strategy aimed at reducing this sensitivity—potentially lowering the distance uncertainty to the percent level. Our analysis focuses on the systemic maser features, as they are most directly linked to the geometric distance measurement.

\begin{figure}
    \centering
    \includegraphics[width=0.9\linewidth]{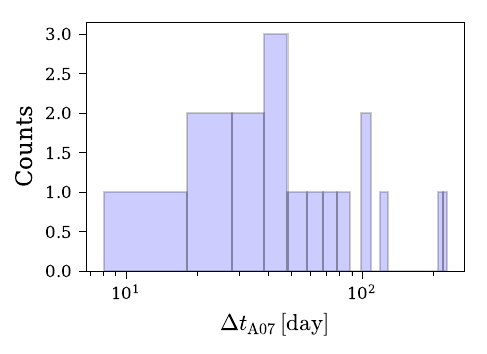}
    \caption{Histogram of time gaps between systemic maser observations in the \citetalias{Argon_2007} dataset, plotted on a logarithmic $x$-axis with a bin width of 10 days.}
    \label{fig:hist_dt}
\end{figure}

\begin{figure*}
    \centering
    \includegraphics[width=0.9\linewidth]{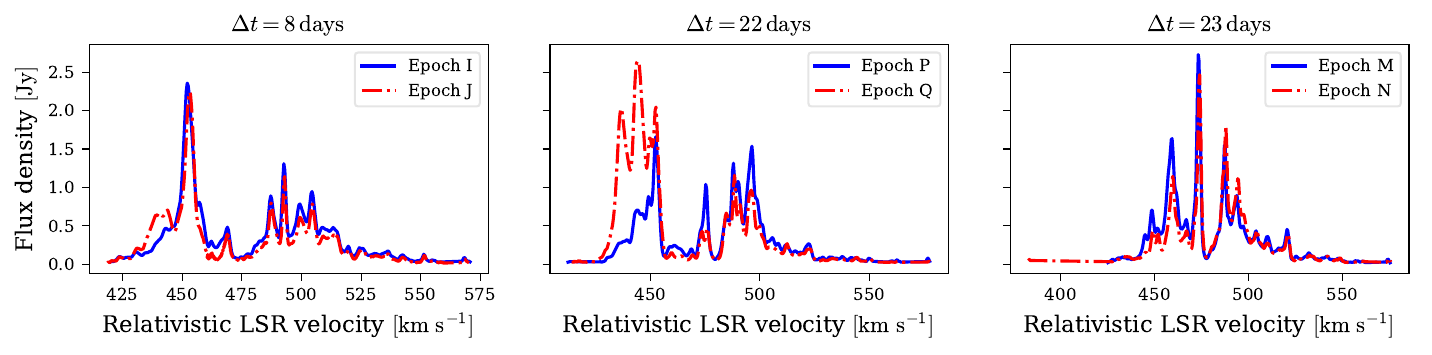}
    \caption{Comparison of velocity-integrated spectra showing flux density as a function of relativistic LSR velocity for three epoch pairs with different time separations: 8 days (left; epochs I and J), 22 days (middle; epochs P and Q), and 23 days (right; epochs M and N).}
    \label{fig:compare_spec}
\end{figure*}

Figure~\ref{fig:hist_dt} shows the distribution of time gaps between adjacent epochs in the \citetalias{Argon_2007} dataset, ranging from 8 to 209 days. Inspection of the \citetalias{Argon_2007} spectra reveals that significant spectral changes can occur between epochs—likely due to the fact that the systemic masers lie directly in front of the supermassive black hole, where local conditions may evolve rapidly. In Figure~\ref{fig:compare_spec}, we compare three pairs of epochs with separations of 8 days (left; epochs I and J), 22 days (middle; epochs P and Q), and 23 days (right; epochs M and N). The 8-day pair shows clear spectral coherence, while the 22-day pair exhibits substantial differences. Interestingly, the 23-day pair retains coherence, suggesting that the threshold for reliable spectral tracking lies near $\mysim$22 days. These comparisons indicate that a time gap of approximately 8 days is reliably short enough to preserve spectral coherence for tracking individual maser peaks, whereas gaps of $\mysim$22 days may coincide with significant spectral evolution, making reliable identification more challenging.

\begin{figure*}
    \centering
    \includegraphics[width=0.9\linewidth]{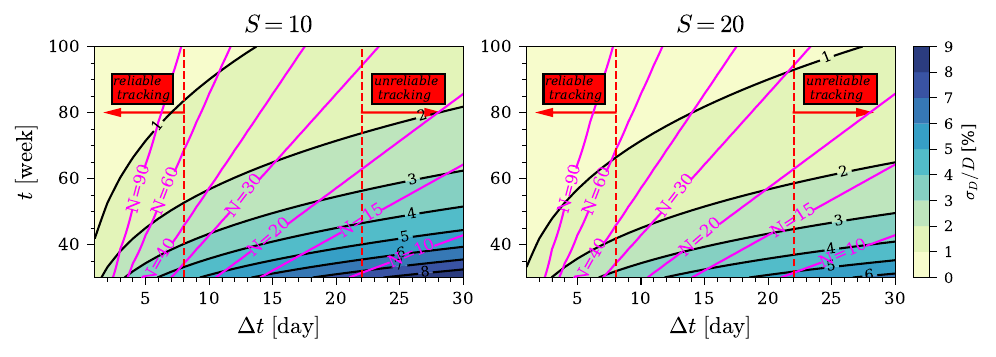}
    \caption{Estimated relative uncertainty in the distance, $\sigma_D / D$, as a function of observation cadence $\Delta t$ and total time span $t$. The relative uncertainty is computed using Equation~\eqref{eq:sigD_D}, assuming an acceleration of $\hat{a} = 8\,\textrm{km}\,\textrm{s}^{-1}\,\textrm{yr}^{-1}$ and a velocity measurement uncertainty of $\sigma_v = 1\,\textrm{km}\,\textrm{s}^{-1}$. The left panel assumes $S = 10$ tracked maser components; the right assumes $S = 20$. Pink contours indicate the number of observational epochs, $N = t / \Delta t$. Red dashed vertical lines mark cadence thresholds inferred from the \citetalias{Argon_2007} dataset, interpreted as boundaries between "unreliable" and "reliable" tracking. The region between these thresholds represents an intermediate regime where tracking fidelity remains uncertain and warrants further investigation.}
    \label{fig:dD_D_suggestion}
\end{figure*}

Next, we apply the statistical framework developed in Section~\ref{subsec:stat_errs2} to evaluate how the expected distance uncertainty depends on the observational cadence and total monitoring duration. As shown in Figure~\ref{fig:dD_D}, Equation~\eqref{eq:sigma_D_D} provides an adequate approximation for the relative distance uncertainty, and we adopt it here for analytic estimates. To estimate the relative uncertainty in acceleration, we assume a typical value of $\hat{a} \approx 8\,\mathrm{km}\,\textrm{s}^{-1}\,\textrm{yr}^{-1}$. We further assume that $S$ maser components can be reliably identified and tracked over a total timespan $t$, with an observational cadence of $\Delta t$ days. The uncertainty in the LOS velocity measurements is taken to be $\sigma_v = 1\,\mathrm{km}\,\textrm{s}^{-1}$. Under these assumptions, the relative uncertainty in the distance can be approximated using the standard expression for the variance of a slope in linear regression:
\begin{equation}
\label{eq:sigD_D}
    \frac{\hat{\sigma}_D}{\hat{D}} = \frac{1}{\hat{a}} \sqrt{\frac{12 \sigma_v^2 \Delta t}{St^3}}.
\end{equation}
Figure~\ref{fig:dD_D_suggestion} illustrates the estimated relative error in the distance as a function of observational cadence and total timespan. The figure includes two panels: one assuming that $S = 10$ maser components are tracked, and the other assuming $S = 20$. Pink contour lines indicate the number of epochs required to span the total duration for each cadence ($N = t / \Delta t$). Red dashed vertical lines mark empirical thresholds for reliable spectral tracking, as discussed earlier. For example, if a cadence of 10 days allows reliable tracking of $S=10$ systemic maser components, then $\myapprox40$ epochs over a total duration of $t\approx60$ weeks would achieve a distance uncertainty at the $\mysim$2\% level. If a year-long campaign is not feasible, the program can be split into segments. For example, split into two segments ($k=2$) of duration $t'=30$ weeks ($t'=t/k$) each; thus, we effectively track twice as many systemic maser components $S'=20$ ($S'=kS$). If we keep the same cadence ($\Delta t' = \Delta t$), Equation~\eqref{eq:sigD_D} implies that the relative precision degrades by a factor of $k = 2$, yielding an overall uncertainty of $\mysim4\%$. Further study is needed to determine the optimal cadence required to robustly track individual maser components and achieve a target distance precision.

\section{Summary and Discussion}
\label{sec:summary}

In this work, we presented a reanalysis of the water maser system in NGC~4258 to reassess its geometric distance, commonly reported as $\mysim7.6\,\mathrm{Mpc}$ with percent-level accuracy. We revisited simplifying assumptions in previous approaches, which involved averaging of the data and did not account for the intrinsic linewidths of maser features, factors that might influence the interpretation of the disk’s geometry and dynamics. To avoid these assumptions, we adopted a method that tracks individual maser components across multiple epochs. Our main conclusion is that the available observational cadence is insufficient to reliably track individual maser components, given their high intrinsic variability. As a result, we showed that different, yet plausible, tracking strategies can yield significantly different distance estimates—even when each method produces small formal uncertainties. Finally, we proposed an observational strategy to reduce these systematic uncertainties and improve the robustness of future distance measurements.

We reviewed prior distance estimates, particularly those by \citetalias{Humphreys_2013} and \citetalias{Reid_2019}, which rely on an averaged dataset constructed from heterogeneous observations (Section~\ref{sec:related_work}). By reproducing their fits under several configurations (Section~\ref{subsec:reproduced results}), we validated our independent fitting implementation.

As detailed in Section~\ref{sec:caveats}, the averaging procedure introduces a critical assumption, the single-orbit approximation, which is not supported by the raw data or physical modeling. It also neglects the line broadening of maser features, which can reach up to FWHM of $2\,\text{km\,s}^{-1}$ due to thermal broadening and hyperfine structure \citep{Varshalovich_2006}, potentially leading to oversampling and underestimated uncertainties (Section~\ref{subsec:thermal_width}). To address these limitations, we construct a new dataset (Section~\ref{sec:D_estimation}) by directly identifying and tracking individual maser components, explicitly avoiding the averaging procedure and its associated assumptions. This new dataset incorporates measured linewidths into the error model, assigns one position per maser feature, and ensures consistency across epochs. Additionally, we introduced an efficient method to marginalize over the nuisance parameters $(r, \phi)$, reducing the number of free parameters from many hundreds to 14. This reduction significantly improves sampling efficiency and enhances convergence reliability. A full catalog of all detected peaks, refined maser trajectories, fitting results, and sub-datasets corresponding to all tested configurations is made publicly available at \url{bit.ly/400pIcn}.

We explore combinations of maser selection and model fitting parameters, of which 17 yield viable datasets (Section~\ref{subsec:obtained_trajectories}). Among these, 15 pass a stringent goodness-of-fit criterion. These accepted configurations result in inferred distances ranging from approximately $\myapprox6.7$ to $\myapprox8.1\,\text{Mpc}$, with distance uncertainties spanning $\myapprox0.07$ to $\myapprox0.19\,\rm{Mpc}$. These results underscore the sensitivity of the disk parameters and inferred distances to the adopted selection criteria in our methodology. Even when fits are statistically strong and visually compelling (Appendix~\ref{appendix:traj_results}), differences at the level of several $\sigma$ persist across configurations—demonstrating the ambiguity inherent in maser tracking over long and sparsely sampled time baselines. We did not attempt to identify or optimize a “best” tracking method, but rather aimed to show that the distance inference is highly sensitive to the underlying assumptions in component identification and tracking.

To reduce this sensitivity, we investigate the effect of observational cadence on maser tracking fidelity (Section~\ref{sec:obs_suggestion}). Analysis of the \citetalias{Argon_2007} spectra shows that epochs spaced by $\lesssim8$ days preserve spectral coherence, while separations of $\gtrsim22$ days can introduce substantial variability (Figure~\ref{fig:compare_spec}). We estimate that tracking 10 systemic maser components over $\mysim$40 epochs with 10-day cadence ($\myapprox14$-month span) would enable distance measurements with $\sim$2\% uncertainty (Figure~\ref{fig:dD_D_suggestion}). 

Within the limits of current data and our methodology approach, NGC 4258 does not yet provide a sufficiently precise distance to serve as a high-precision anchor for the distance ladder. In particular, the method proposed by \citet{Kushnir_2024}, which uses NGC 4258 as a sole anchor to substantially reduce Cepheid-related systematic uncertainties, would suffer from insufficient accuracy in the current geometric distance measurement—thereby limiting its utility for a precise determination of $H_0$. Alternative approaches that rely on other anchors, such as the Large Magellanic Cloud (LMC) and the Milky Way (MW), would be less affected by this limitation, although their reliance on objects at very different distances may introduce other (as yet unidentified) systematics, as emphasized by \citet{Kushnir_2024}. A comprehensive assessment of how reduced accuracy in the NGC 4258 distance impacts various determinations of $H_0$ is beyond the scope of this work. Nonetheless, for our approach, higher-cadence monitoring of NGC 4258 would provide a clear path toward achieving percent-level distance precision, thereby enhancing its potential as a robust anchor in the extragalactic distance scale. 

\section*{Acknowledgements}

We thank Dominic W. Pesce for his insightful discussion and valuable feedback throughout this work. We are also grateful to Adam G. Riess for his thoughtful comments and suggestions. We thank Mark J. Reid for providing the averaged dataset and for his helpful comments, and Elizabeth M. L. Humphreys for sharing Figure~\ref{fig:coor_sys_HUM13}. We are grateful to Boaz Katz, Barak Zacky, Eran Ofek, and Reinhard Genzel
for valuable discussions. We thank Avshalom Badash, Tal Wasserman, Jonathan Mushkin, Amir Sharon, Ido Irani and Guy Trostianetsky for helpful input. DK is supported by a research grant from The Abramson Family Center for Young Scientists, an ISF grant, the Minerva Stiftung, and the Pazi foundation.

\section*{Data Availability}
All materials are available at \url{bit.ly/400pIcn}. This includes a complete summary of all identified peaks as described in Section~\ref{subsubsec:identify_peaks}, covering both the initial detections and the refined catalog, together with their associated positional information. It also includes the full sets of fitted parameters, posterior samples, and derived quantities for each dataset listed in Table~\ref{tab:results_newdata}.



\bibliographystyle{mnras}
\bibliography{example}



\appendix


\section{The Warped Disk Model}
\label{appendix:warped}
\begin{figure}
    \centering
    \includegraphics[width=0.9\linewidth]{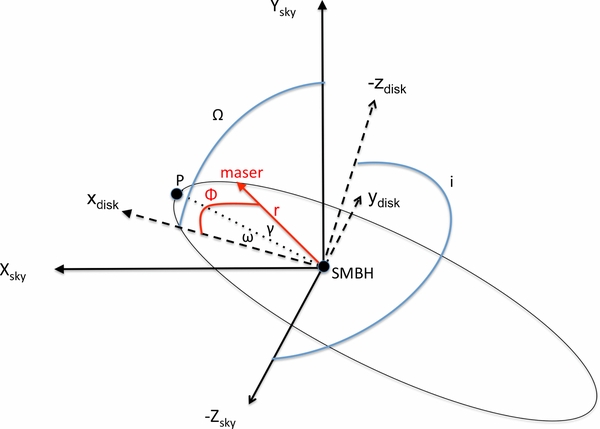}
    \caption{Figure adapted from \citetalias{Humphreys_2013}. The geometry of the maser disk projected onto the sky plane ($X_\text{sky}-Y_\text{sky}$) is shown alongside the intrinsic disk plane coordinates ($x_{\text{disk}}-y_{\text{disk}}$). Both coordinate systems are centered on the supermassive black hole (SMBH).}
    \label{fig:coor_sys_HUM13}
\end{figure}

The model presented below is based on \citetalias{Humphreys_2013}, with minor modifications proposed by \citetalias{Reid_2019}. Figure~\ref{fig:coor_sys_HUM13} illustrates the geometry of the maser disk and its coordinate system projected onto the sky plane. The east–west and north–south sky positions are modeled as:
\begin{equation}
\begin{split}
    X_{\text{model}}&=r(\sin\Omega(r)\cos\phi-\cos\Omega(r)\cos i(r)\sin\phi)+X_{0}, \\
    Y_{\text{model}}&=r(\cos\Omega(r)\cos\phi+\sin\Omega(r)\cos i(r)\sin\phi)+Y_{0},
\end{split}
\end{equation}
where $X_0$ and $Y_0$ denote the coordinates of the disk center, $\Omega(r)$ is the position angle, and $i(r)$ is the inclination angle. Both angles are Taylor-expanded about a fiducial radius $r_0 = 6.1\,\mathrm{mas}$:
\begin{equation}
\begin{split}
    \Omega(r)&=\Omega_{0}+\Omega_1\left(r-r_0\right)+\Omega_2\left(r-r_0\right)^{2}, \\
    i(r)&=i_{0}+i_1\left(r-r_0\right)+i_2\left(r-r_0\right)^{2}.
\end{split}
\end{equation}
The parameters $\Omega_1$, $\Omega_2$, $i_1$, and $i_2$ describe the first- and second-order warping of the disk orientation. The non-relativistic LOS velocity is given by:
\begin{equation}
v_{\text{los,\ensuremath{\text{model}}}}=v_{r}(r,\phi)\sin i(r)\sin\phi+v_{\gamma}(r,\phi)\sin i(r)\cos\phi+v_{0},
\end{equation}
where $v_0$ is the systemic velocity of the galaxy, and $v_r(r,\phi)$ and $v_\gamma(r,\phi)$ are the radial and tangential velocity components, respectively:
\begin{equation}
\begin{split}
        v_{r}(r,\phi)&=\sqrt{\frac{GM}{rD(1+e\cos\gamma(r,\phi))}}e\sin\gamma,\\v_{\gamma}
        (r,\phi)&=\sqrt{\frac{GM(1+e\cos\gamma(r,\phi))}{rD}}.
\end{split}
\end{equation}
Here, $e$ is the orbital eccentricity, and $\gamma(r,\phi)$ is the angular separation from periapsis:
\begin{equation}
    \gamma(r,\phi)=\phi-\omega(r)=\phi-\left(\omega_{0}+r\omega_1\right).
\end{equation}
where $\omega(r)$ is the argument of periapsis, Taylor-expanded to first order.

The relativistic correction to the LOS velocity is modeled as:
\begin{equation}
\begin{split}
    &v'_{\mathrm{los,model}}(r,\phi) = c\left[\beta_0 \cdot \beta_{||}\cdot \beta_{\text{GR}} -1 \right], \\
    &\beta_0 = \left({1+\frac{v_{0}}{c}}\right)/{\sqrt{1-\frac{v_{0}^{2}}{c^{2}}}},\\ 
    &\beta_{||} = \left({1+\frac{v_{\mathrm{los,model}}(r,\phi)-v_{0}}{c}}\right)/{\sqrt{1-\frac{v_{r}^{2}(r,\phi)+v_{\gamma}^{2}(r,\phi)}{c^{2}}}}, \\
    &\beta_{\text{GR}} = 1/\sqrt{1-\frac{2GM}{c^{2}rD}},
\end{split}
\end{equation}
where the corrections respectively account for: (i) recession of the galaxy, (ii) orbital motion of the masers, and (iii) gravitational redshift. Finally, the non-relativistic LOS acceleration is modeled as:
\begin{equation}
    a_{\text{los,\ensuremath{\text{model}}}}=-\frac{GM}{r^{2}D^{2}}\sin i(r)\sin\phi.
\end{equation}


\section{Fitting Validation}
\label{appendix:fitting_results}

This section presents a comparison of disk model fitting results for the averaged dataset, including those reported by \citetalias{Humphreys_2013}, \citetalias{Reid_2019}, and our own analysis using different error floor assumptions. Details of our fitting approach are provided in Section~\ref{subsec:fitting_method}. A summary of our fitting results, alongside values from previous studies, is given in Table~\ref{tab:disk_fit_results}. The table compares the warped disk parameters, the adopted or inferred error floors, and fitting metrics such as the reduced chi-squared ($\chi^2_\nu$) and the smallest tail probability associated with the observed $\chi^2_\nu$.

\begin{table*}
\begin{minipage}{0.8\textwidth}
\caption{Comparison of Disk Fitting Results}
\centering
\renewcommand{\arraystretch}{1.25} 
\setlength{\tabcolsep}{5pt}
\begin{threeparttable}
\label{tab:disk_fit_results}
\begin{tabular}{lccccc}
\toprule
& \textbf{H13\tnote{*}} & \textbf{This Work} & \textbf{R19\tnote{*}} & \multicolumn{2}{c}{\textbf{This Work}} \\
& & (H13 error floors) & & (R19 error floors) & (fitted error floors) \\
\midrule
\multicolumn{6}{c}{\textbf{Disk Fitting Parameters}} \\
\midrule
$D$ {[Mpc]} & $7.596 \pm 0.141$ & $7.277^{+0.168}_{-0.164}$ & $7.576 \pm 0.075$ & $7.551^{+0.071}_{-0.072}$ & $7.547^{+0.079}_{-0.078}$ \\
$M_{\text{BH}}$ [$10^7\,M_\odot$] & $4.00 \pm 0.08$ & $3.83^{+0.09}_{-0.09}$ & $3.97 \pm 0.04$ & $3.97^{+0.04}_{-0.04}$ & $3.97^{+0.04}_{-0.04}$ \\
$v_0$ {[$\mathrm{km\,s^{-1}}$]} & $474.2 \pm 0.4$ & $473.2^{+0.4}_{-0.5}$ & $473.3 \pm 0.4$ & $472.8^{+0.4}_{-0.4}$ & $472.6^{+0.5}_{-0.6}$ \\
$X_0$ {[mas]} & $-0.204 \pm 0.004$ & $-0.154^{+0.006}_{-0.006}$ & $-0.152 \pm 0.003$ & $-0.158^{+0.004}_{-0.004}$ & $-0.157^{+0.005}_{-0.005}$ \\
$Y_0$ {[mas]} & $0.560 \pm 0.005$ & $0.558^{+0.006}_{-0.006}$ & $0.556 \pm 0.004$ & $0.555^{+0.004}_{-0.004}$ & $0.554^{+0.004}_{-0.004}$ \\
$i_0$ {[deg]} & $86.93 \pm 0.19 ^{\S}$ & $86.76^{+0.25}_{-0.25}$ & $87.05 \pm 0.08$ & $86.05^{+0.11}_{-0.13}$ & $85.99^{+0.10}_{-0.10}$ \\
$\Omega_0$ {[deg]} & $88.43 \pm 0.12^{\S}$ & $88.53^{+0.06}_{-0.06}$ & $88.43 \pm 0.04$ & $88.39^{+0.04}_{-0.04}$ & $88.40^{+0.05}_{-0.05}$ \\
$e$ & $0.006 \pm 0.001$ & $0.008^{+0.001}_{-0.001}$ & $0.007 \pm 0.001$ & $0.006^{+0.001}_{-0.001}$ & $0.006^{+0.001}_{-0.001}$ \\
$\omega_0$ {[deg]} & $294 \pm 54$ & $275^{+22}_{-18}$ & $318 \pm 12$\tnote{1} & $222^{+18}_{-23}$ & $225^{+19}_{-21}$ \\
$i_1$ {[$\mathrm{deg\,mas^{-1}}$]} & $2.49 \pm 0.09$ & $2.33^{+0.22}_{-0.23}$ & $2.59 \pm 0.06$ & $1.61^{+0.09}_{-0.11}$ & $1.54^{+0.09}_{-0.09}$ \\
$i_2$ {[$\mathrm{deg\,mas^{-2}}$]} & --- & $-0.029^{+0.068}_{-0.070}$ & $0.041 \pm 0.016$ & $-0.189^{+0.022}_{-0.025}$ & $-0.203^{+0.022}_{-0.022}$ \\
$\Omega_1$ {[$\mathrm{deg\,mas^{-1}}$]} & $2.30 \pm 0.05$ & $2.31^{+0.04}_{-0.04}$ & $2.21 \pm 0.02$ & $2.19^{+0.02}_{-0.02}$ & $2.18^{+0.02}_{-0.02}$ \\
$\Omega_2$ {[$\mathrm{deg\,mas^{-2}}$]} & $-0.24 \pm 0.02$ & $-0.21^{+0.03}_{-0.03}$ & $-0.13 \pm 0.01$ & $-0.11^{+0.02}_{-0.02}$ & $-0.11^{+0.02}_{-0.02}$ \\
$\omega_1$ {[$\mathrm{deg\,mas^{-1}}$]} & $60 \pm 8$ & $116^{+3}_{-4}$ & $123 \pm 6$ & $124^{+5}_{-4}$ & $123^{+5}_{-4}$ \\
\midrule
\multicolumn{6}{c}{\textbf{Error Floors}\tnote{\textdagger}} \\
\midrule
$\Delta X$ {[mas]} & $[0.02]$ & $[0.02]$ & $0.0016 \pm 0.0005$ & $[0.0016]$ & $0.0018^{+0.0003}_{-0.0003}$ \\
$\Delta Y$ {[mas]} & $[0.03]$ & $[0.03]$ & $0.0041 \pm 0.0005$ & $[0.0041]$ & $0.0037^{+0.0004}_{-0.0004}$ \\
$\Delta v_{\text{sys}}$ {[$\mathrm{km\,s^{-1}}$]} & $[1.00]$ & $[1.00]$ & $0.31 \pm 0.18$ & $[0.31]$ & $0.25^{+0.13}_{-0.09}$ \\
$\Delta v_{\text{high}}$ {[$\mathrm{km\,s^{-1}}$]} & $[1.00]$ & $[1.00]$ & $2.25\pm 0.28$ & $[2.25]$ & $2.99^{+0.27}_{-0.24}$ \\
$\Delta a$ {[$\mathrm{km\,s^{-1}\,yr^{-1}}$]} & $[0.30]$ & $[0.30]$ & $0.46 \pm 0.04$ & $[0.46]$ & $0.48^{+0.05}_{-0.04}$ \\
\midrule
\multicolumn{6}{c}{\textbf{Fitting Score}} \\
\midrule
$dof$ & 509\tnote{2} & 495\tnote{3} & 483\tnote{4} & 495\tnote{3} & 490\tnote{5} \\ 
$\chi^2_\nu$ & $\approx 1.4$ & $\approx 0.4$ & $\approx 1.2$ & $\approx 1.4$ & $\approx 1.2$ \\
             & (reported) & & (reported) & & \\
min($p_>$, $p_<$)\tnote{‡}  & $\approx 5.8\times10^{-8}$ & $\approx 4.2\times10^{-36}$ & $\approx 1.6\times10^{-3}$ & $\approx 9.1\times10^{-9}$ & $\approx 1.5\times10^{-3}$\\ 
\bottomrule
\end{tabular}

\begin{tablenotes}
    \small
    \item[*] For full comparison, reported errors were deflated by their respective $\sqrt{\chi^2_\nu}$ values to recover the statistical uncertainties.
    \item[\S] Disk inclination and position angle values from \citetalias{Humphreys_2013} have been re-referenced from $r=0$ to $r=6.1\,\text{mas}$. 
    \item[\textdagger] Brackets indicate that error floor values were added in quadrature and were not treated as free parameters in the model.
    \item[‡] Smallest tail probability for the observed $\chi^2_\nu$; quantifies how unlikely the fit is under the assumed model and uncertainties.
    \item[1] The discrepancy in $\omega_0$ comparing our analyses to the results of \citetalias{Reid_2019} arises from differing azimuthal angle definitions: we use $0^\circ, 180^\circ, 270^\circ$ for redshifted, blueshifted, and systemic masers, respectively, while they appear to use $90^\circ, -90^\circ, 0^\circ$.
    \item[2] Calculated using the full averaged dataset, which includes 370 data entries and 13 warped disk parameters.
    \item[3] Calculated using the clipped averaged dataset, which includes 358 data entries and 14 warped disk parameters.
    \item[4] As reported. Used a clipped averaged dataset (number of data entries not specified), with 14 warped disk parameters and 5 free error floor parameters.
    \item[5] Used the clipped averaged dataset, including 358 data entries, 14 warped disk parameters, and 5 free error floor parameters.
\end{tablenotes}
\end{threeparttable}
\end{minipage}
\end{table*}

Our reanalysis reveals significant issues with the statistical validity of fitting the averaged dataset, regardless of the error floor approach:
\begin{itemize}
\item \textbf{Fixed H13 Error Floors:} Using the fixed error floors adopted by \citetalias{Humphreys_2013}, our reanalysis yields $D = 7.277 \pm 0.166\, \mathrm{Mpc}$, compared to their original result of $7.596 \pm 0.167\,\mathrm{Mpc}$ ($\myapprox1.3\sigma$ difference). The corresponding black hole mass in our analysis is $M_{\mathrm{BH}} = (3.83 \pm 0.09)\times10^{7}\,M_\odot$, slightly lower than the \citetalias{Humphreys_2013} value of $(4.00 \pm 0.09)\times10^{7}\,M_\odot$ ($\myapprox0.1\%$ difference in the ratio $M_{\mathrm{BH}}/D$ between the two analyses). We also note discrepancies in several geometric parameters: the disk center $X_0 = -0.204 \pm 0.005\,\mathrm{mas}$ in \citetalias{Humphreys_2013} fit versus $X_0 = -0.154 \pm 0.006\,\mathrm{mas}$ in our analysis, and the first-order warp in the periapsis angle, $\omega_1 = 60 \pm 10^{\circ}$ in their model compared to our result of $116 \pm 4^{\circ}$. While \citetalias{Humphreys_2013} reported $\chi^2_\nu \approx 1.4$, we obtain a much lower reduced chi-squared value of $\chi^2_\nu \approx 0.4$ for 495 degrees of freedom. The probability of obtaining $\chi^2_\nu < 0.4$ by chance is approximately $4 \times 10^{-36}$, making such a low value statistically implausible. This suggests that \citetalias{Humphreys_2013} adopted error floors may be overly conservative. We believe that the discrepancies in the geometric parameters arise from the limited sampling of the full parameter space in the \citetalias{Humphreys_2013} analysis, which involved $\myapprox730$ free parameters with strong degeneracies. To reproduce a similar situation, in which the nuisance parameters are not fully explored, we evaluated for each data entry the values $(r_{i,\mathrm{max}}, \phi_{i,\mathrm{max}})$ that maximize the individual likelihood for a given set of global disk parameters. This approach yields $D = 7.58 \pm 0.21\,\mathrm{Mpc}$, $M_{\mathrm{BH}} = (4.01 \pm 0.11) \times 10^{7}\,M_\odot$, and geometric parameters consistent with \citetalias{Humphreys_2013}, indicating that their MCMC likely failed to fully sample the degenerate nuisance-parameter space.
\item \textbf{Fixed R19 Error Floors:} Using the best-fit error floors from \citetalias{Reid_2019}, our analysis yields a distance of $D=7.551\pm0.072\,\text{Mpc}$, consistent with \citetalias{Reid_2019} distance estimate of $7.576 \pm 0.082\,\mathrm{Mpc}$. However, the corresponding reduced chi-squared of $\chi^2_\nu\approx1.4$ (495 degrees-of-freedom) is somewhat higher than the value of $\chi_\nu^2\approx1.2$ reported by \citetalias{Reid_2019}. The probability of obtaining $\chi^2_\nu > 1.4$ by chance is approximately $9.1 \times 10^{-9}$, indicating that the model is statistically incompatible with the data.  
\item \textbf{Free Error Floors:} Using error floors as free parameters, we obtain a distance of $7.547\pm0.079\,\text{Mpc}$, consistent with the distance reported by \citetalias{Reid_2019}, along with the additional disk parameters. We result in a reduced chi-squared value of $\chi^2_\nu \approx 1.2$ (490 degrees of freedom)--comparable to their reported value. Although this configuration yields the $\chi^2_\nu$ value closest to unity among the three tested, indicating modest improvement in the overall fit quality; it remains, however, slightly higher than typically expected for an ideal noise model. Nevertheless, the interpretation of the reduced chi-squared statistic as a measure of goodness-of-fit is non-trivial in this case. In such fits, the residuals and the total uncertainties (which include the fitted error floor) are no longer independent, which changes the behavior of the likelihood and invalidates the standard assumptions that give a chi-squared distribution \citep[see sections~6.3 and 7.4 in][]{Cowan_1998}. Additionally, the fitted error floors show some internal inconsistency: the fitted velocity error floor for high-velocity masers exceeds the bin width, while for systemic masers, it is smaller than the bin width. This likely reflects the model’s attempt to accommodate residual systematics or data characteristics, rather than inaccuracies in the observational error estimates themselves.
\end{itemize}

\begin{figure*}
    \centering
    \includegraphics[width=0.70\linewidth]{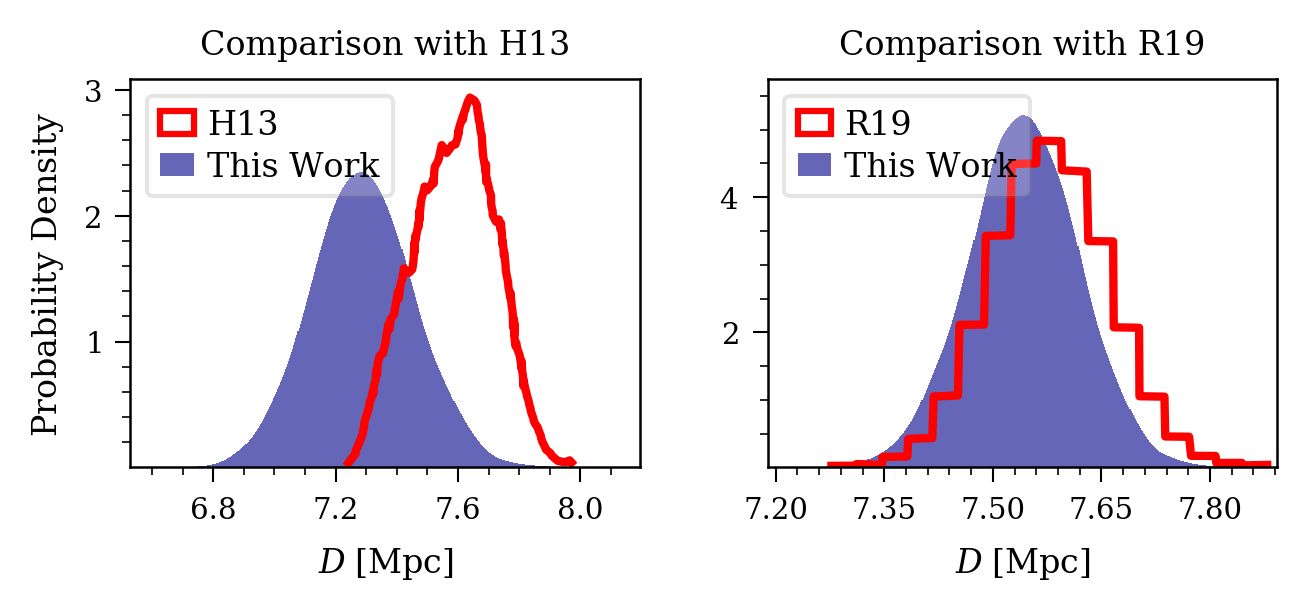}
    \caption{Posterior distributions of the distance $D$, comparing results from this work (blue) with those of \citetalias{Humphreys_2013} (left) and \citetalias{Reid_2019} (right). In both panels, the red curve represents the published result, while the blue filled distribution shows our reanalysis using the corresponding error floor assumptions.
    }
    \label{fig:compareH13H19}
\end{figure*}

A visual comparison of the distance estimates is presented in Figure~\ref{fig:compareH13H19}, which shows the posterior distributions from our reanalysis (blue) alongside the published results (red) of \citetalias{Humphreys_2013} (left) and \citetalias{Reid_2019} (right), using the respective error floor treatments adopted in those studies.

\begin{figure*}
    \centering
\includegraphics[width=0.90\linewidth]{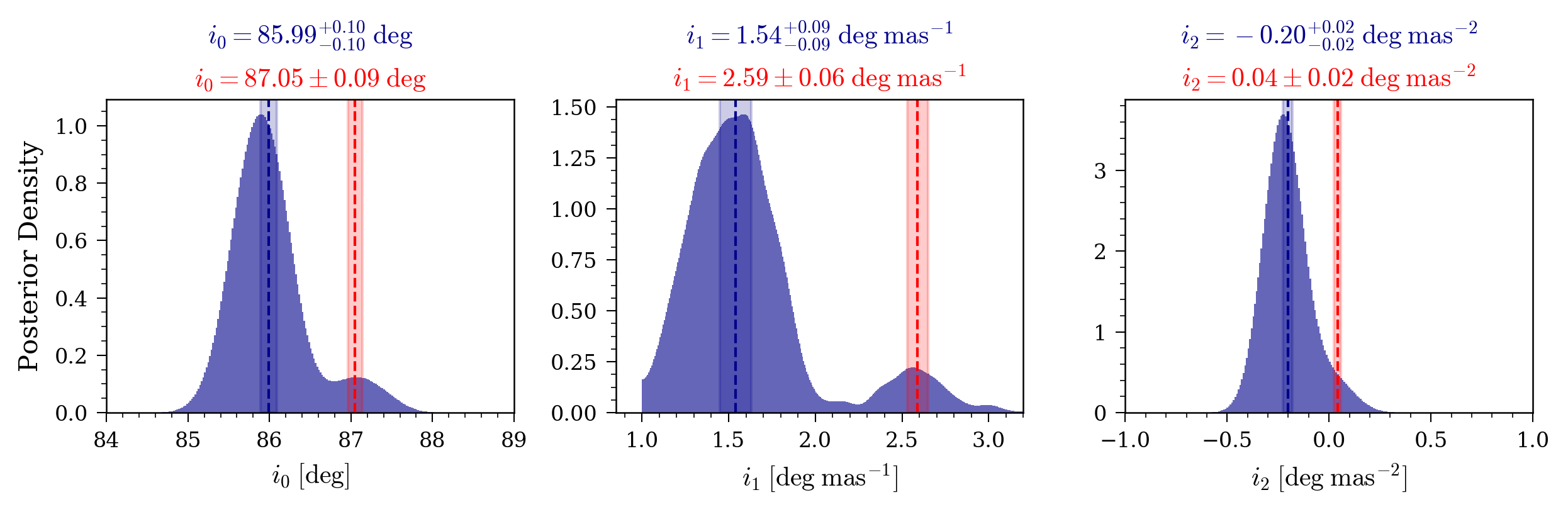}
    \caption{Posterior distributions of the inclination parameters $i_0$ (left), $i_1$ (middle), and $i_2$ (right) from our reanalysis of the averaged dataset treating the error floors as free parameters. This likelihood evaluation of this reanalysis was restricted to tighter $(r,\phi)$ bounds as described in Section~\ref{subsec:fitting_method}. In addition, it does not correspond to a fully converged nested-sampling run: the posterior shown here includes the first 45{,}000 samples (out of 77{,}470). The blue shaded regions denote the values from our full analysis as reported in Table~\ref{tab:disk_fit_results}. For comparison, the red shaded regions and dashed red lines indicate the values reported by \citetalias{Reid_2019}. Subplot titles show the values from our analysis (blue, top) and the published best-fit values (red, bottom)}.
    \label{fig:post_inc}
\end{figure*}

Inspection of Table~\ref{tab:disk_fit_results} reveals a discrepancy in the inclination warping parameters, $i_0$, $i_1$, and $i_2$, between our analysis and previous works. As illustrated in Figure~\ref{fig:post_inc}, the posterior distributions from our reanalysis suggest the presence of a degeneracy among the inclination parameters. We emphasize that the posterior shown in Figure~\ref{fig:post_inc} corresponds to an intermediate stage of the nested-sampling run and is not fully converged. Within this partially converged posterior, our best-fit values (blue shaded regions) coincide with the highest-probability region, while the values reported by \citetalias{Humphreys_2013} and \citetalias{Reid_2019} (red shaded regions) fall along a lower-probability shoulder of the distribution rather than being strongly excluded. This degeneracy appears to be confined to the inclination warping terms and does not substantially impact the global disk parameters.

\section{Line Widths}
\label{appendix:line_widths}
In this section, we examine the impact of maser line width broadening. As a starting point, Figure~\ref{fig:vF_spec} presents integrated spectra of normalized flux density as a function of relativistic LSR velocity for three representative epochs from \citetalias{Argon_2007}, each corresponding to one of the systemic (top), redshifted (middle), and blueshifted (bottom) velocity regimes. Maser components can be approximately identified by locating local maxima in the spectra (with an additional quality check, see Section~\ref{subsubsec:identify_peaks} for details, marked with ‘x’), which correspond to individual emission features. These spectra reveal fewer than five blueshifted masers, approximately ten redshifted masers, and at most 40 systemic masers—significantly fewer than the $\mysim$370 independent data points included in the averaged dataset. This discrepancy suggests that averaging may artificially inflate the apparent number of independent measurements. Next, we show that the line broadening of maser lines causes adjacent velocity channels to contain emission from the same maser components. Treating those channels as independent in the averaging process leads to oversampling, as maser components cannot be considered statistically independent at arbitrarily high velocity resolution.

In the simple theory of masers, the velocity linewidth $\Delta v$ (FWHM) is primarily determined by thermal broadening, where typical water maser environments have temperatures between $\myapprox 400-1000\,\mathrm{K}$ \cite[]{Neufeld_1990}. \cite{Varshalovich_2006} demonstrated that thermal broadening together with the hyperfine structure of the $6_{16} - 5_{23}$ $\mathrm{H_2O}$ molecular transition can broaden the line profile by up to $\Delta v  \lesssim2~\mathrm{km\,s^{-1}}$. The observed maser linewidth also depends on the optical depth; as the optical depth increases, the effective linewidth becomes narrower due to amplification effects.
These masers also have intrinsically small physical sizes ($\leq$$10^{14}\,\mathrm{cm}$, corresponding to $\leq$$1\,\mu\mathrm{as}$ at a distance of $7.6\,\mathrm{Mpc}$; \citealt{Herrnstein_1999}), such that positional linewidths are dominated by instrumental uncertainty. In VLBI, the effective resolution is set by the synthesized beam ($\myapprox0.5 \times 0.3\,\mathrm{mas}$ at $22\,\mathrm{GHz}$; \citetalias{Argon_2007}). Maser positions are determined by fitting two-dimensional Gaussians in each velocity channel, with centroiding uncertainties of  $\theta_\text{beam}/(2\,\times\,\text{S/N} )$, typically in the range of $2\text{--}10\,\mu\text{as}$.

Consider a spectral peak in velocity space with a linewidth $\Delta v$ of a few $\textrm{km}\,\textrm{s}^{-1}$, spanning multiple velocity resolution elements. We can examine two extreme cases. In the first case, each resolution element corresponds to an independent maser component, i.e., there is no intrinsic broadening, and the measurements in adjacent resolution elements can be treated as independent measurements of different maser features. In this scenario, each maser follows a Keplerian orbit (see Equation~\eqref{eq:v_x_relation}). This relation allows us to connect $\Delta v$ and the corresponding positional spread $\Delta X$:
\begin{equation} 
    \Delta X_{\text{Kepler}} = \sqrt{\frac{r^3 D}{GM_{\text{BH}}}} \Delta v \approx 18.5\,\left(\frac{\Delta v}{5\,\textrm{km}\,\textrm{s}^{-1}}\right)\mu\textrm{as},
\label{eq:dxdv}
\end{equation}
where we adopt $D=7.6\,\text{Mpc}$, $M_{\text{BH}} = 4\times10^{7}\, \text{M}_\odot$, and $r = 4\,\text{mas}$ --- a typical radius for systemic masers. In this case, a measurable positional drift along the spectral peak is expected. In the second extreme, the entire $\Delta v$ arises from line broadening of a single maser component. In this case, the observed positional spread $\Delta X$ should be consistent with instrumental uncertainty alone (as discussed above), and any drift would be smaller than predicted by the Keplerian relation. In what follows, we examine the empirical relationship between $\Delta v$ and $\Delta X$ to assess which of these scenarios is more consistent with the data.

We begin by identifying maser components in the integrated flux–velocity spectra. Each component is assigned a weight reflecting its prominence—i.e., how strongly it stands out relative to neighboring features (see Section~\ref{subsubsec:identify_peaks} for details)—and a linewidth $\Delta v$, defined as the FWHM measured at 50\% of the prominence height. To relate $\Delta v$ to $\Delta X$, we isolate the region surrounding each peak based on its central velocity and associated $\Delta v$ in both the \citetalias{Argon_2007} and averaged datasets. Within each region, we fit a linear relation of the form $X = m \cdot v_{\text{los}} + b$, applying covariance rescaling such that the reduced chi-squared is unity. The corresponding positional spread is then computed as $\Delta X = m \cdot \Delta v$.

\begin{figure*}
    \centering
    \includegraphics[width=1\linewidth]{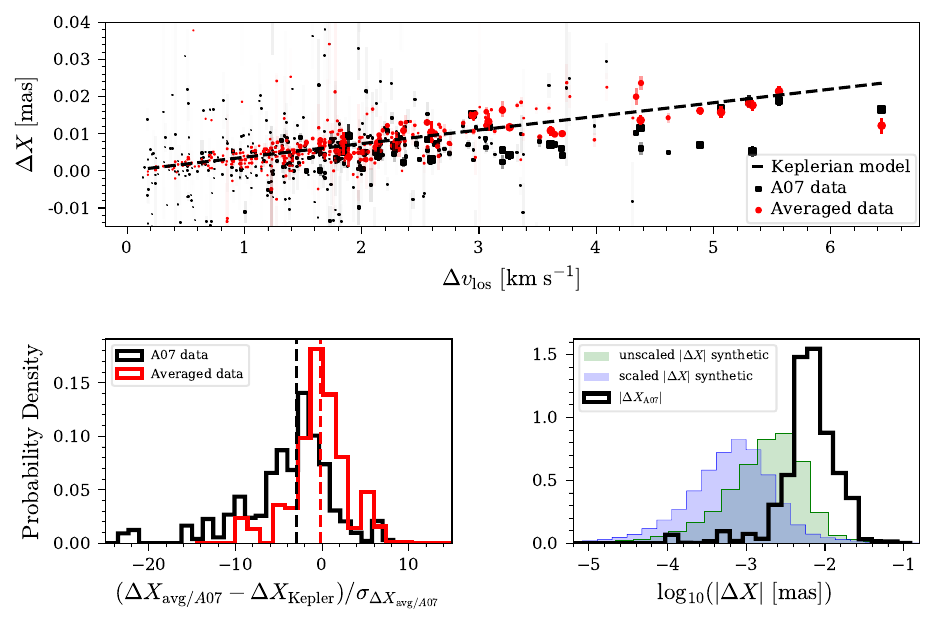}
    \caption{Upper panel: Relationship between velocity width $\Delta v_{\text{los}}$ and positional spread $\Delta X$ for maser components, derived from the \citetalias{Argon_2007} dataset (black squares) and the averaged dataset (red circles). The dashed line represents the Keplerian model prediction from Equation~\eqref{eq:dxdv}. Marker size and error bar transparency are scaled by peak prominence. Lower left panel: Prominence-weighted histograms of normalized residuals in $\Delta X$ relative to the Keplerian model. While the averaged data (red) are broadly consistent with the model, the \citetalias{Argon_2007} data (black) show a median shift of $-2.92\sigma_{\Delta X_\text{A07}}$, indicating a systematic offset below the Keplerian trend. Lower right panel: Prominence-weighted histograms of $\log_{10} |\Delta X|$ for the \citetalias{Argon_2007} data (black) and synthetic distributions generated under the assumption of no intrinsic positional spread, except for measurement noise. The synthetic datasets use unscaled (green) and scaled (blue) positional uncertainties, where the scaling is derived from a reduced-$\chi^2$ fit to the $X$–$v_{\text{los}}$ relation in the \citetalias{Argon_2007} data. Both synthetic distributions show a clear offset relative to the observed data, suggesting that masers are not observed as strictly point-like features.}
    \label{fig:dxdv}
\end{figure*}

Figure~\ref{fig:dxdv} (upper panel) shows the relationship between $\Delta X$ and $\Delta v$ for the \citetalias{Argon_2007} (black squares) and averaged (red circles) datasets. Marker size and error bar transparency scale with peak prominence. The black dashed line represents the theoretical Keplerian prediction from Equation~\eqref{eq:dxdv}. The averaged data closely follow this line, consistent with the first extreme case discussed above, suggesting that thermal broadening has been effectively neglected in the averaging procedure. In contrast, the \citetalias{Argon_2007} data exhibit systematic deviations below the Keplerian line, indicating a contribution from thermal broadening. In total, 400 maser components are identified in the \citetalias{Argon_2007} sample, of which 146 lie above and 254 lie below the Keplerian line, corresponding to a binomial probability of $p=1.6\times10^{-8}$ for such an imbalance under the null hypotheses of equal likelihood. Restricting the comparison to components with $\Delta v < 2\,\text{km s}^{-1}$, 114 peaks are found above and 169 below the Keplerian line with $p=2.2\times10^{-4}$. For the averaged dataset, 154 features lie above and 167 below, yielding a probability of $p=3.4\times10^{-2}$. We note that in the averaged analysis, features represented by fewer than two binned data points were excluded. To quantify this asymmetry effect, the left lower panel presents a prominence-weighted histogram of normalized residuals of $\Delta X$ with respect to the Keplerian model, defined as $(\Delta X_{\text{avg}/\text{A07}} - \Delta X_{\text{Kepler}})/\sigma_{\Delta X_{\text{avg}/\text{A07}}}$, where $\sigma_{\Delta X}$ is computed from the uncertainty in the linear fit described above. 
The \citetalias{Argon_2007} data (black) show a median shift of $-2.92\sigma_{\Delta X_\text{A07}}$, indicating a systematic offset below the Keplerian trend, while
the median residual for the averaged data (red) is $-0.23\sigma_{\Delta X_\text{avg}}$, indicating good agreement with the Keplerian expectation. 
To further investigate the origin of each maser component’s positional spread, the lower-right panel presents a prominence-weighted histogram of $\log_{10} |\Delta X|$, comparing the \citetalias{Argon_2007} dataset (black) to synthetic distributions. For the synthetic tests, we retain the observed velocity of each maser peak and perform 100 independent realizations of $X$ per peak by drawing from a Gaussian centered at zero with a standard deviation equal to either the unscaled (green) or scaled (blue) observational $X$ uncertainties. The scaling is based on the prior linear fit, in which the covariance matrix is rescaled to yield a reduced chi-squared of unity. This procedure reflects the null hypothesis that any apparent slope arises purely from measurement noise, without intrinsic spatial structure. For each realization, we repeat the linear fit using the same procedure as for the observed data and compute the corresponding $\Delta X$. We then aggregate the resulting $\log_{10}|\Delta X|$ values over all peaks and all realizations to construct the synthetic histograms. The resulting synthetic distributions are similar, indicating that covariance matrix scaling has little effect on the outcome. However, both synthetic distributions show a clear offset relative to the \citetalias{Argon_2007} data. This discrepancy—along with the asymmetry of $\Delta X$ in the upper panel, where most values are positive—suggests that masers are not observed as strictly point-like features, as would be expected in the second extreme case. The spatial observed spread may arise from unresolved blending of nearby maser spots or from the intrinsic finite size of the maser emission regions themselves. 

These results demonstrate that maser features exhibit non-negligible broadening in velocity, inconsistent with the first extreme case assumed by the averaging procedure—namely, that each resolution element corresponds to an independent maser with no internal broadening.


\section{Masers Trajectory}
\label{appendix:trajectories}

Throughout this work, we compute the trajectories of maser components in position, velocity, and acceleration space using globally defined disk parameters. Each trajectory is initialized with an angular radius $r_0$ and azimuthal angle $\phi_0$ at a reference epoch $t_0$. These initial conditions are converted into Cartesian coordinates and velocities, $(x_0, y_0, v_{x,0}, v_{y,0})$, which are then evolved by solving the following system of differential equations:
\begin{equation}
\frac{dx}{dt} = v_x, \quad \frac{dy}{dt} = v_y, \quad \frac{dv_x}{dt} = -a_r \frac{x}{r}, \quad \frac{dv_y}{dt} = -a_r \frac{y}{r},
\end{equation}
where the radial distance and azimuthal angle are defined as
\begin{equation}
r = \sqrt{x^2 + y^2}, \quad \phi = \tan(y/x),
\end{equation}
and the radial gravitational acceleration $a_r$ includes the first post-Newtonian correction:
\begin{equation}
a_r = \frac{GM}{r^2} \left(1 + \frac{3GM}{rc^2}\right).
\end{equation}
From the solutions for $(r(t), \phi(t))$, we compute the corresponding sky-plane positions $(X, Y)$, LOS velocities, and LOS accelerations at each observational epoch.


\section{Synthetic Data}
\label{appendix:synthetic_data}
To validate the fitting program, we generate synthetic datasets to test whether it can reliably recover a known input distance. We begin by selecting the set of global disk parameters from \citetalias{Reid_2019} (using rounded versions of their reported values) and choosing one of the datasets from Table~\ref{tab:results_newdata}.  For each maser group, we adopt the fitted $(r, \phi)$ values and identify initial conditions $(r_0, \phi_0)$ corresponding to the epoch nearest the temporal midpoint of the group’s observational span. These midpoints, along with the global disk parameters, are used to model the orbital motion of each maser component by numerically integrating the equations of motion over the group’s time range, as described in Appendix~\ref{appendix:trajectories}.

From the resulting trajectories $(r(t), \phi(t))$, we compute synthetic observables—sky-plane positions $(X(t), Y(t))$ and line-of-sight velocities $v_{\text{los}}(t)$, and add Gaussian noise consistent with the original measurement uncertainties of the selected dataset. This procedure is repeated multiple times to produce an ensemble of simulated datasets. For each realization, we estimate the LOS acceleration of each maser group by fitting a linear trend to the velocity as a function of time, and compute both the associated uncertainties and the velocity–acceleration cross-correlation. Notably, no RANSAC algorithm is applied during acceleration estimation in the synthetic analysis. This approach enables us to assess the robustness of the distance inference under controlled, noise-perturbed conditions.

\begin{figure}
    \centering
    \includegraphics[width=0.9\linewidth]{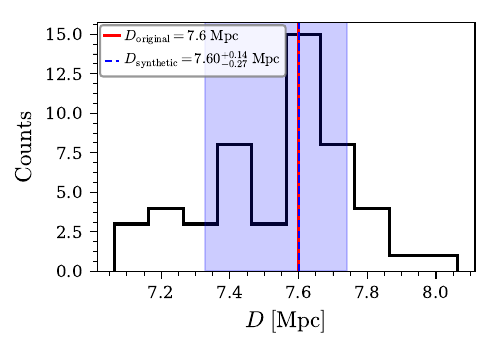}
    \caption{Distribution of recovered distances $D$ derived from synthetic dataset realizations used to validate the results in Table~\ref{tab:results_newdata}. The synthetic data are based on entry \#16 in Table~\ref{tab:results_newdata}, generated using the disk parameters from \citetalias{Reid_2019} and the inferred $(r, \phi)$ values. The histogram shows the distribution of recovered distances across all realizations. The blue shaded region denotes the 68\% credible interval, yielding $D=7.60^{+0.14}_{-0.28}\;\text{Mpc}$. The red dashed line marks the true input distance of $D = 7.6$~Mpc.}
    \label{fig:D_posterior_synthetic}
\end{figure}

As an example, Figure~\ref{fig:D_posterior_synthetic} shows the posterior distribution of the recovered distance for a synthetic ensemble based on entry \#16 from Table~\ref{tab:results_newdata}. Using a true distance of $D=7.6\,\text{Mpc}$, we obtain a recovered value of $D=7.60^{+0.14}_{-0.28}\, \text{Mpc}$, demonstrating consistency between input and inferred values.

\section{Synthetic Averaged Data}
\label{appendix:synthetic_binned}
The purpose of the synthetic tests in this appendix is to assess whether the averaging procedure itself introduces systematic effects in the recovered distance or the resulting $\chi^2$ values. The averaging procedure implicitly imposes a single–orbit approximation that may bias the results. To isolate this effect from any unmodeled structure in the averaged dataset, we generate synthetic datasets from a perfectly known warped–disk model and apply the same binning and fitting procedures to these controlled realizations, directly evaluating whether averaging alone alters the inferred distance or distorts the expected $\chi^2$ distribution.
As described in Appendix~\ref{appendix:synthetic_data}, we generate $\myapprox50$ realizations of a synthetic dataset based on our largest sample, entry~\#17
\footnote{We emphasize that although this dataset yields poor-fitting results, none of them are used in constructing the synthetic realizations; the realizations eventually resemble the original only in observational noise, sample size, and cadence.} from Table~\ref{tab:results_newdata} adopting the global disk parameters of \citetalias{Reid_2019} (rounded versions of their reported values) as inputs to the Keplerian dynamics. Each synthetic realization is then processed following the procedure of \citetalias{Humphreys_2013}: the simulated data are first binned to a velocity resolution of $1\,\mathrm{km\,s^{-1}}$, after which Gaussian uncertainties are assigned to the binned positions and accelerations. We note that, in the averaged synthetic realizations, the binning procedure is applied to substantially fewer measurements per bin than in the full observational dataset.
Each averaged synthetic realization is fitted twice using two distinct error–floor prescriptions. The first resembles the \citetalias{Humphreys_2013} methodology, which we adopt a fixed velocity error floor equal to half the bin width ($0.5\,\mathrm{km\,s^{-1}}$). The second follows \citetalias{Reid_2019}, treating the error floors as free parameters that are optimized simultaneously with the warped–disk model parameters.  Details of the fitting scheme are provided in Section~\ref{subsec:fitting_method}. This latter prescription is performed only as a complementary analysis, as the synthetic data have fully known uncertainties and follow purely Keplerian trajectories, leaving no motivation to treat the error floors as free parameters.
These tests allow us to isolate the impact of averaging on parameter recovery and on the resulting $\chi^2$ statistics.
Unlike a linear and symmetric inference problem, such as estimating the PSF centroid, where poor $\chi^2_\nu$ does not bias the estimator \citep{Lindegren_1978}, the warped–disk model is nonlinear, and data modifications, such as averaging, can introduce systematic errors.

\begin{figure*}
    \centering
    \includegraphics[width=0.9\linewidth]{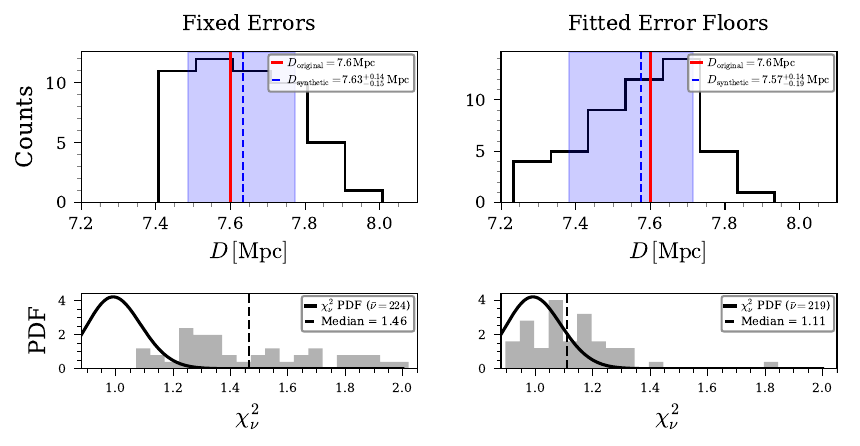}
    \caption{The upper left panel shows the distribution of recovered distances for $\myapprox50$ averaged synthetic realizations fitted using a fixed velocity error floor of $0.5\,\mathrm{km\,s^{-1}}$, with a median distance of $D = 7.63^{+0.14}_{-0.15}\,\mathrm{Mpc}$ (blue dashed line). The upper right panel shows the corresponding distance distribution when the error floors are treated as free parameters, with a median distance of $D = 7.57^{+0.14}_{-0.19}\,\mathrm{Mpc}$ (blue dashed line). In both upper panels, the blue-shaded region denotes the 68\% credible interval for the recovered distance distribution. The input distance used to generate the synthetic datasets is $D = 7.60\,\mathrm{Mpc}$ (solid red line). The lower left and lower right panels display the distributions of the reduced chi-squared values, $\chi^2_\nu$, for the fixed–error–floor and free–error–floor prescriptions, respectively, with median values of $\chi^2_\nu \approx 1.5$ and $\chi^2_\nu \approx 1.1$ (black dashed lines). For clarity, the histograms in the lower panels are truncated at $\chi^2_\nu \lesssim 2$. Because the number of degrees of freedom, $\nu$, varies slightly among realizations, the expected $\chi^2_\nu$ probability density function (solid black curve) is computed using the mean value of $\nu$ across the ensemble.}
    \label{fig:synthetic_binned}
\end{figure*}

The underlying model is exactly correct, and the synthetic datasets were generated from a known disk configuration. As shown in Figure~\ref{fig:synthetic_binned}, with the upper left panel displaying the distribution of recovered distances obtained using a fixed velocity error floor, yielding $D = 7.63^{+0.14}_{-0.15}\,\mathrm{Mpc}$, and the upper right panel showing the corresponding distribution when the error floors are treated as free parameters, giving $D = 7.57^{+0.14}_{-0.19}\,\mathrm{Mpc}$, consistent with the input distance used to generate the synthetic datasets, $D = 7.60\,\mathrm{Mpc}$. However, the distance distribution recovered under the fixed–error–floor prescription is notably non-Gaussian, appearing substantially flatter and more uniform than expected for statistically well-behaved noise, suggesting a high probability for a biased distance. In addition, the synthetic fits do not recover a statistically consistent solution: the resulting reduced chi-squared values remain significantly larger than unity. The lower left panel of Figure~\ref{fig:synthetic_binned} shows that the fixed–error–floor methodology yields a median reduced chi-squared of $\chi^2_\nu \approx 1.5$, while the lower right panel demonstrates that allowing the error floors to vary reduces this only to $\chi^2_\nu \approx 1.1$. In both cases, the distribution of $\chi^2_\nu$ values deviates strongly from that expected for a correct noise model, demonstrating that the averaging procedure alone introduces systematic distortions in the goodness-of-fit statistic.
In addition, \citetalias{Humphreys_2013} adopts a $1\%$ systematic uncertainty to account for unmodeled spiral–disk structure; although not stated explicitly, such a term may affect both the inferred distance and the resulting $\chi^2_\nu$. Our synthetic tests show that statistically inconsistent fits arise even when the underlying disk model is perfectly known and contains no non-Keplerian structure. Moreover, because these tests are based on a relatively small synthetic maser sample compared to the \citetalias{Argon_2007} dataset, the impact of averaging is likely underestimated. In larger datasets, binning merges physically distinct masers into a single feature, producing a dynamical signal that does not correspond to any individual maser and may further degrade the goodness-of-fit.


\section{Trajectory Results for Representative Dataset}
\label{appendix:traj_results}

This section presents two example maser trajectories constructed for distinct datasets: dataset \#8 (Figure~\ref{fig:trajectory_group}) and dataset \#1 (Figure~\ref{fig:trajectory_group_13_1}) listed in Table~\ref{tab:results_newdata}. The figures display the measured properties of individual maser components—sky-projected positions, LOS velocities, LOS accelerations, and prominence—plotted as a function of time across multiple epochs. Each tracked maser is identified with a group ID as appears in the legend, identical to the public catalog\footref{fn:url} identifier. Masers with colored patches in the legend are also presented in Figure~\ref{fig:ransac} and~\ref{fig:xv_sys_traj}.  Colored lines indicate the fitted trajectories derived from posterior samples of the model parameters: red for redshifted components, blue for blueshifted, and green for systemic features. For each sample from the global disk posterior, we randomly select a maser component from the group and compute its best-fit $(r_0, \phi_0)$ values under that sample. These values are used as initial conditions to integrate the full trajectory, as described in Appendix~\ref{appendix:trajectories}.

\begin{figure*} 
    \centering
    \begin{tabular}{ccc}
        \includegraphics[width=0.3\linewidth]{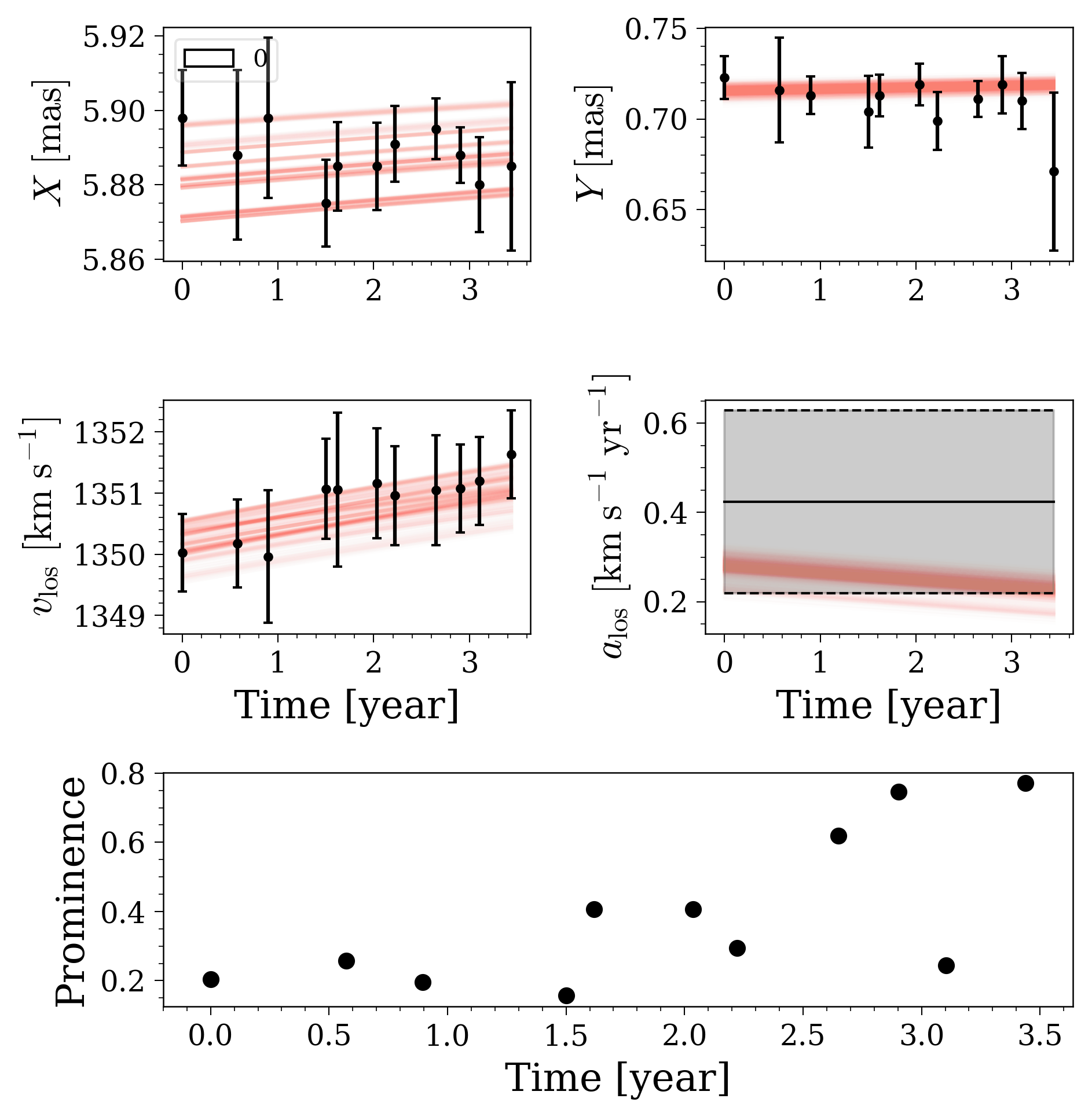} &
        \includegraphics[width=0.3\linewidth]{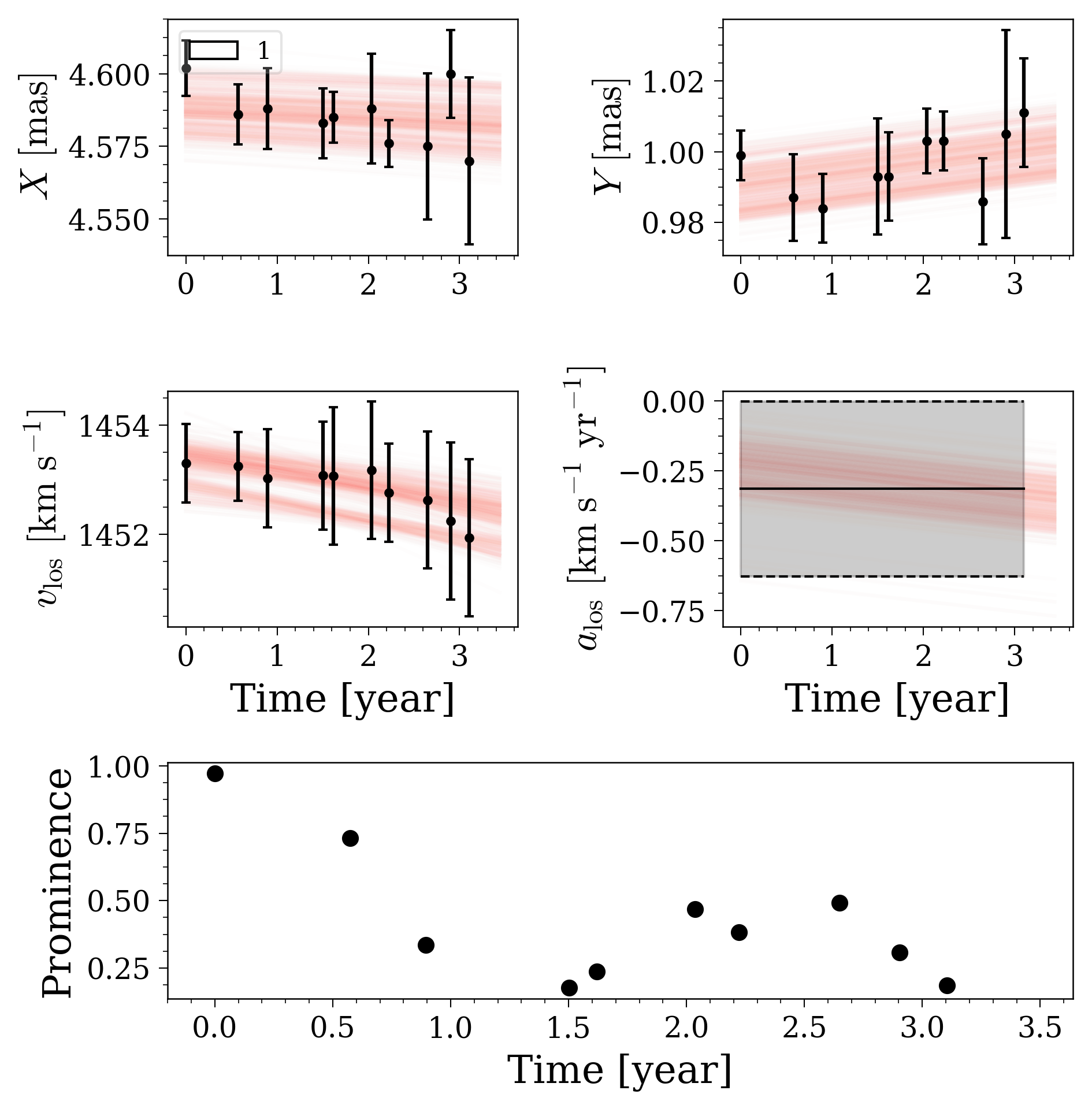} &
        \includegraphics[width=0.3\linewidth]{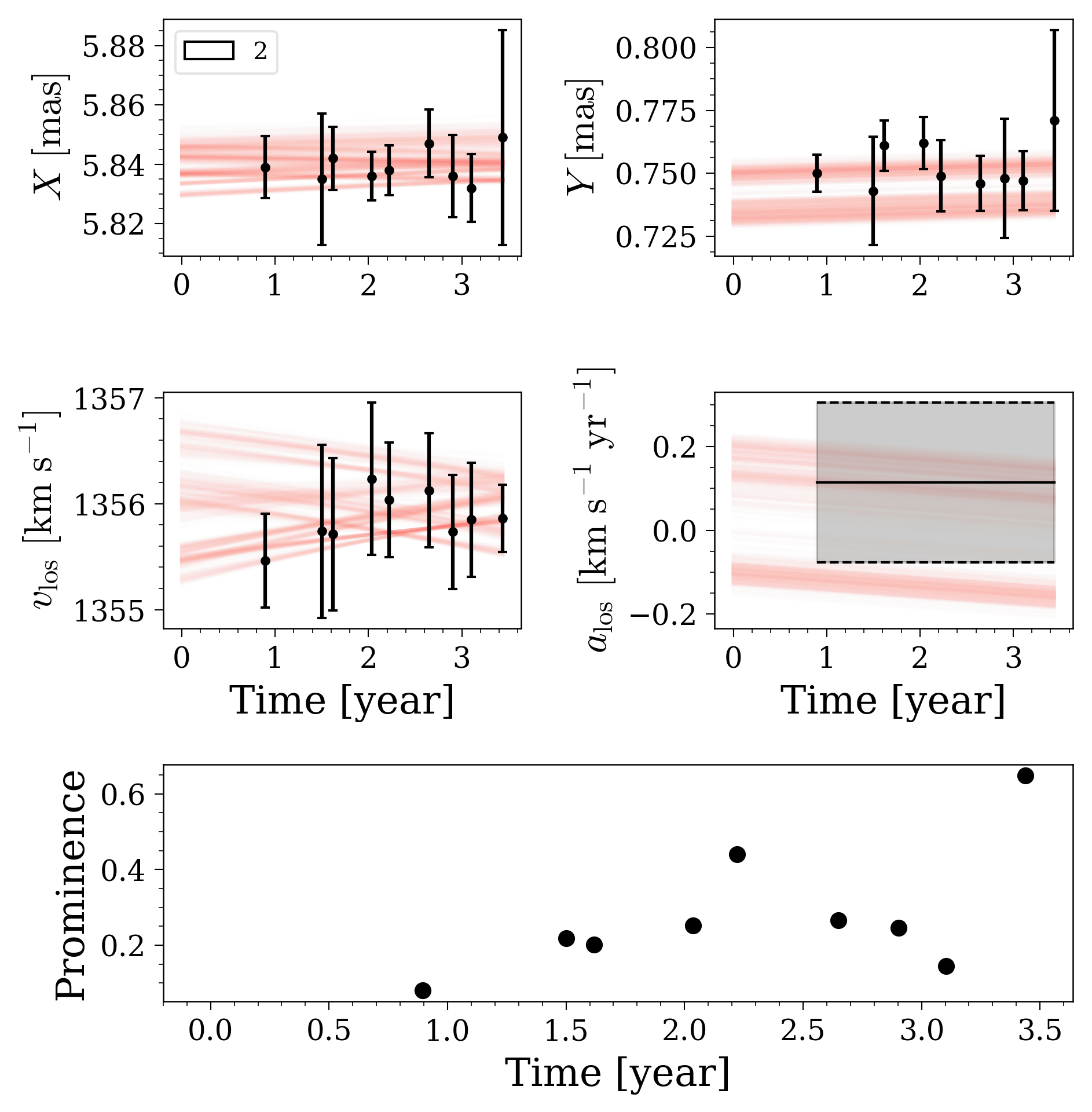} 
    \end{tabular}
    \caption{Maser component measurements—$X$, $Y$, $v_{\text{LOS}}$, $a_{\text{LOS}}$, and prominence—plotted as a function of time for representative dataset \#8, as listed in Table~\ref{tab:results_newdata}. Group IDs shown in the legend correspond to those in the public digital catalog and match the colored patches used in Figures~\ref{fig:ransac} and~\ref{fig:xv_sys_traj}. Colored lines represent fitted trajectories derived from posterior samples: red for redshifted masers, blue for blueshifted, and green for systemic features. Trajectories are shown for $X$, $Y$, $v_{\text{LOS}}$, and $a_{\text{LOS}}$ only. For each global disk posterior sample, a single epoch is randomly selected. The optimal $(r_0, \phi_0)$ values for that epoch component are then used as initial conditions to compute the full trajectory, as described in Appendix~\ref{appendix:trajectories}.}
    \label{fig:trajectory_group}
\end{figure*}

\begin{figure*}
    \centering
    \addtocounter{figure}{-1} 
    \begin{tabular}{ccc}
        \includegraphics[width=0.3\linewidth]{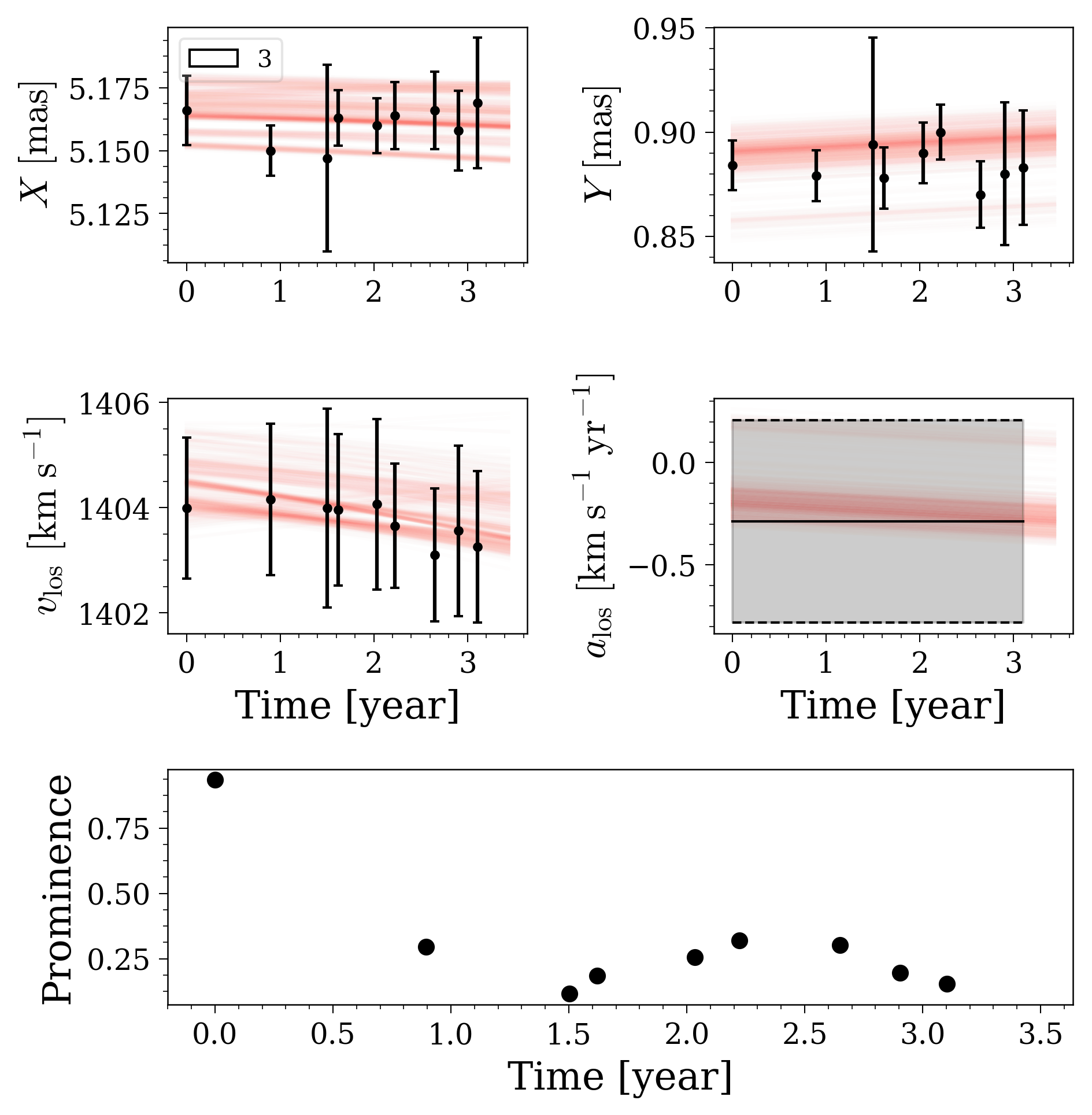} &
        \includegraphics[width=0.3\linewidth]{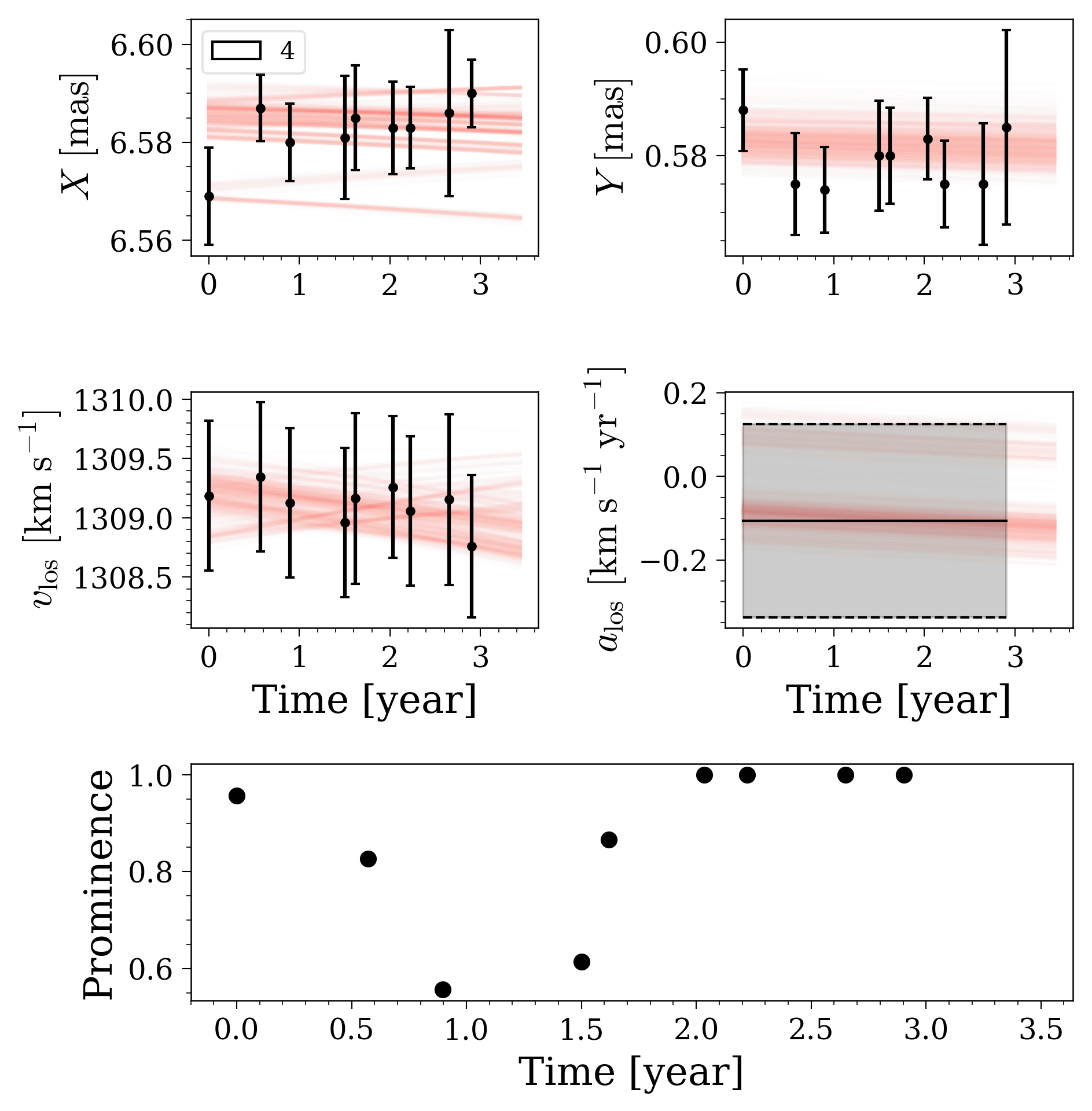} &
        \includegraphics[width=0.3\linewidth]{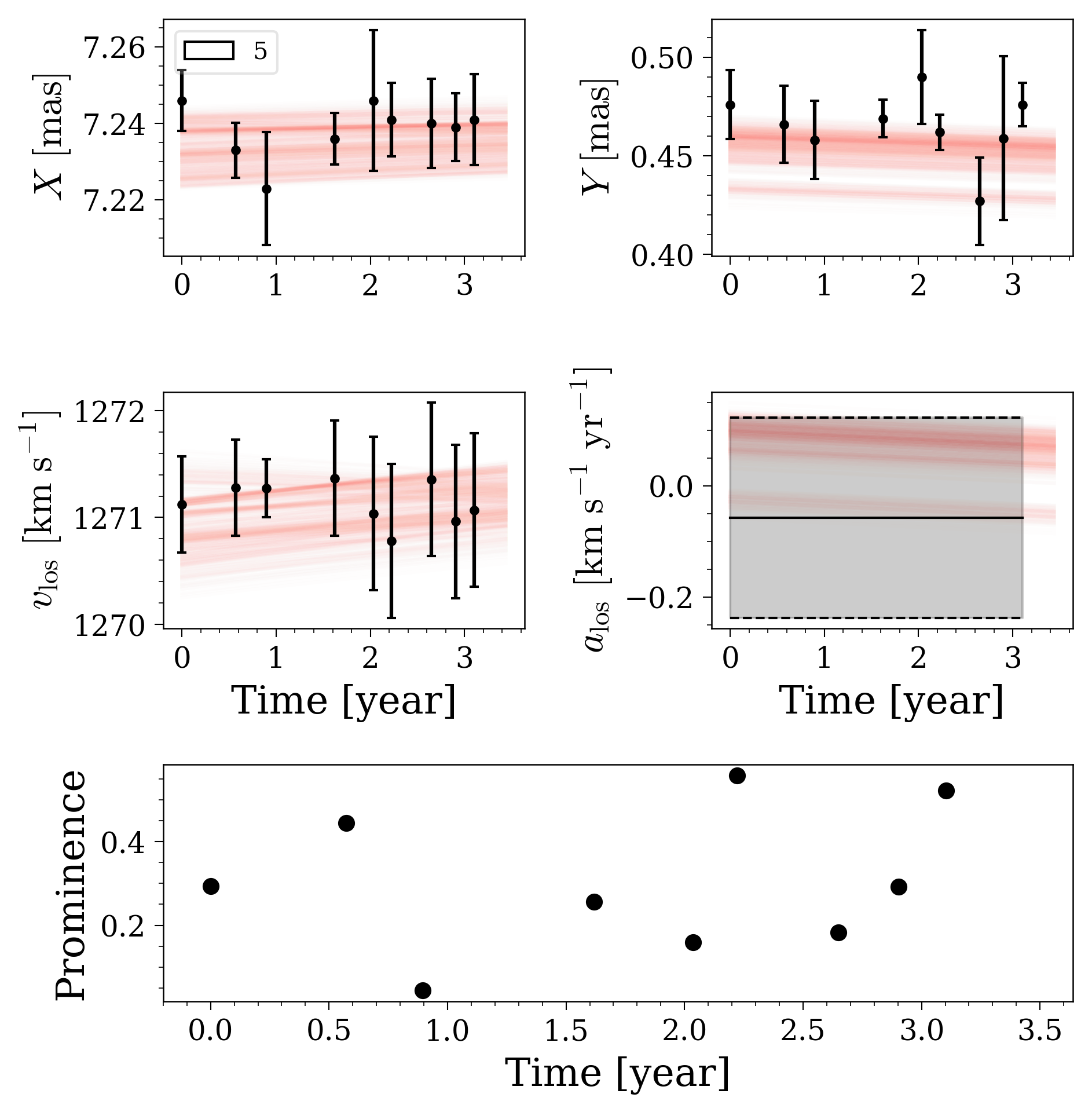} \\
        \includegraphics[width=0.3\linewidth]{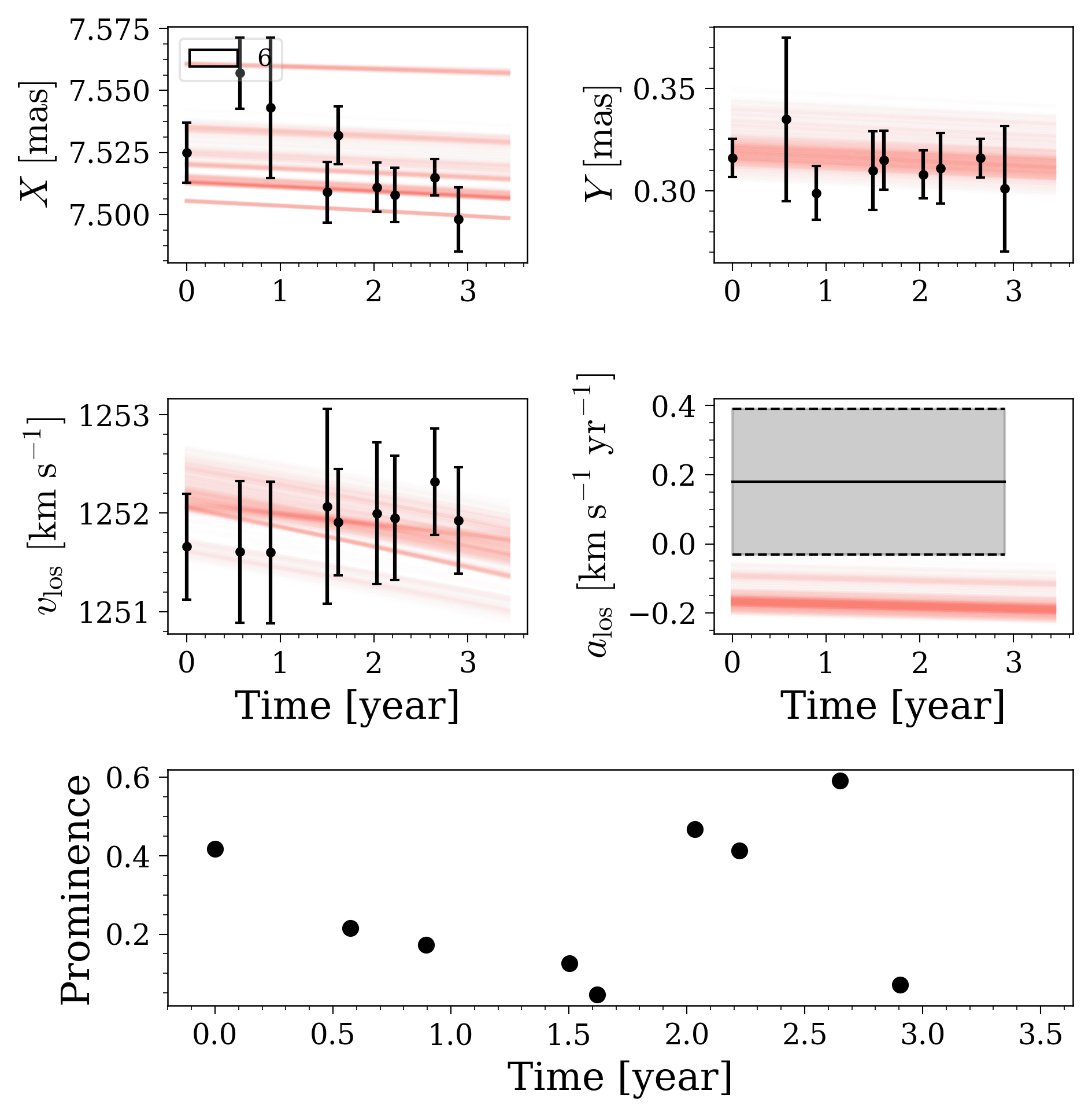} &
        \includegraphics[width=0.3\linewidth]{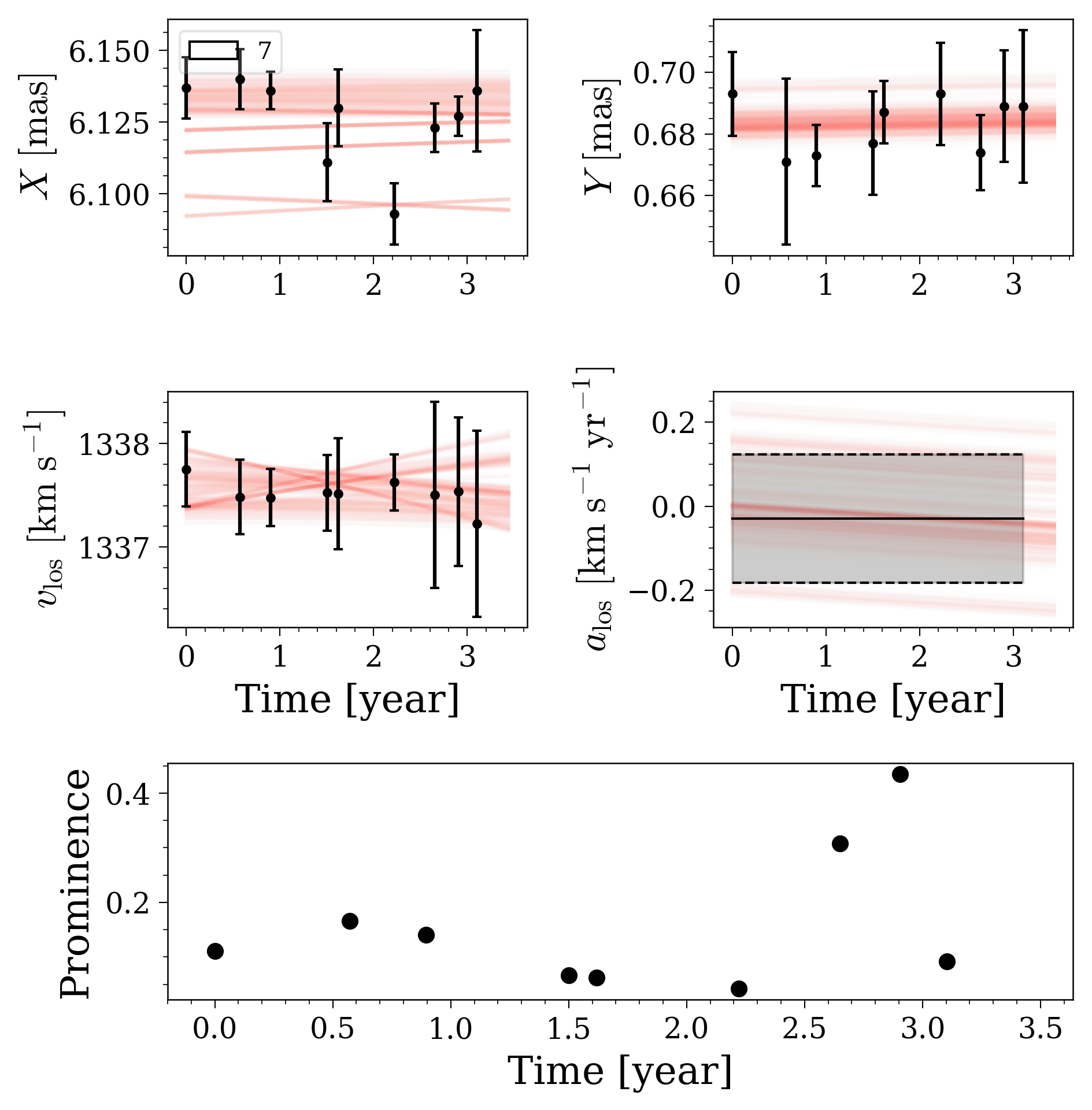} &
        \includegraphics[width=0.3\linewidth]{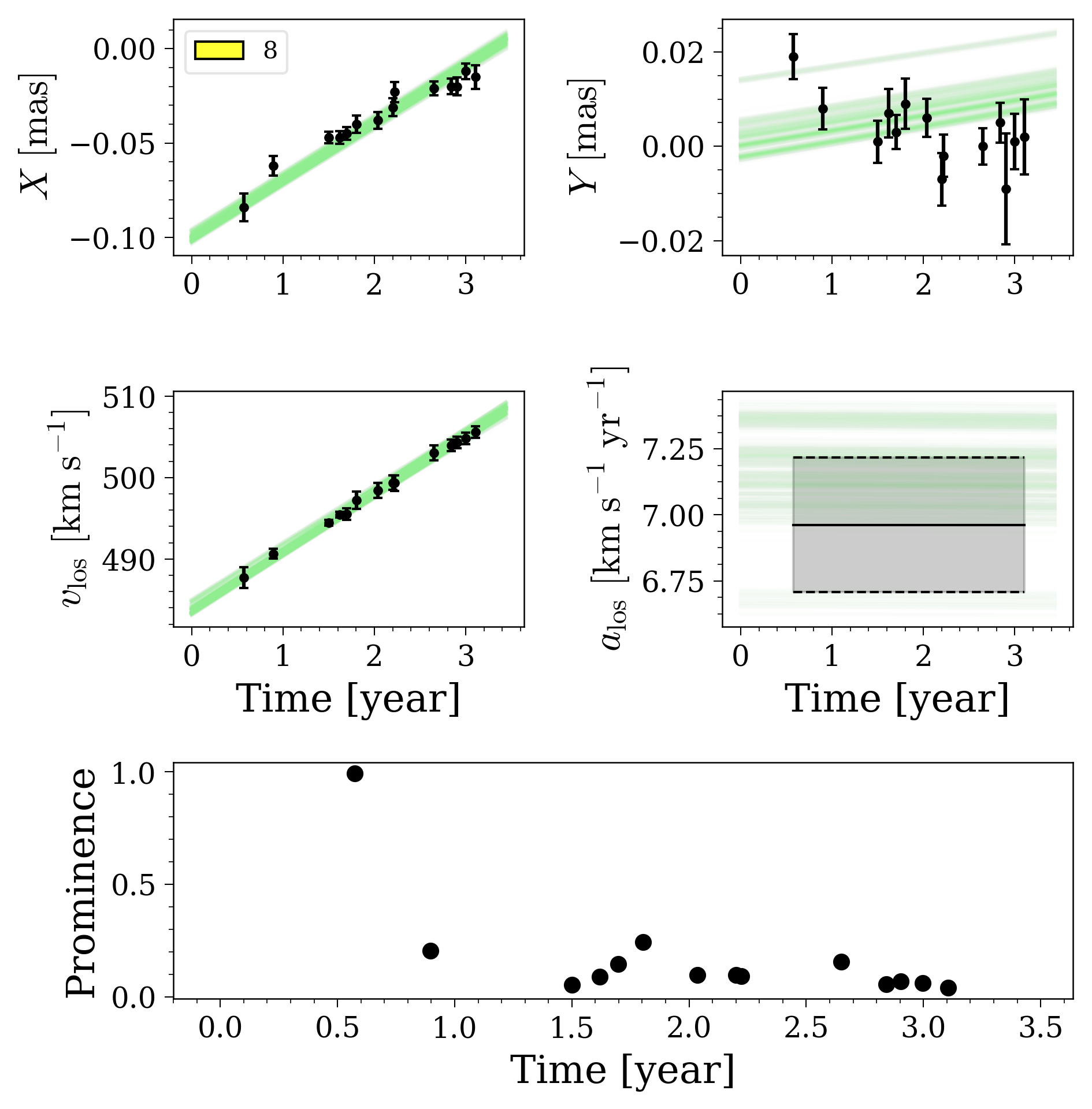} \\
        \includegraphics[width=0.3\linewidth]{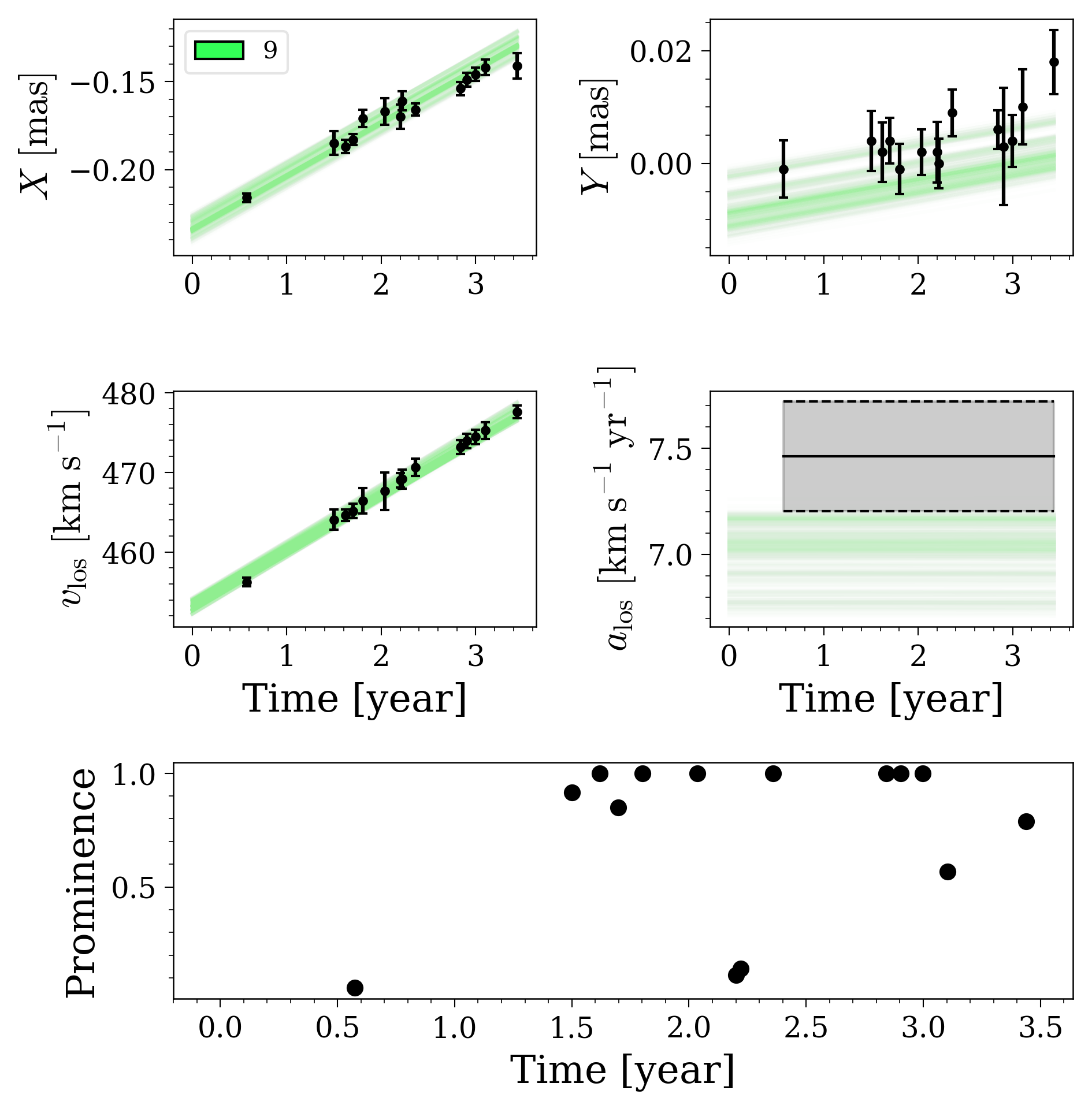} &
        \includegraphics[width=0.3\linewidth]{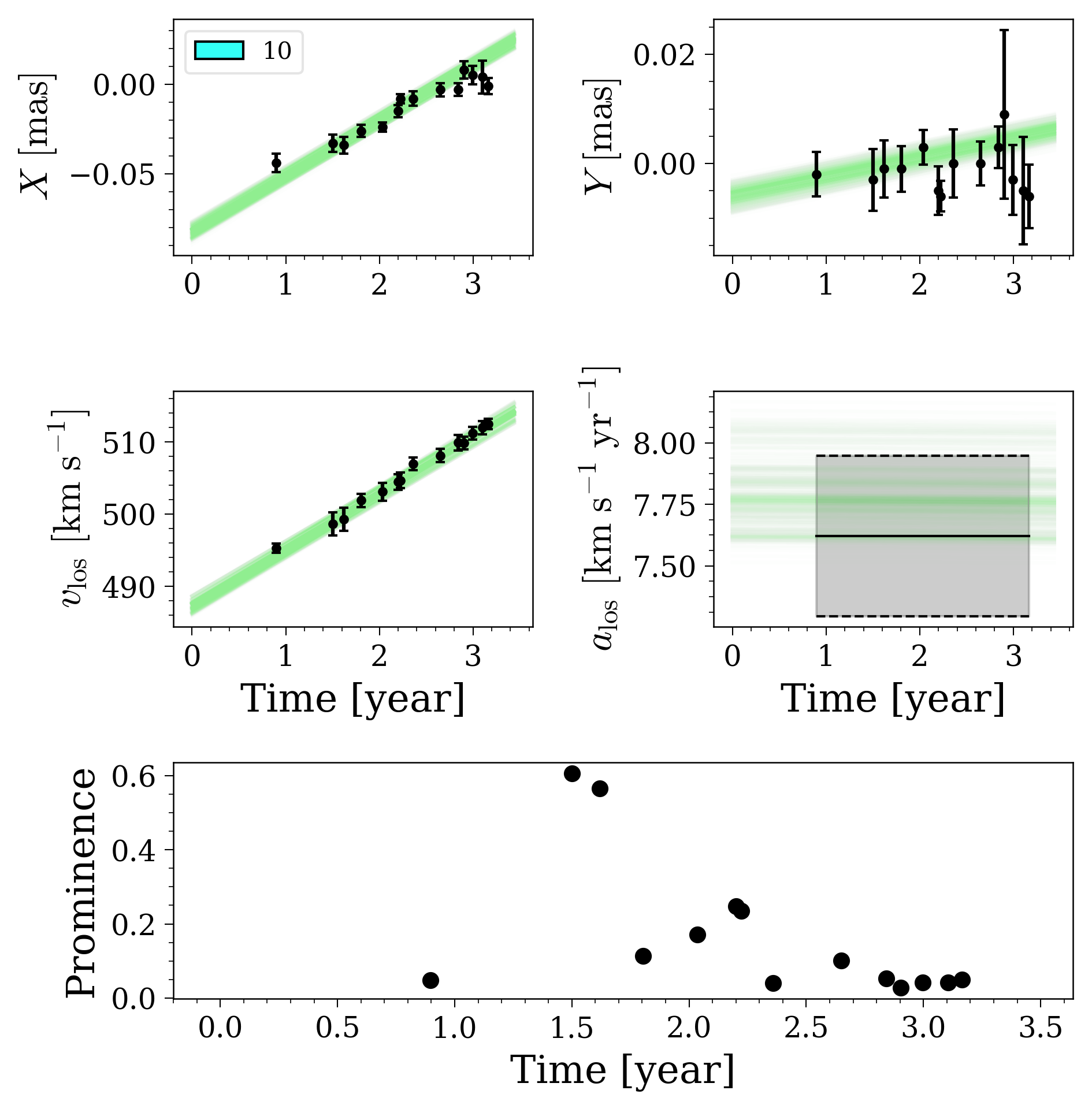}&
        \includegraphics[width=0.3\linewidth]{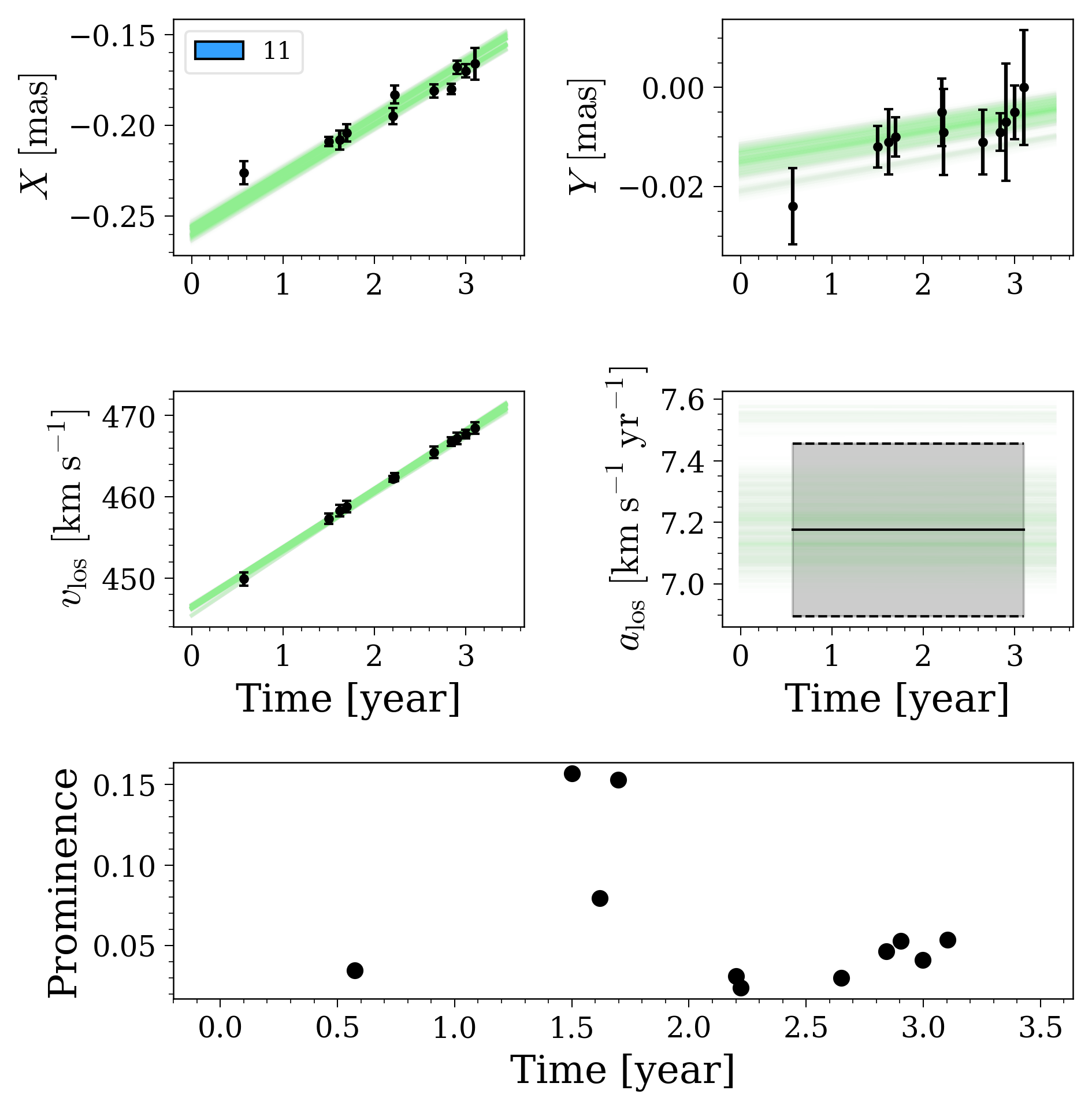} 
        \\ 
        \includegraphics[width=0.3\linewidth]{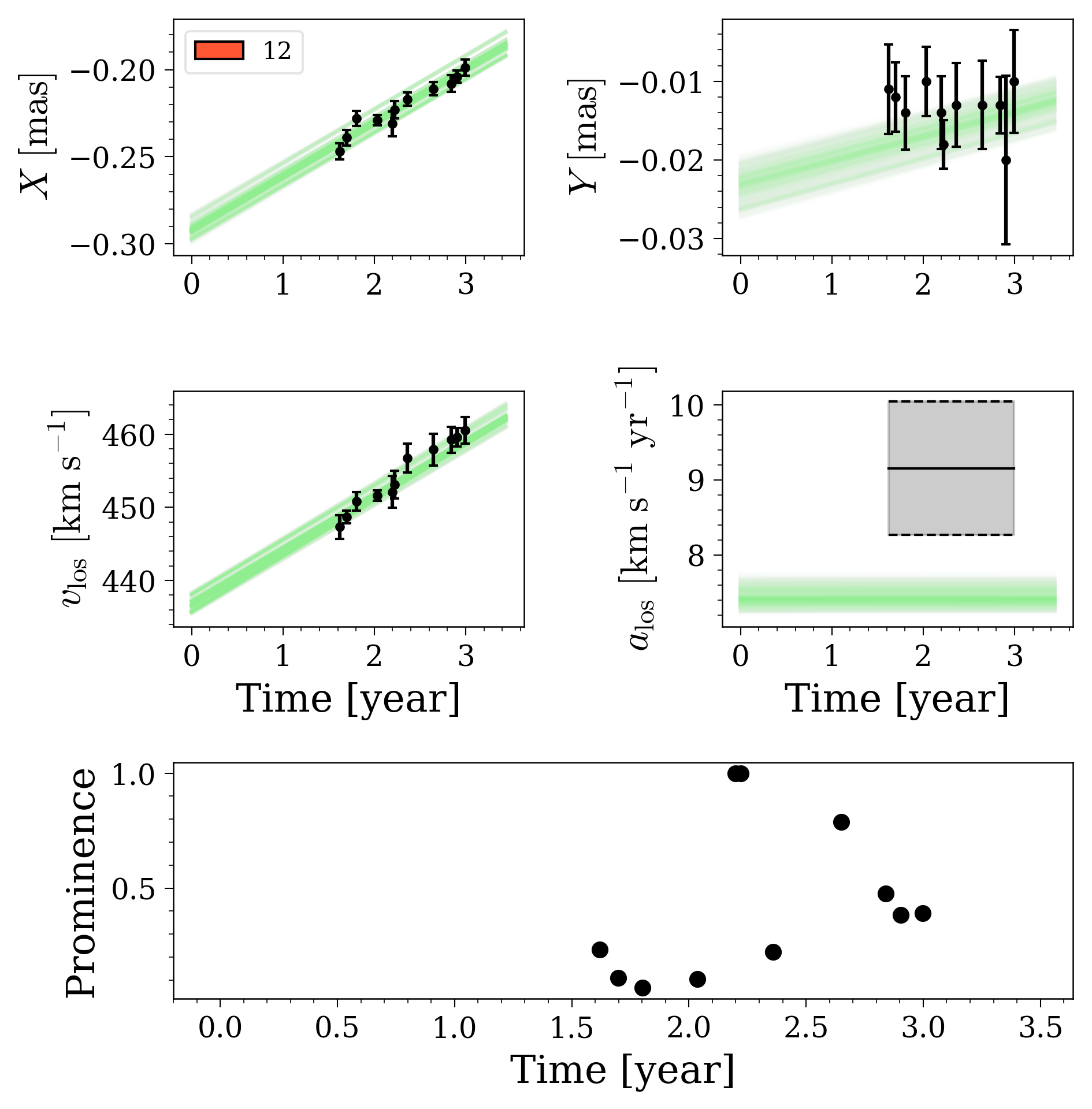}&
        \includegraphics[width=0.3\linewidth]{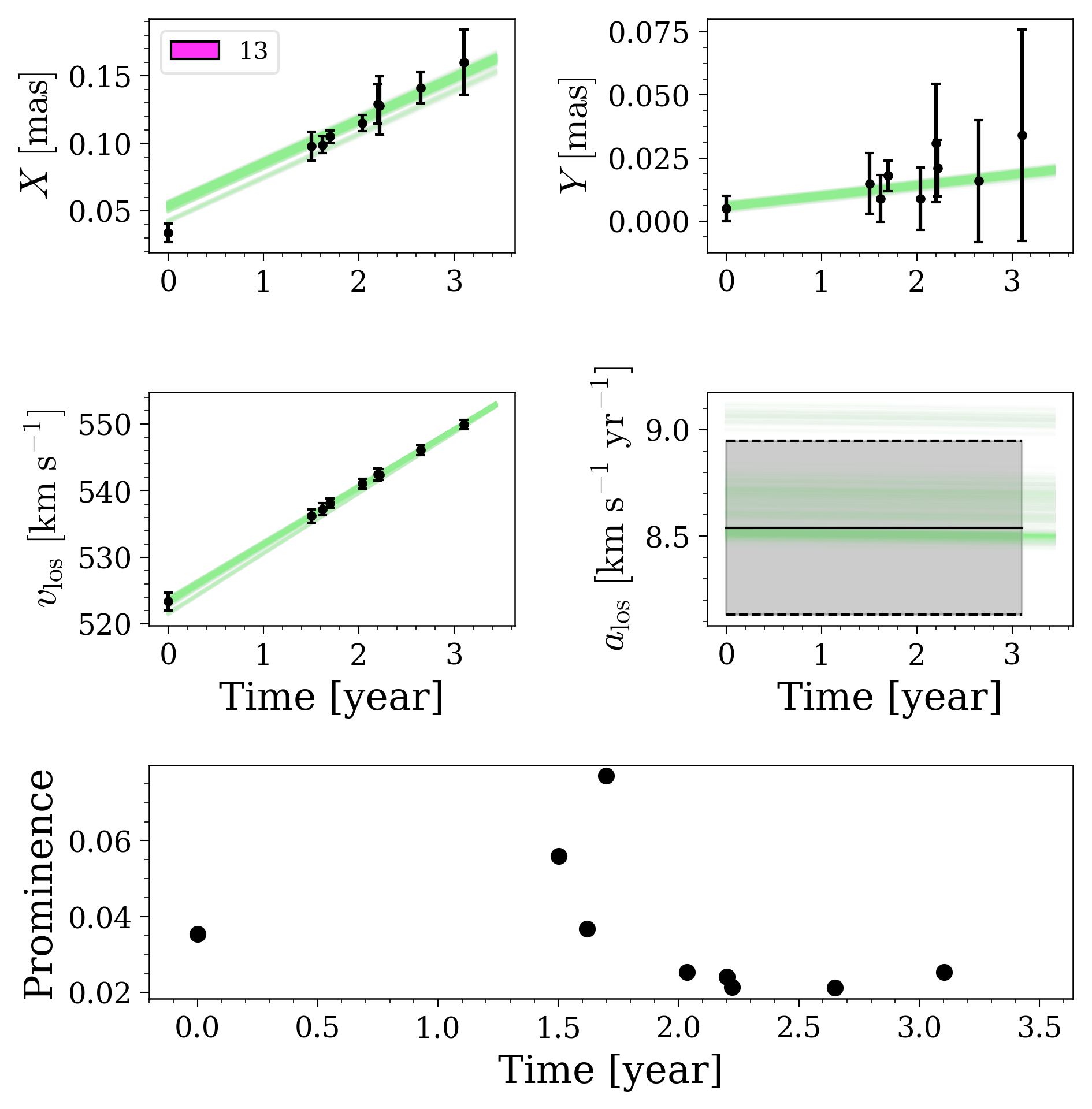}&
        \includegraphics[width=0.3\linewidth]{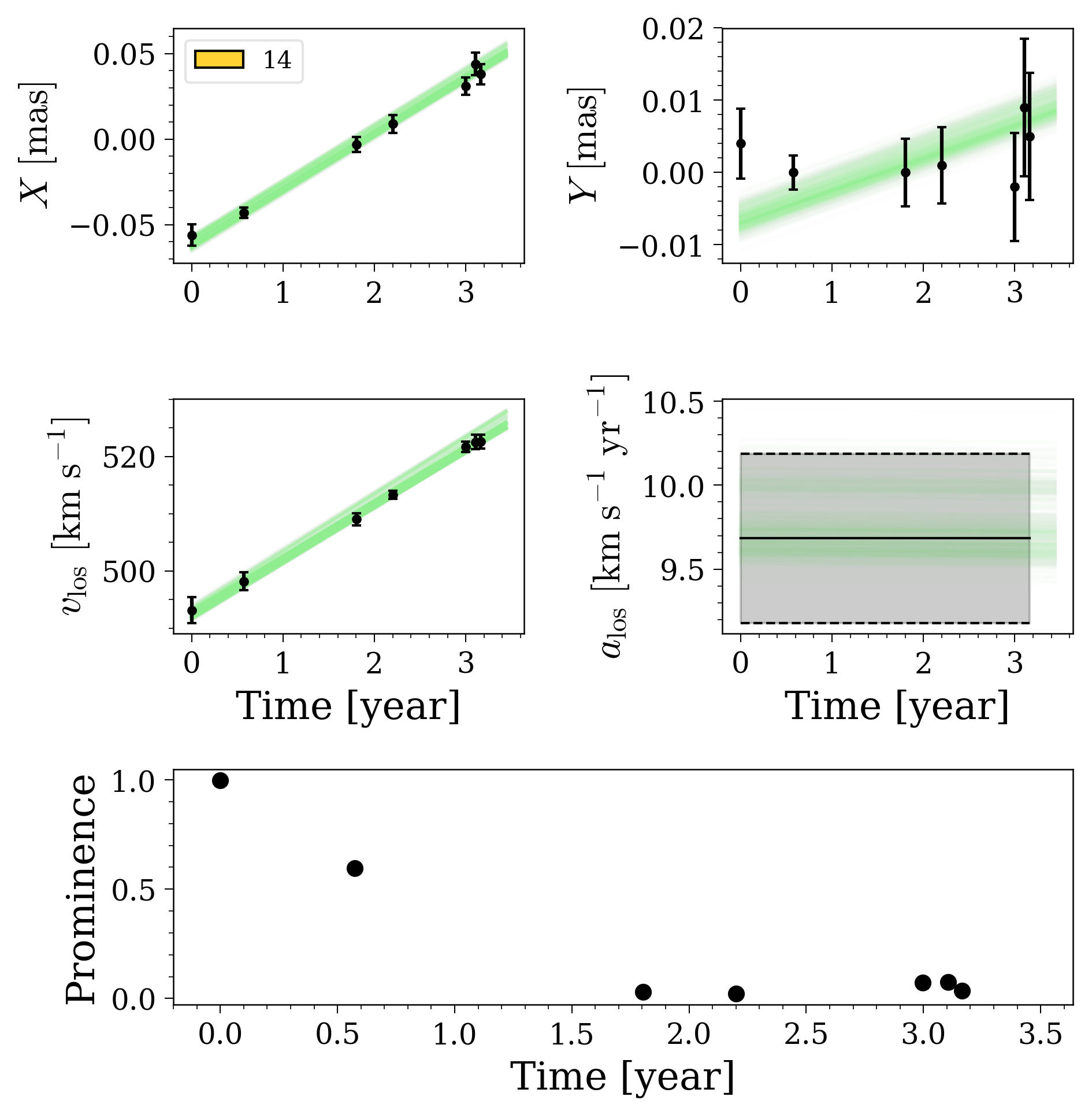} 
    \end{tabular}
    \caption{--- \textit{continued}}
    \label{fig:trajectory_group_2}
\end{figure*}

\begin{figure*}
    \centering
    \addtocounter{figure}{-1} 
    \begin{tabular}{ccc}
        \includegraphics[width=0.3\textwidth]{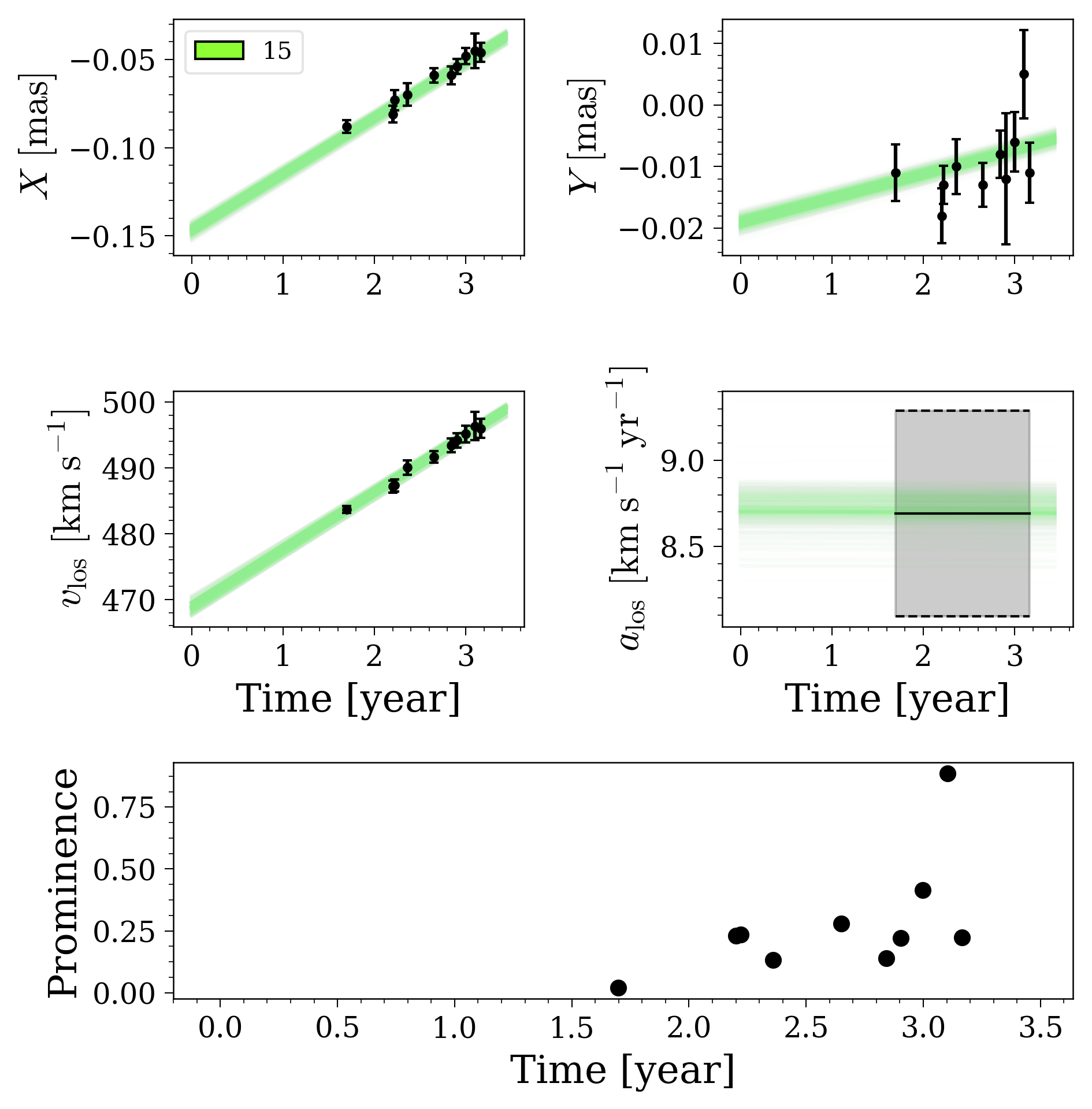} &
        \includegraphics[width=0.3\textwidth]{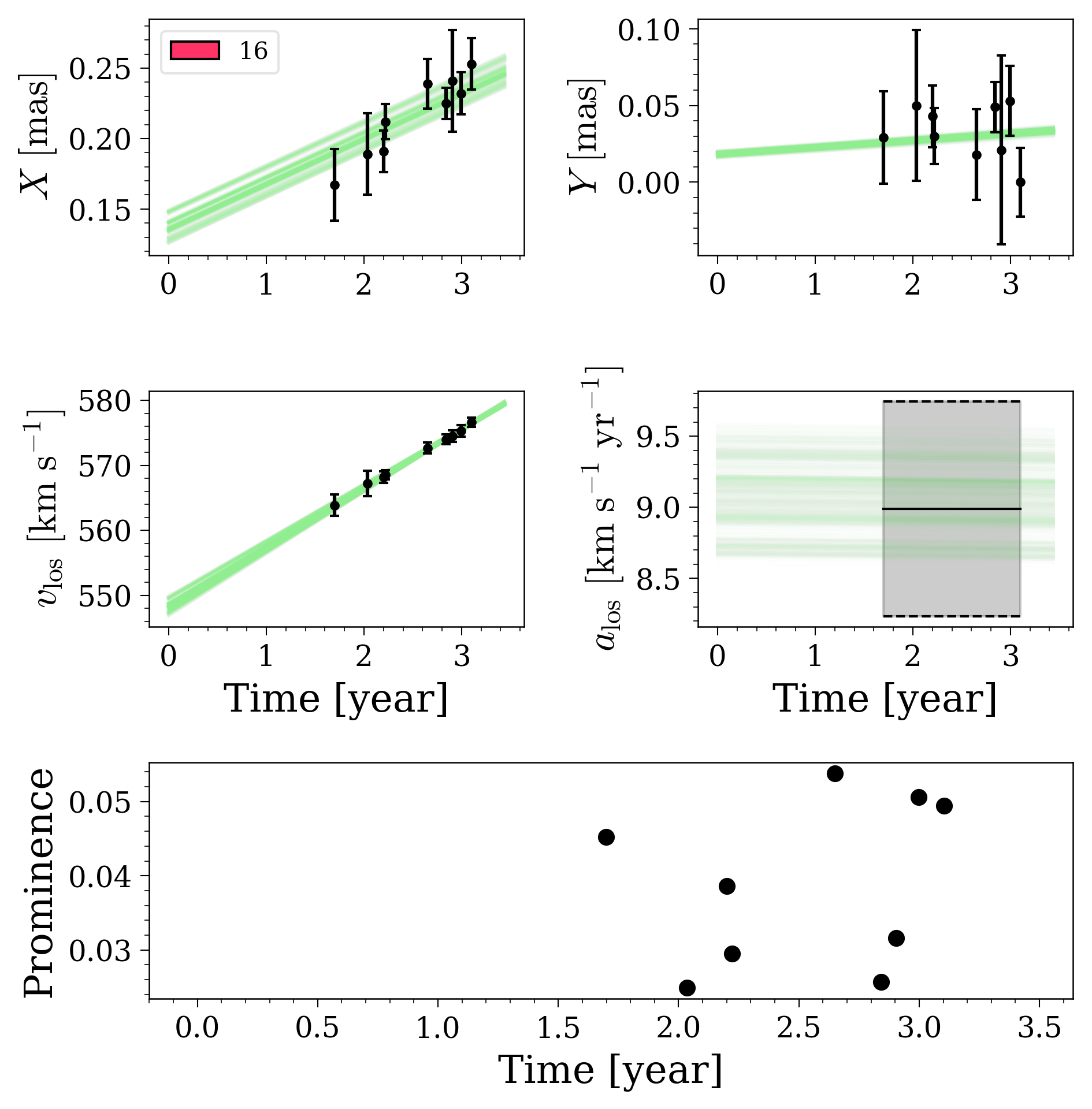} &
        \includegraphics[width=0.3\textwidth]{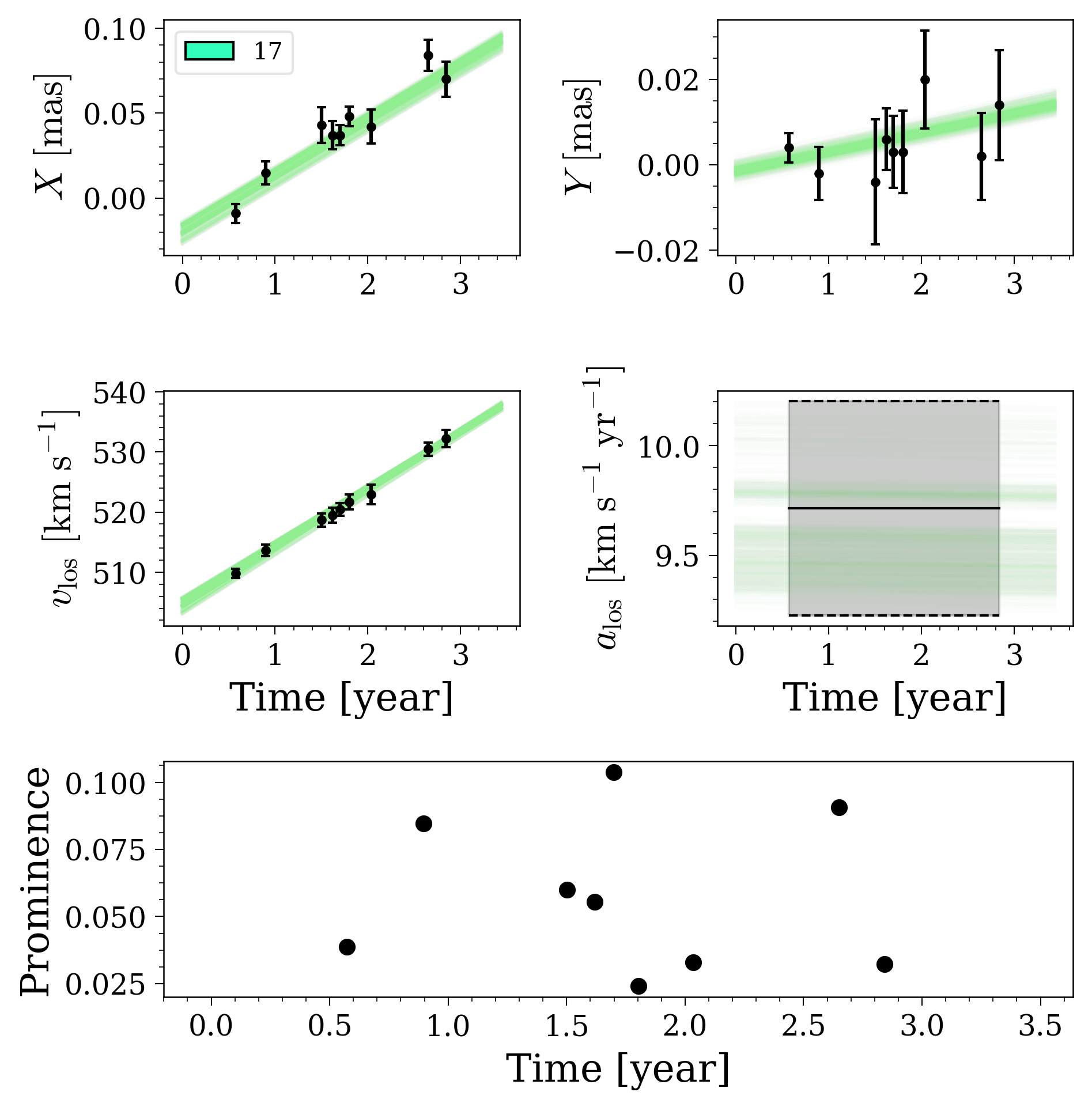} \\
        \includegraphics[width=0.3\textwidth]{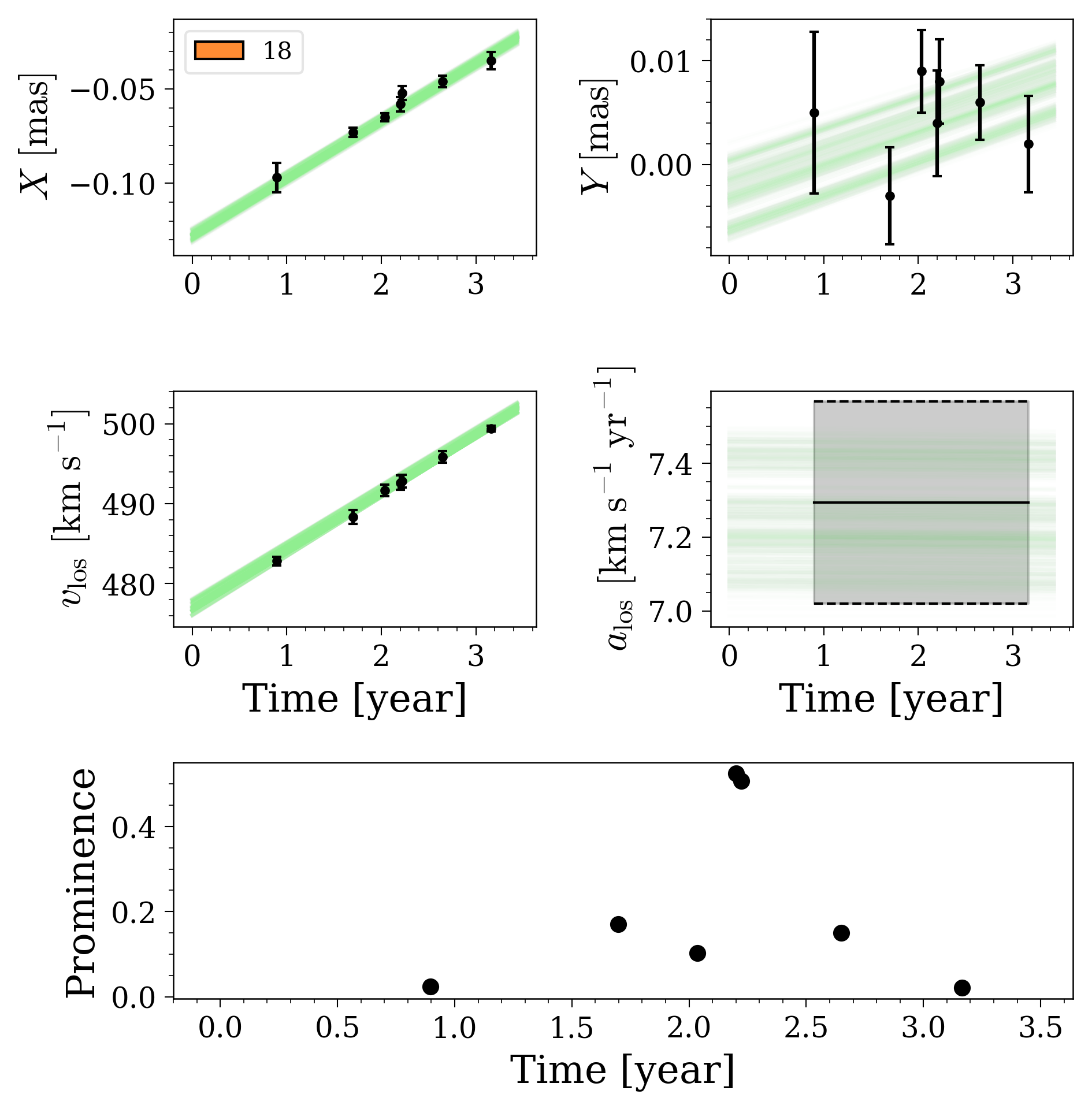} &
        \includegraphics[width=0.3\textwidth]{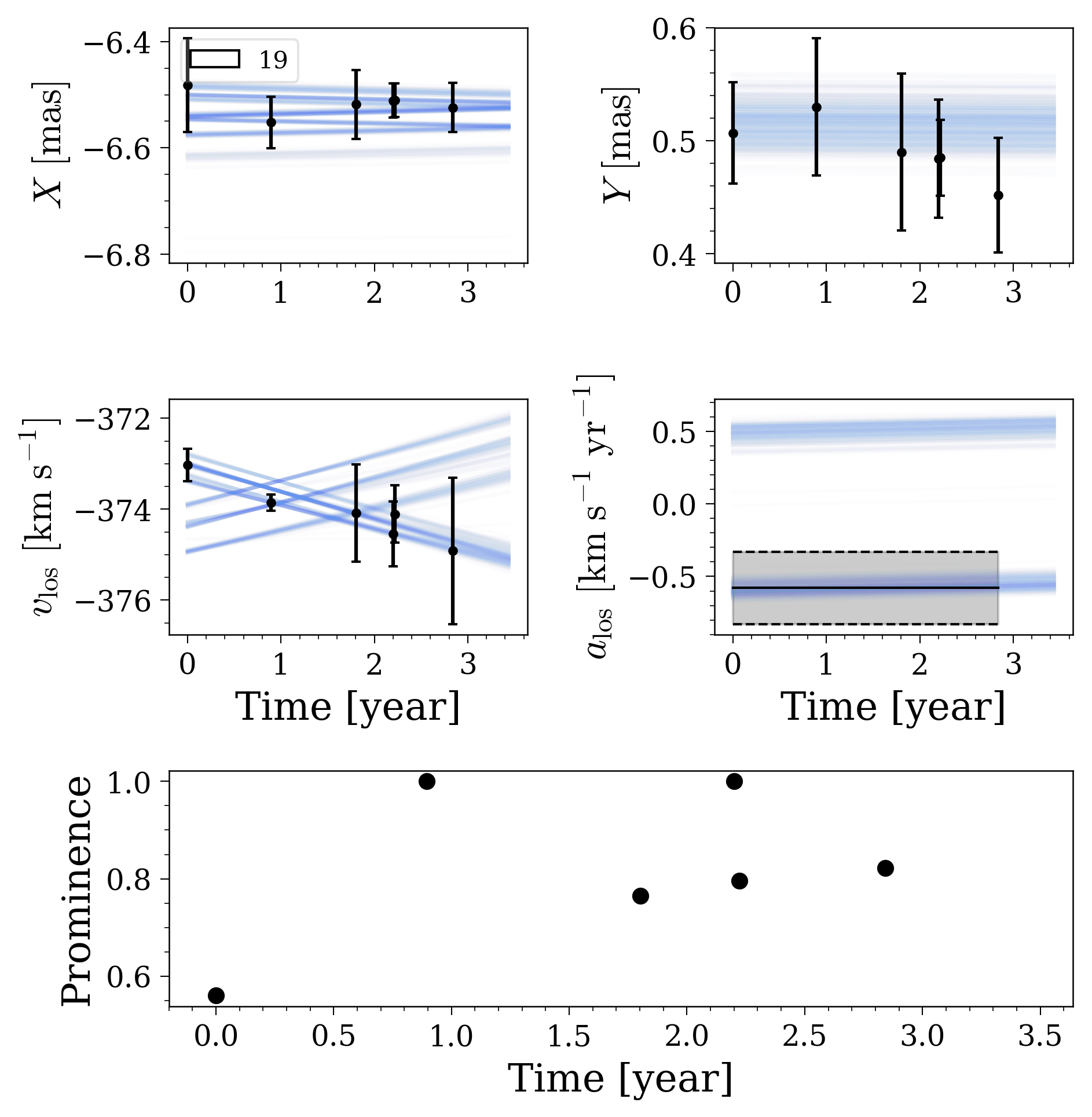} &
        \includegraphics[width=0.3\textwidth]{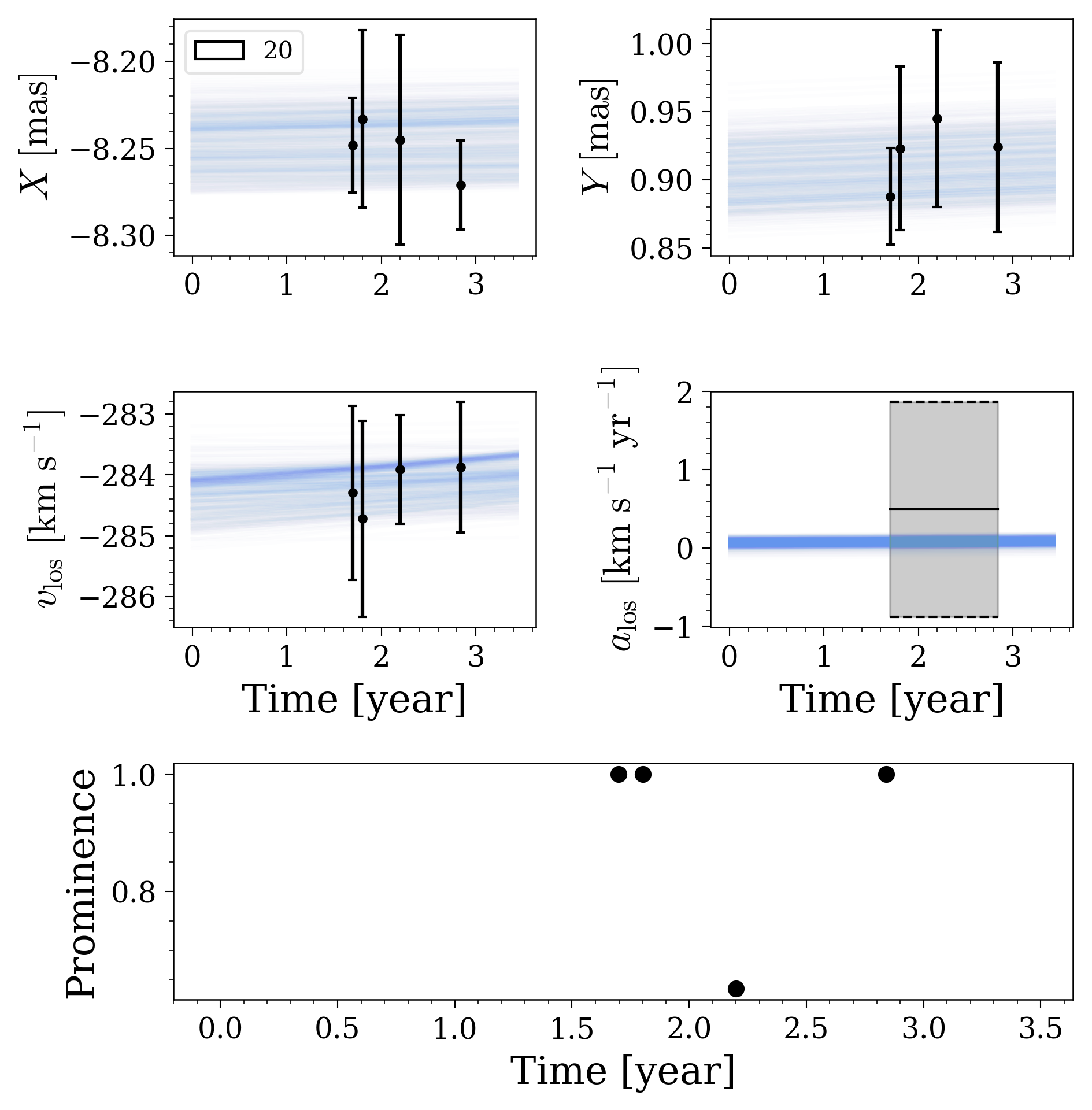} \\ &
        \includegraphics[width=0.3\textwidth]{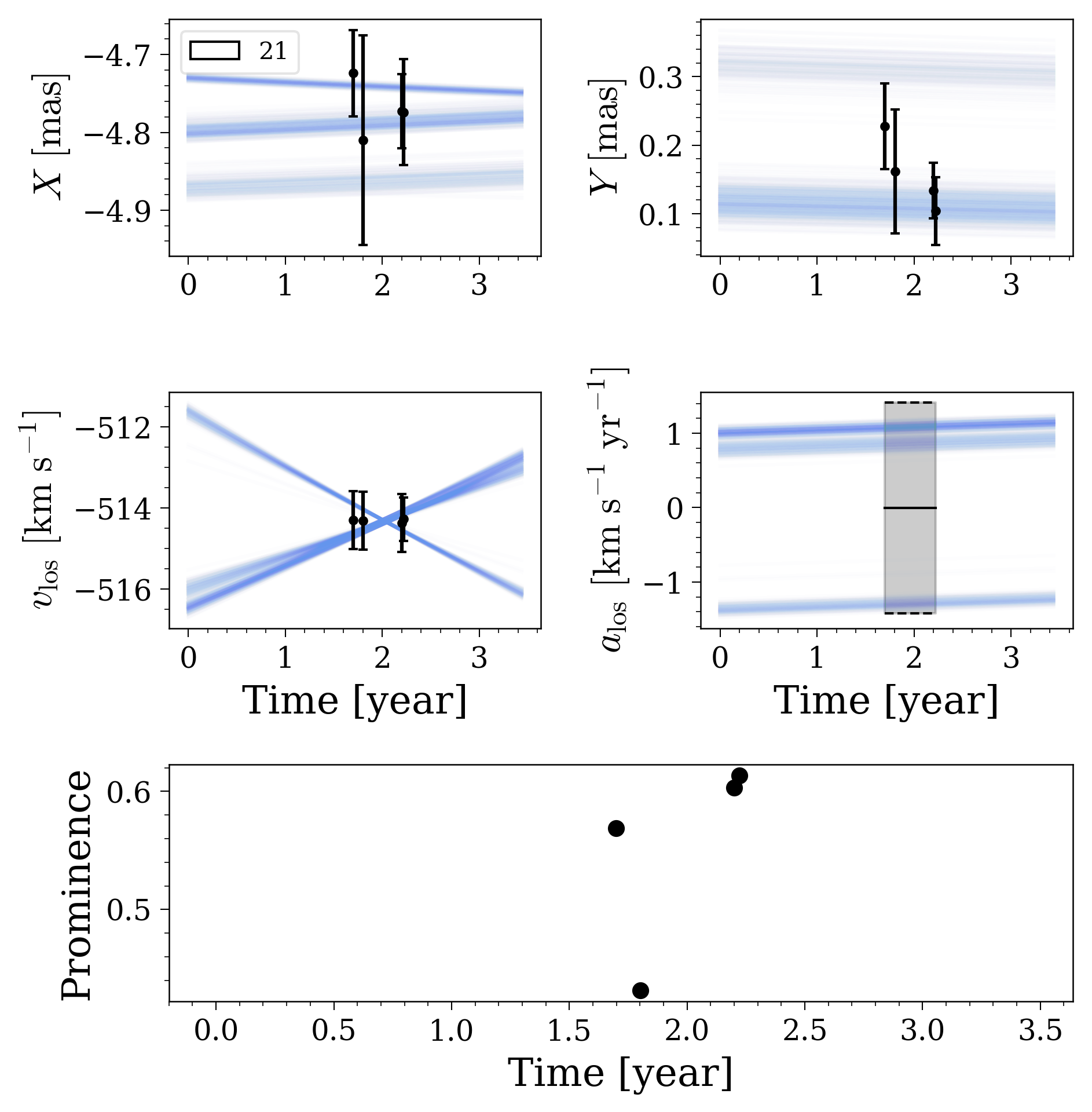}  & \\ 
    \end{tabular}
    \caption{--- \textit{continued}}
    \label{fig:trajectory_group_2}
\end{figure*}
\begin{figure*} 
    \centering
    \begin{tabular}{ccc}
        \includegraphics[width=0.3\textwidth]{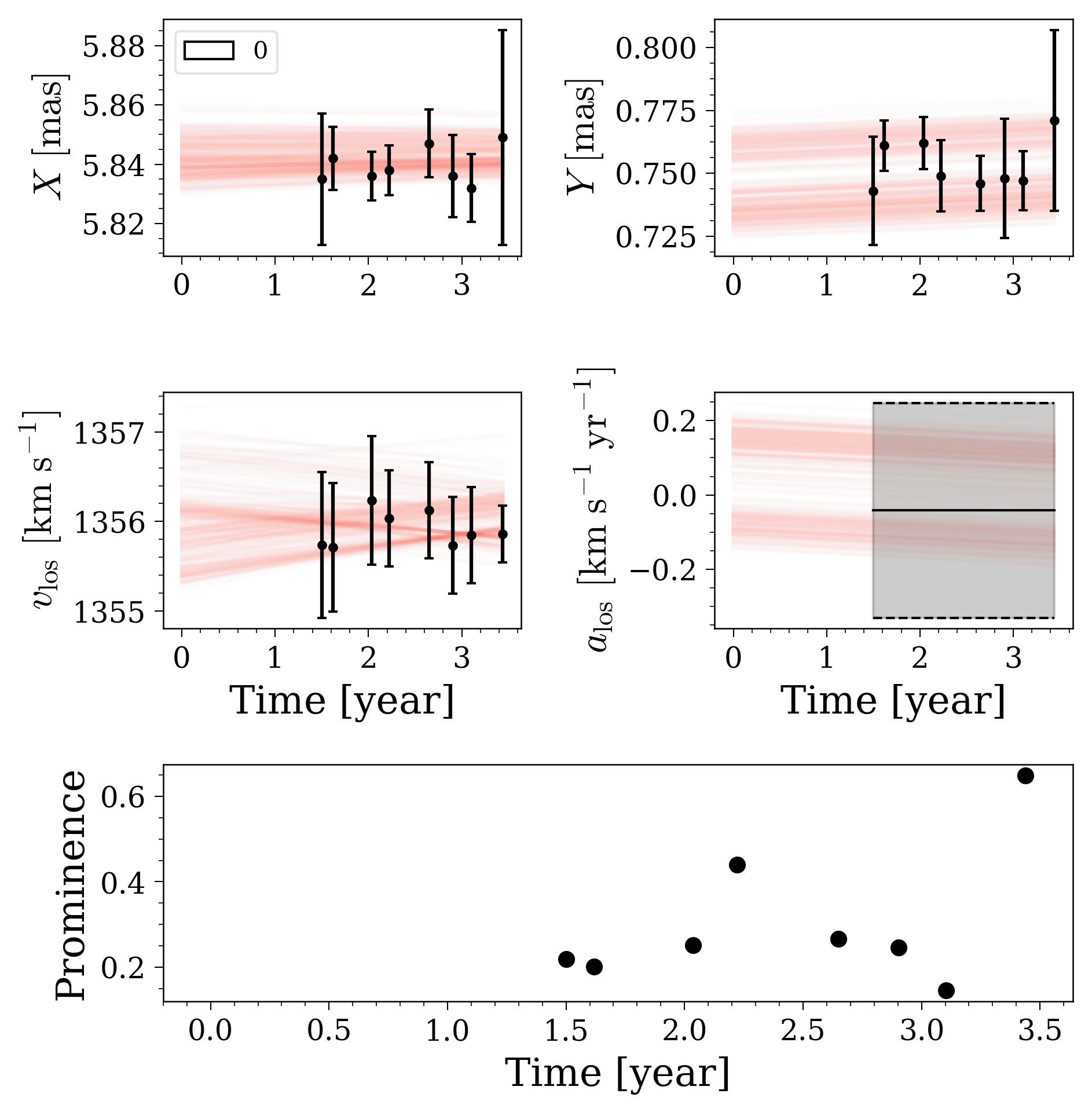} &
        \includegraphics[width=0.3\textwidth]{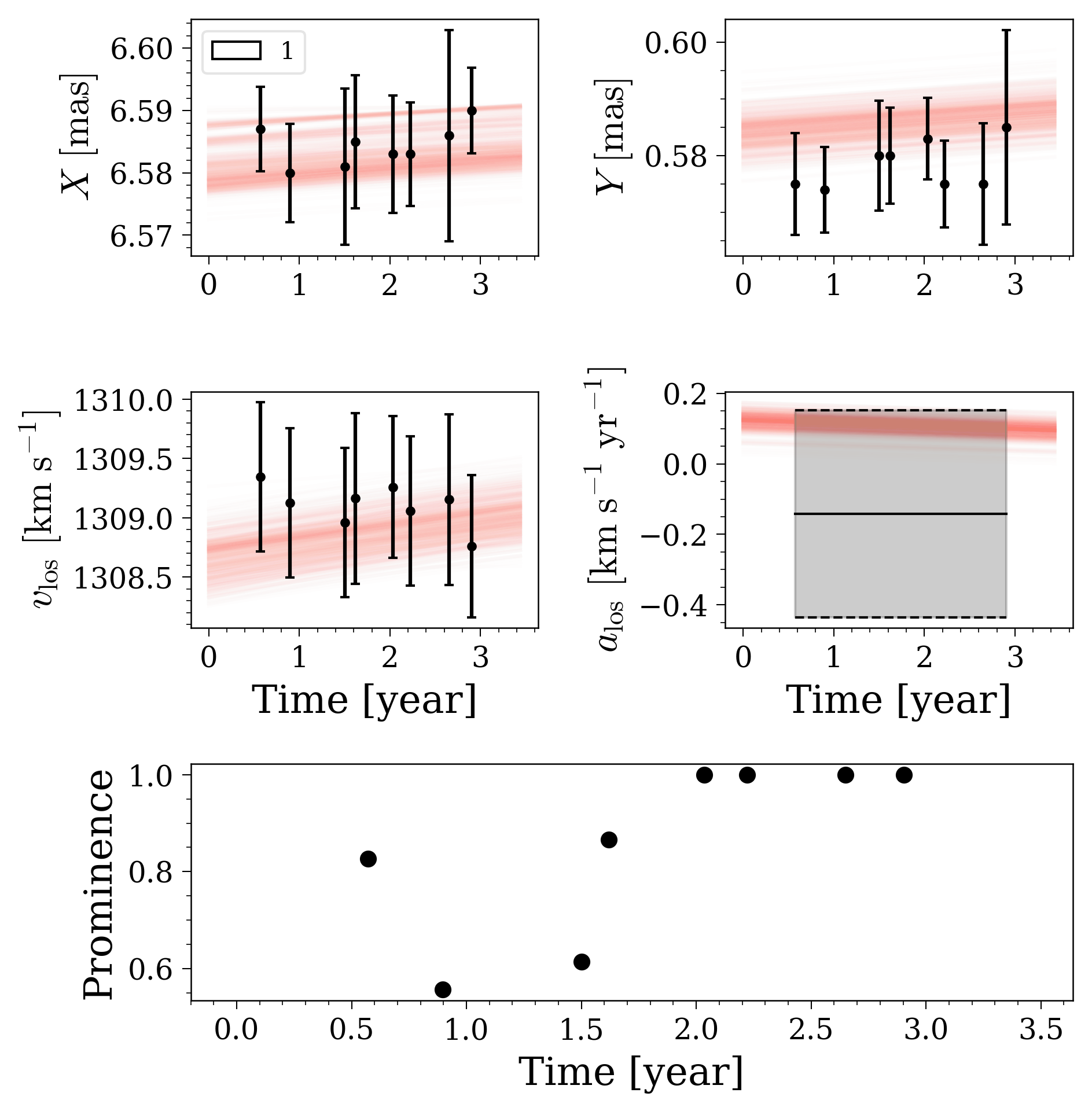} &
        \includegraphics[width=0.3\textwidth]{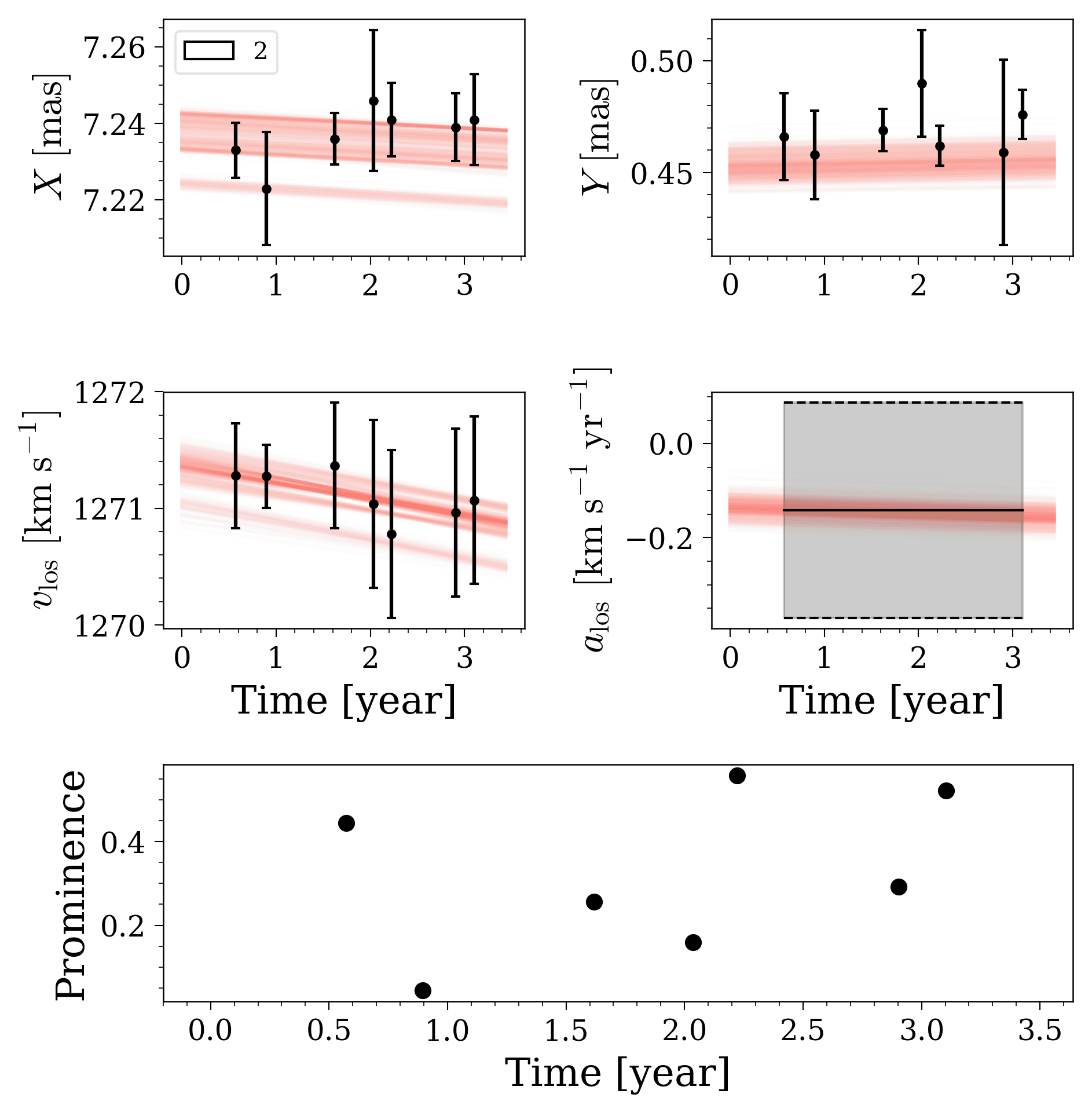} \\
        \includegraphics[width=0.3\textwidth]{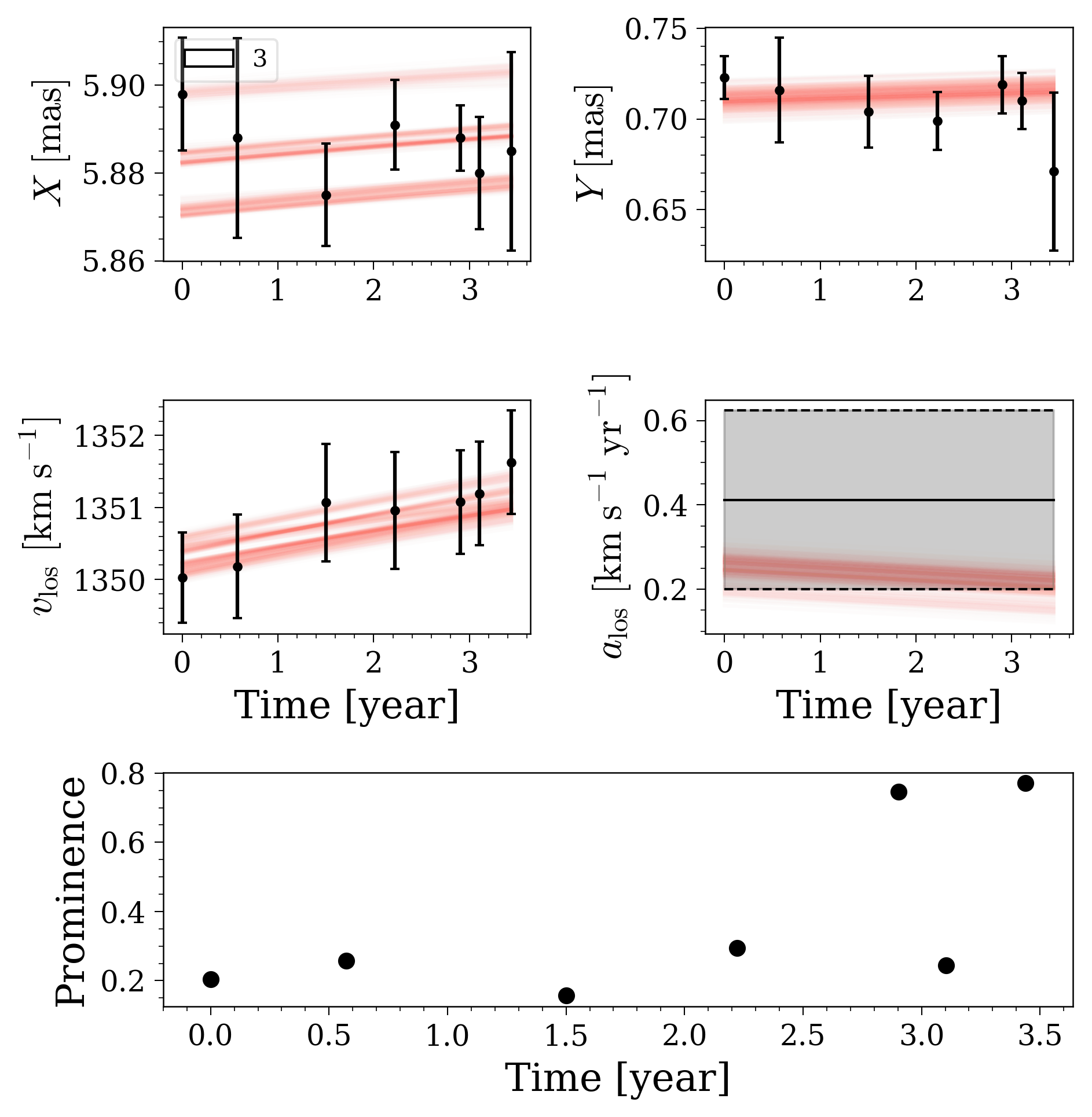} &
        \includegraphics[width=0.3\textwidth]{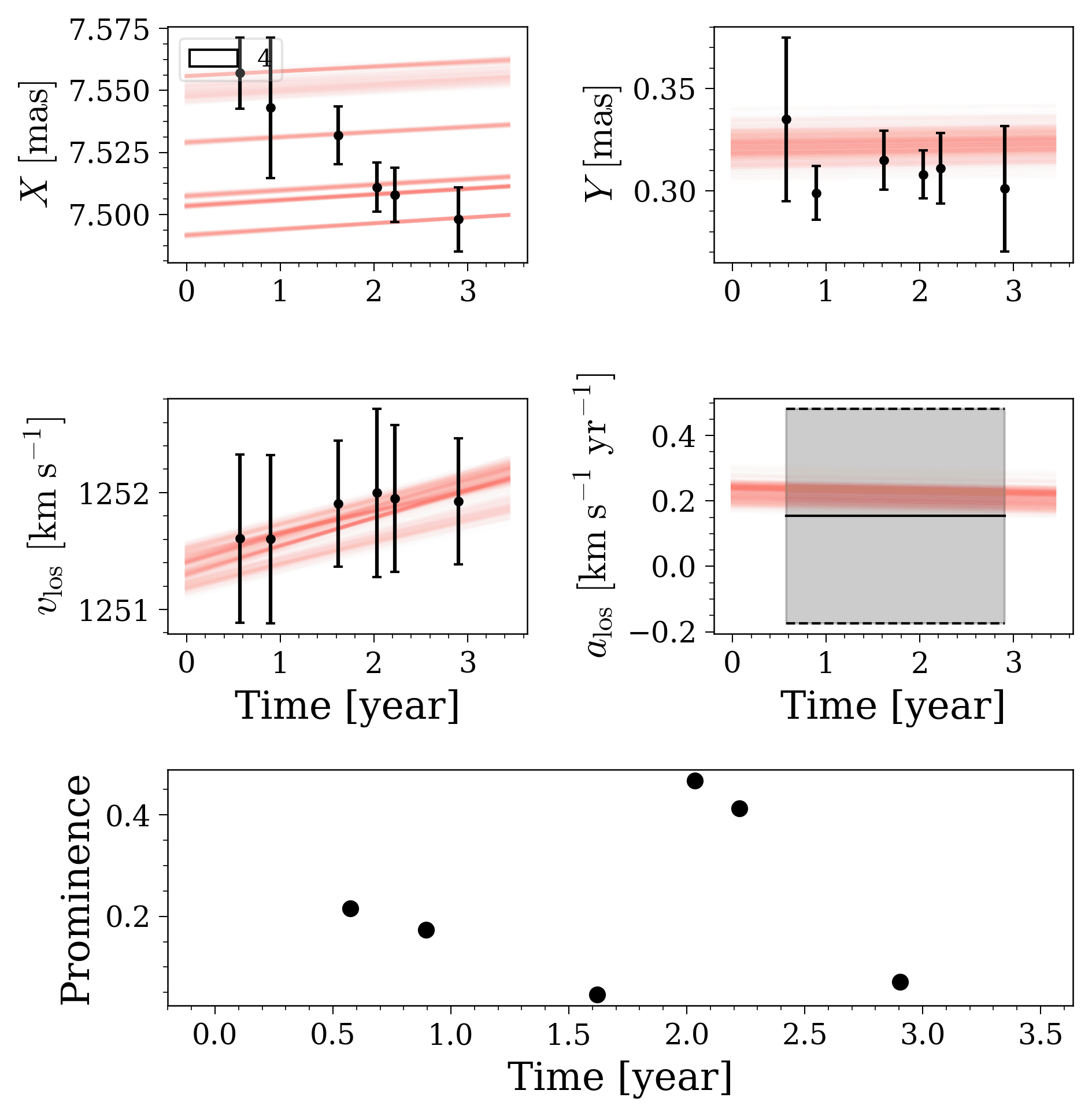} &
        \includegraphics[width=0.3\textwidth]{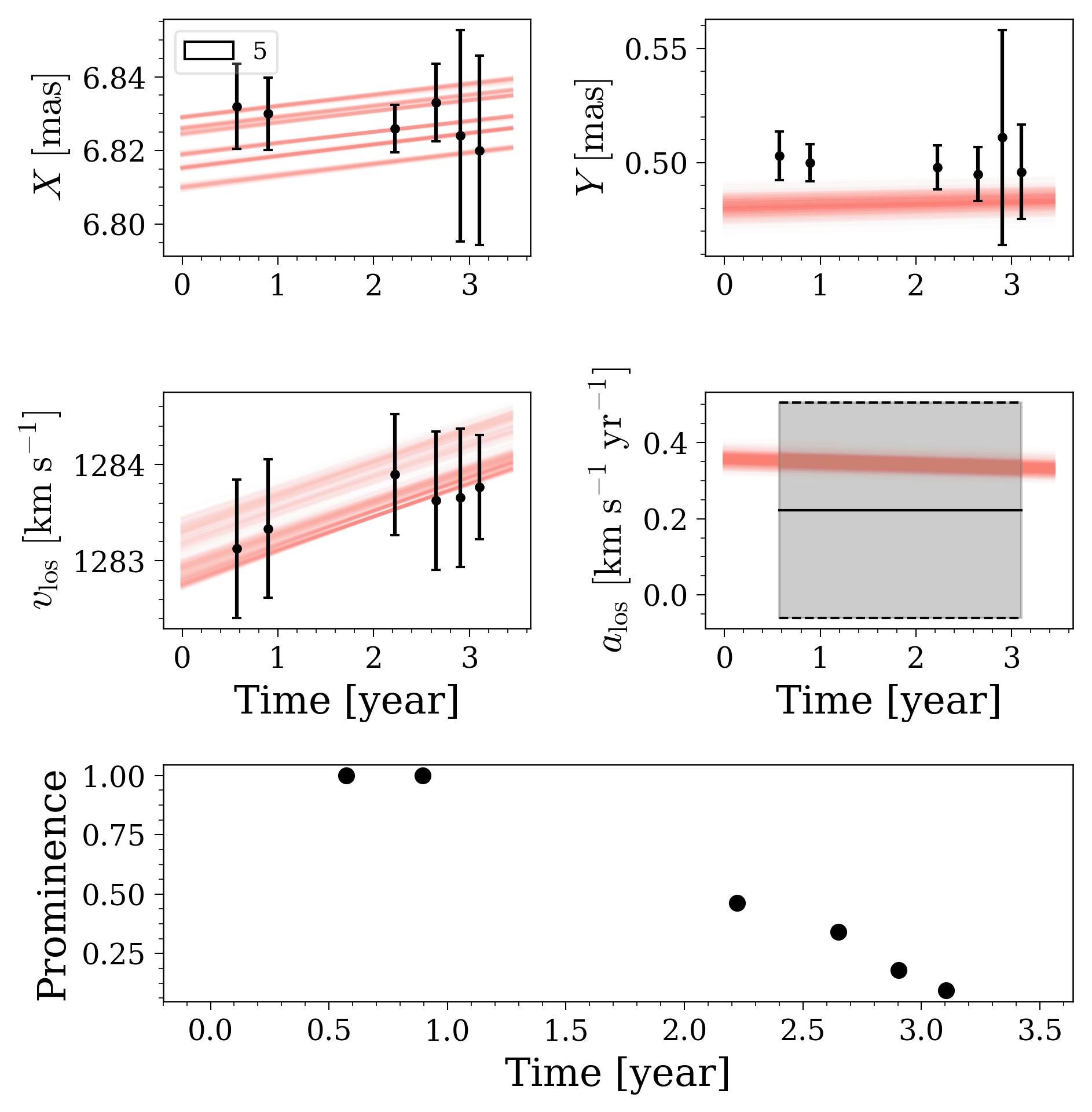} \\
        \includegraphics[width=0.3\textwidth]{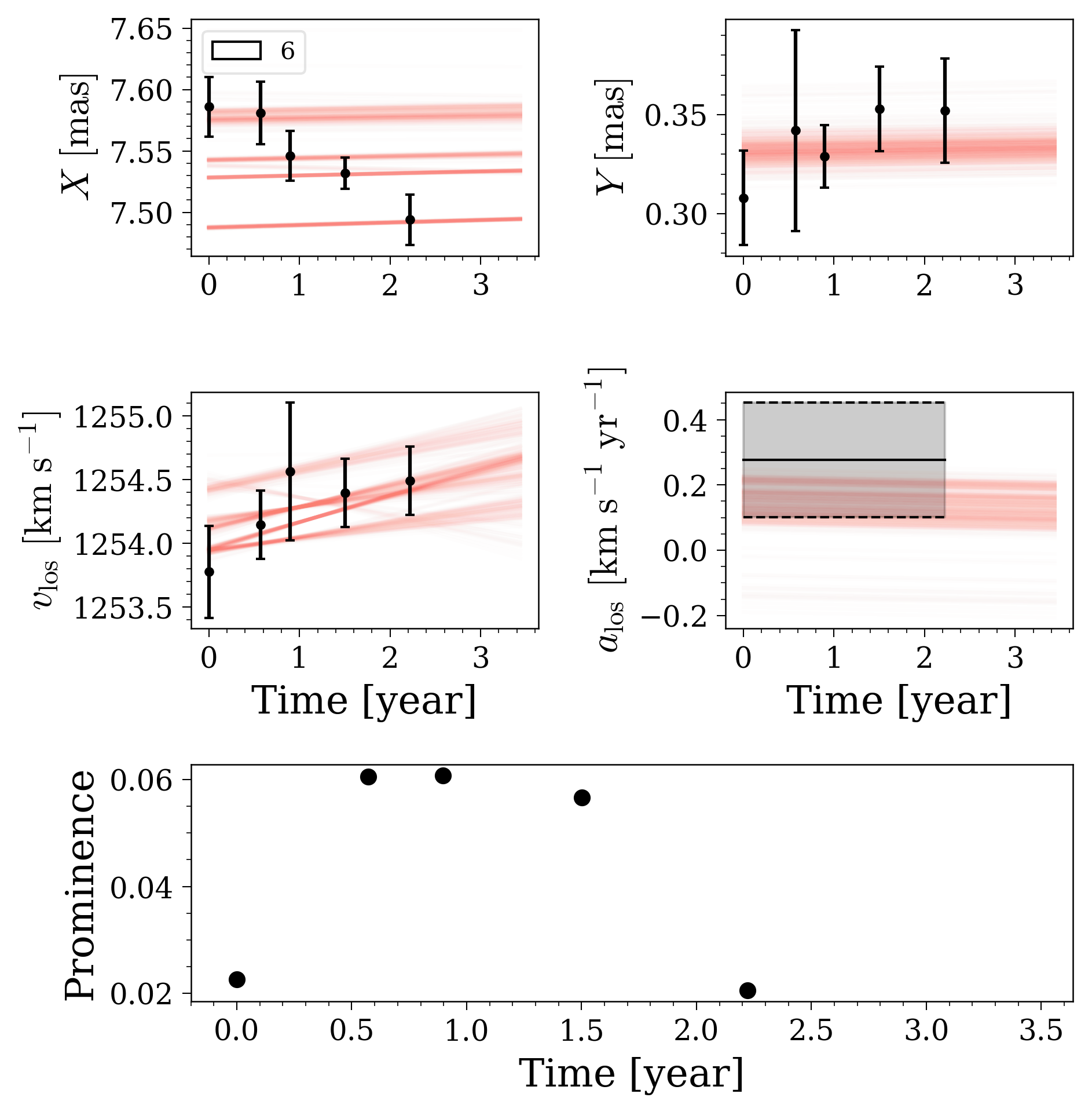} &
        \includegraphics[width=0.3\textwidth]{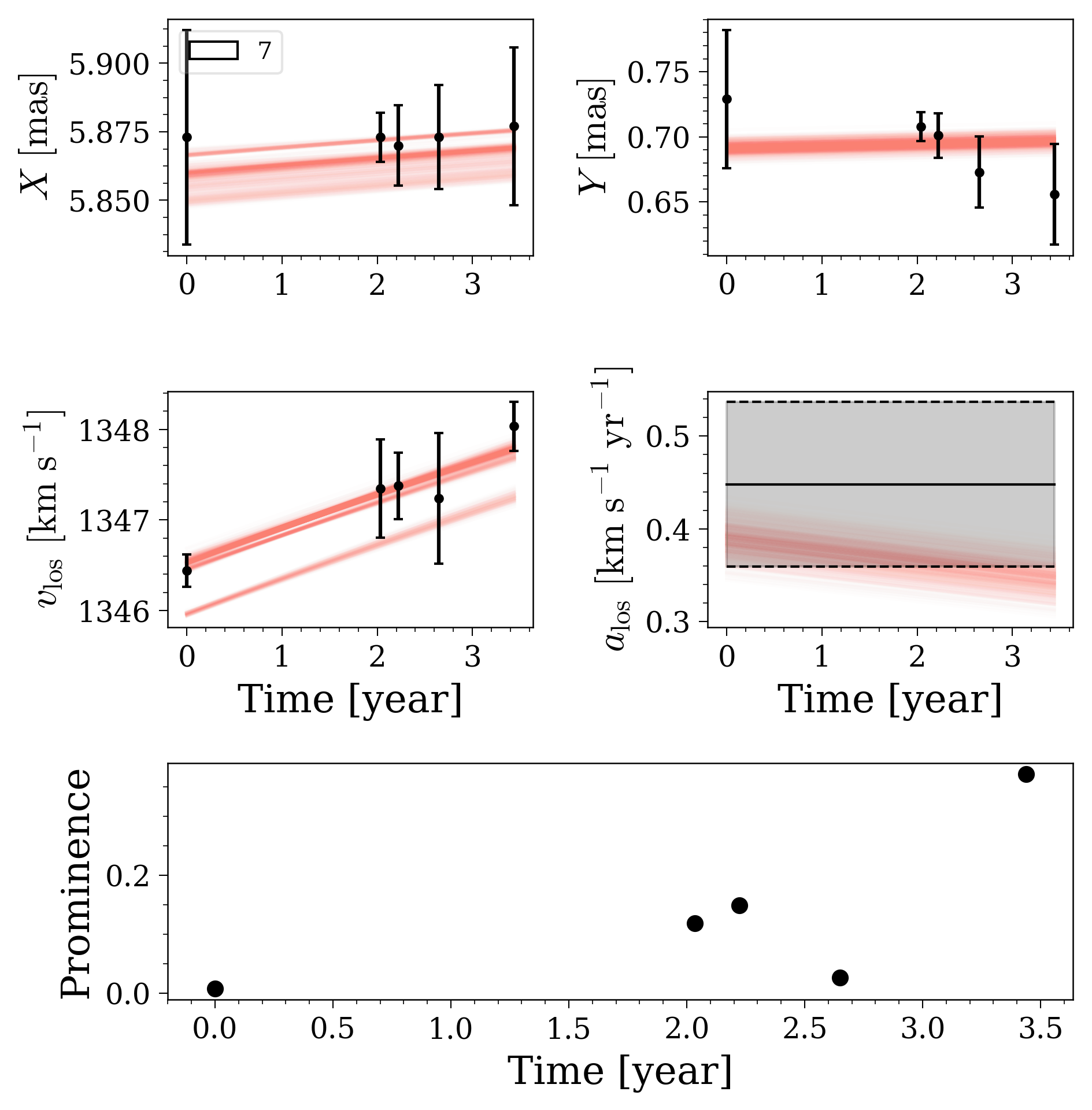} &
        \includegraphics[width=0.3\textwidth]{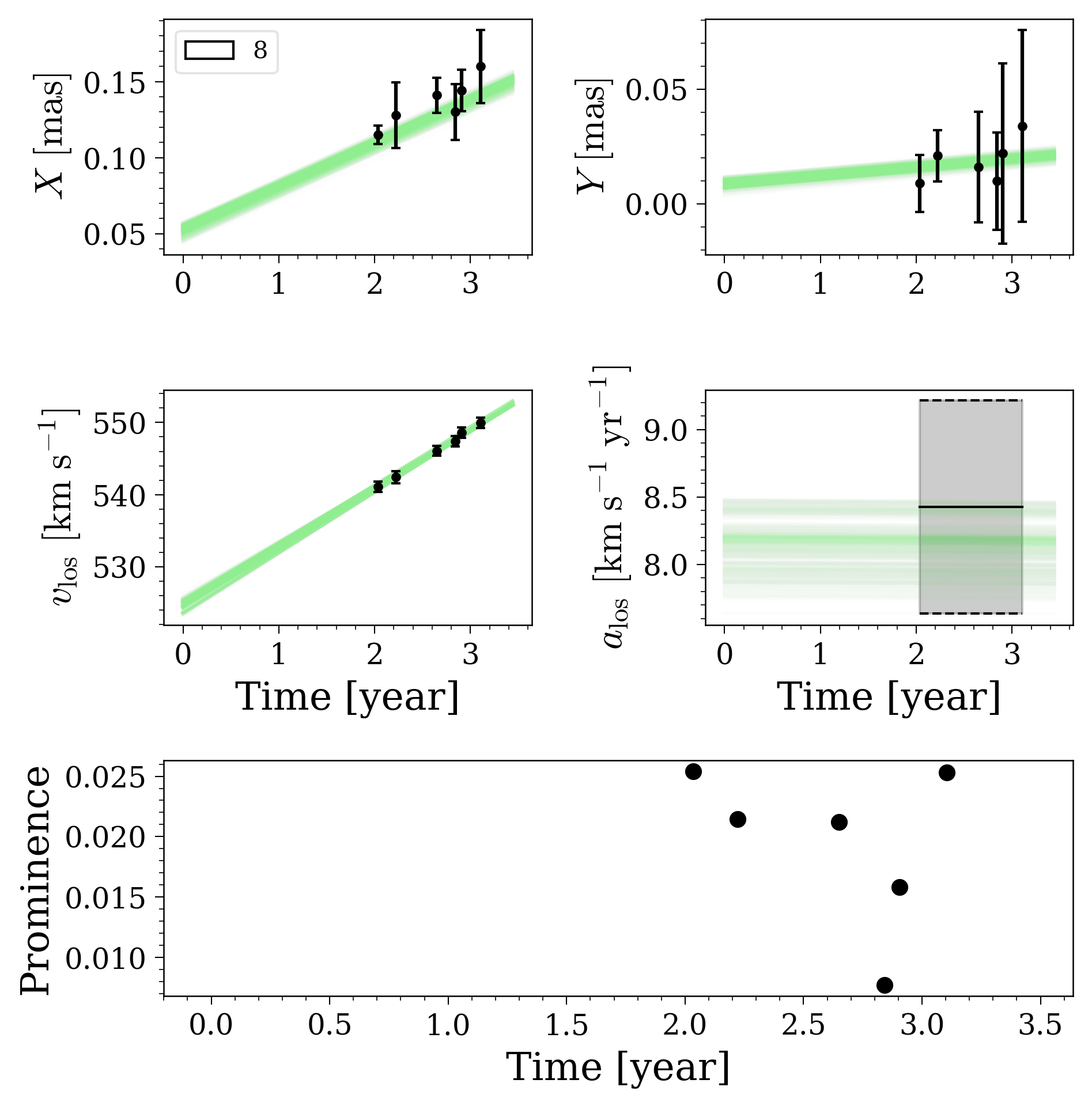} \\
        \includegraphics[width=0.3\textwidth]{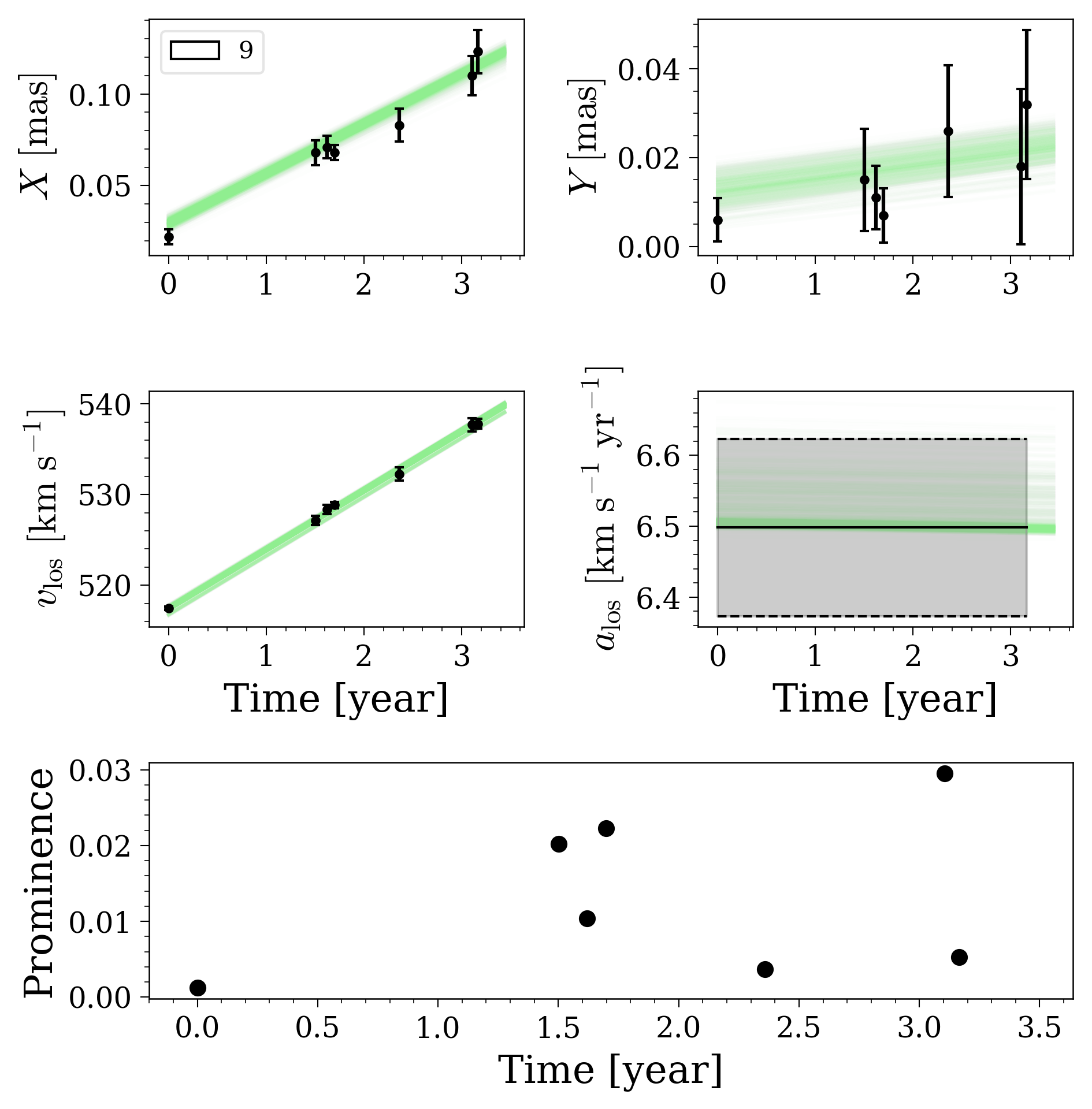} &
        \includegraphics[width=0.3\textwidth]{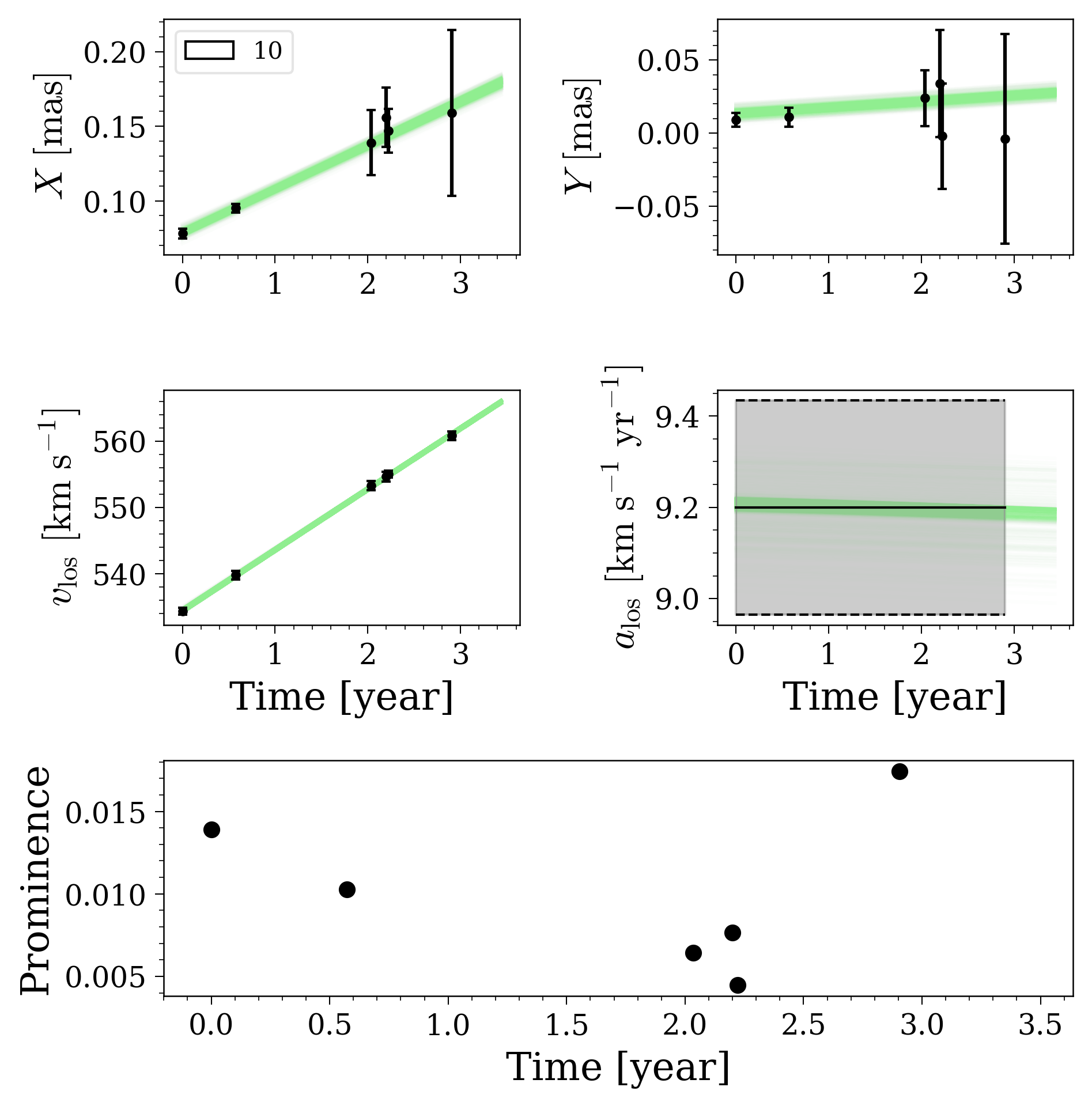} &
        \includegraphics[width=0.3\textwidth]{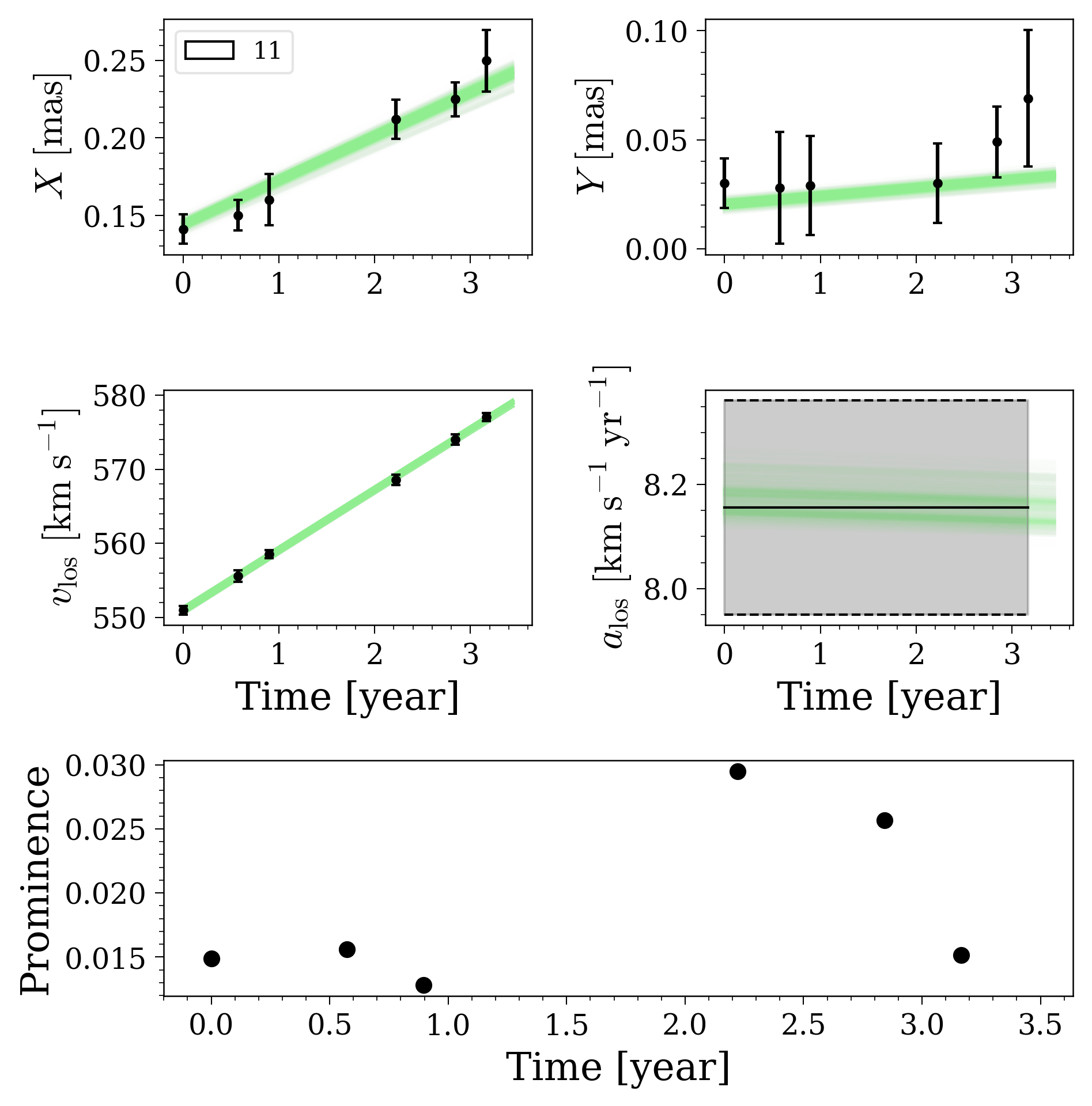}
        
    \end{tabular}
    \caption{Same as Figure~\ref{fig:trajectory_group}, but for dataset \#1.}
    \label{fig:trajectory_group_13_1}
\end{figure*}

\begin{figure*}
    \centering
    \addtocounter{figure}{-1} 
    
    \begin{tabular}{ccc}

        \includegraphics[width=0.3\textwidth]{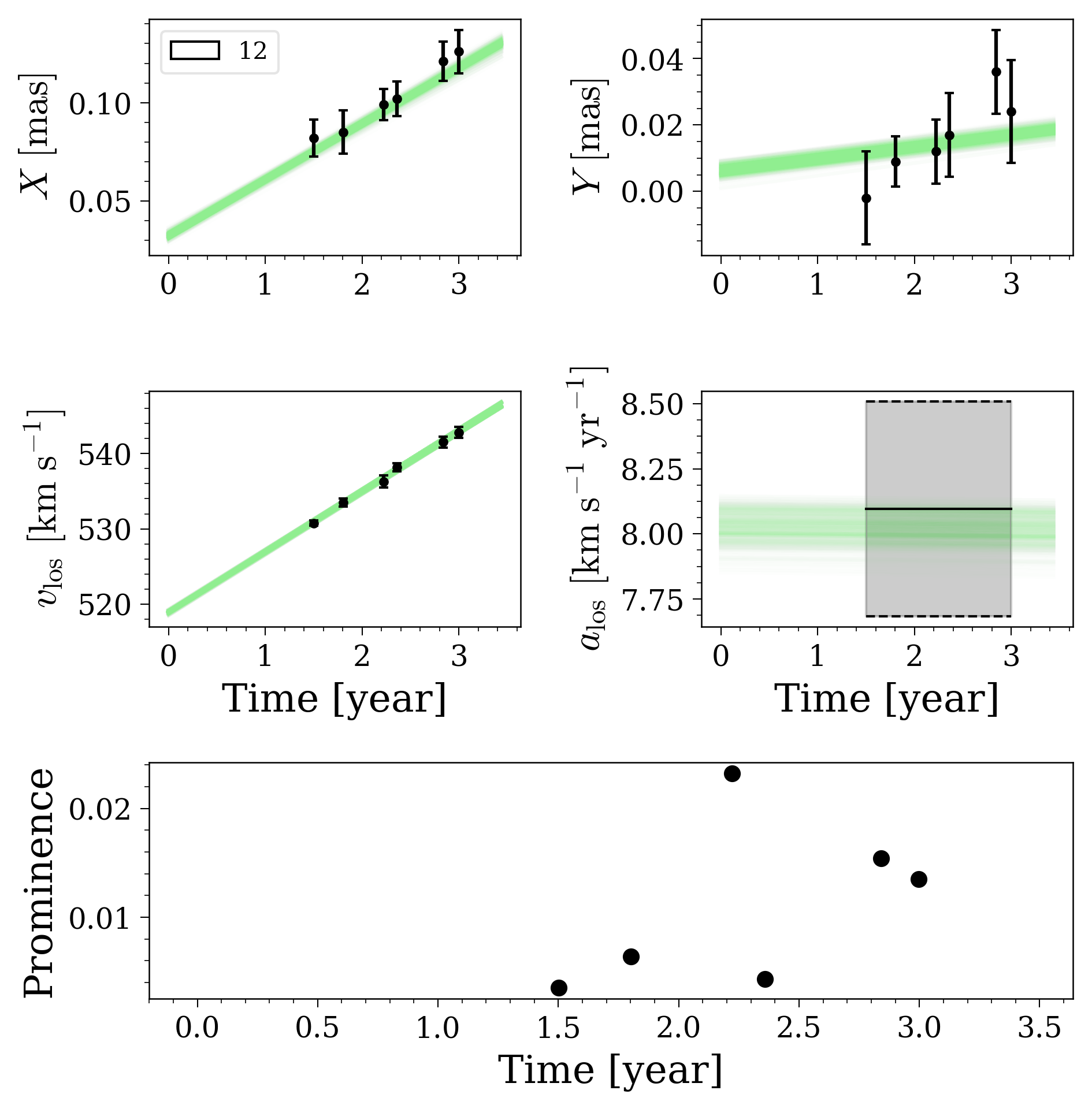} &
        \includegraphics[width=0.3\textwidth]{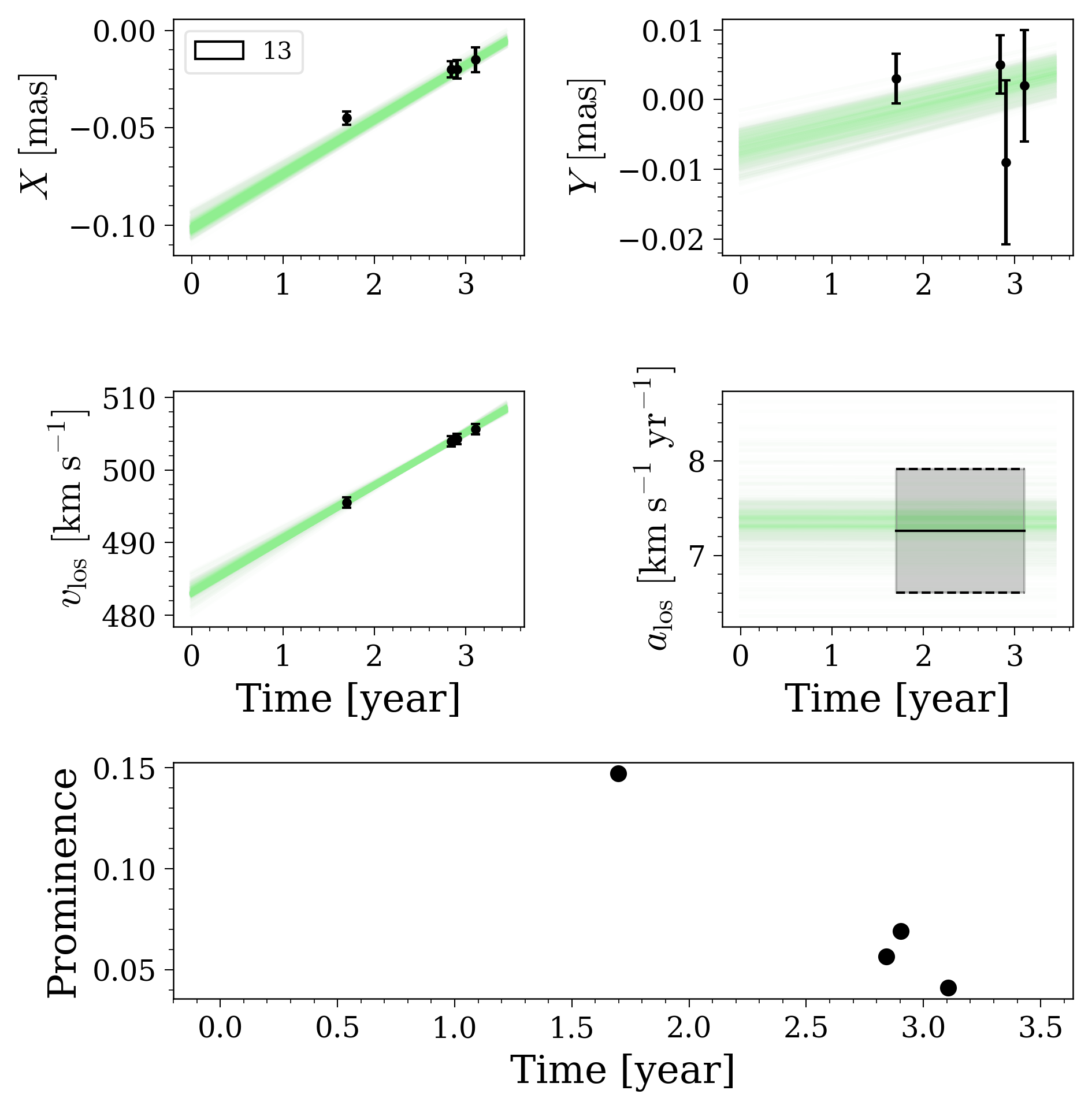} &
        \includegraphics[width=0.3\textwidth]{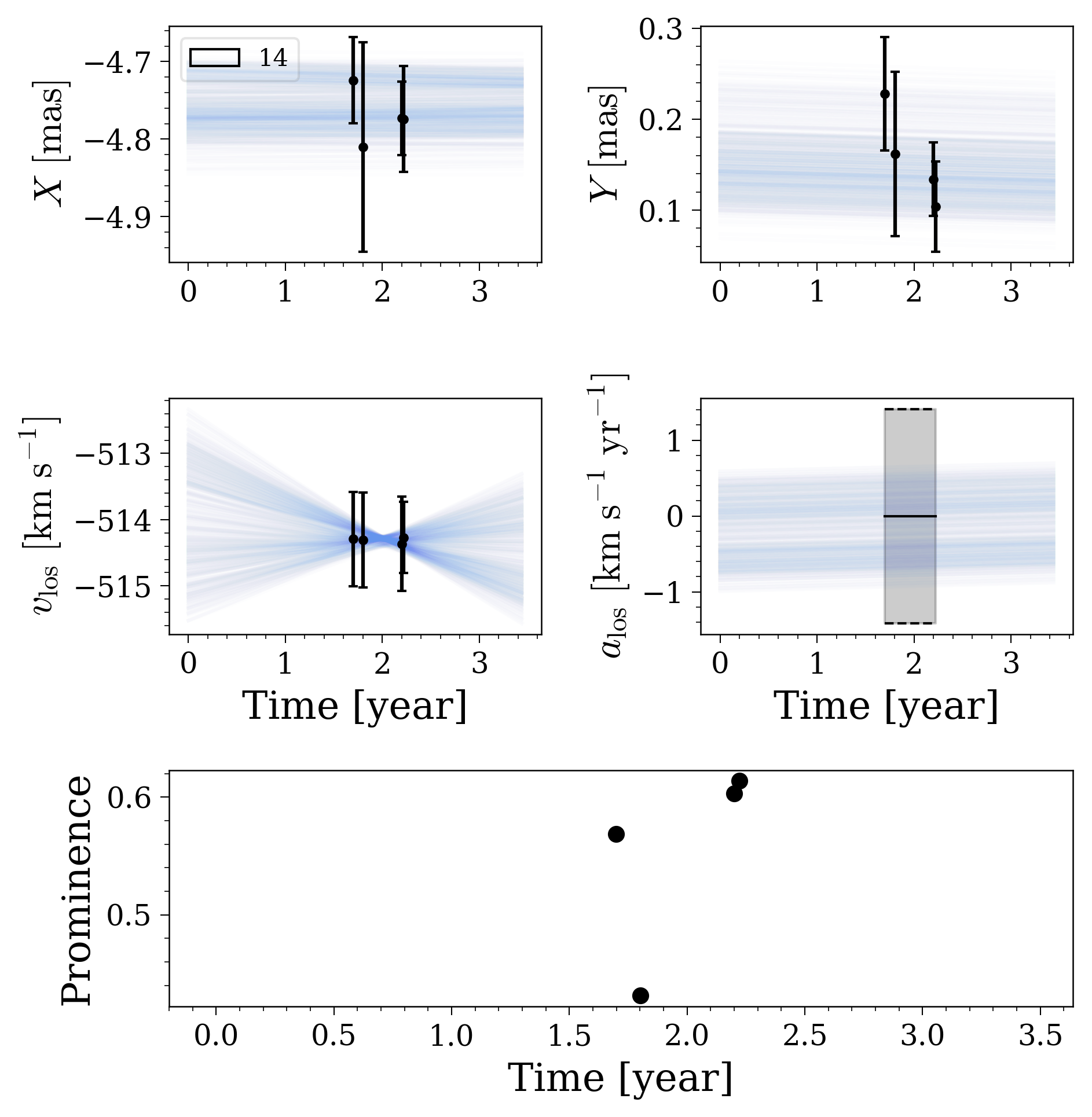} \\
    \end{tabular}
    \caption{--- \textit{continued}}
    \label{fig:trajectory_group_13_2}
\end{figure*}


\bsp	
\label{lastpage}
\end{document}